\documentclass[onecolumn,authoryear]{els-mrw} 

\usepackage{amsmath,amssymb,amsfonts,amsthm,makeidx,graphicx}
\usepackage{txfonts}
\usepackage{helvet}
                             
\usepackage{natbib}    

\newcommand\aap{A\&A}                %
\newcommand\aapr{A\&ARv}             %
\newcommand\aj{AJ}                   
\newcommand\apj{ApJ}                 %
\newcommand\apjl{ApJ}                
\newcommand\apjs{ApJS}              
\newcommand\araa{ARA\&A}             
\newcommand\mnras{MNRAS}           

\newcommand\nat{Nature}              %
\newcommand\pasp{PASP}               %
\newcommand\pasj{PASJ}               %
\newcommand{\apss}{Astrophysics and Space Science}
\newcommand{\pasa}{Publications of the Astronomical Society of Australia}
\newcommand{\rmxaa}{Revista Mexicana de Astronomia y Astrofisica}
\newcommand{\fcp}{Fundamental Cosmic Physics}

\begin{document}

\chapter{Galaxies and Black Holes in the First Billion Years}\label{title}

\author{Richard S Ellis}%

\address[1]{\orgname{University College London}, \orgdiv{Department of Physics \& Astronomy}, \orgaddress{Gower Street, London WC1E 6BT, UK}}


\maketitle

\begin{abstract}

{I present written notes from three lectures given at the 54th Saas-Fee Advanced Course of the Swiss Society of Astrophysics and Astronomy in January 2025 entitled {\it Galaxies and Black Holes in the First Billion Years as seen by the JWST}.  I focused my lectures on progress in studies of cosmic reionisation, the properties of galaxies in the reionisation era, topics related to the redshift frontier and the search for Population III stars. The lectures were given to graduate students in astrophysics and cover both pedagogical material as well as observational results from the first two and half years of JWST science operations. The pace of discovery with JWST is, of course, rapid and so my lectures discuss long-term goals, the analysis methods and their assumptions and limitations in the hope that the underlying material will retain some value in the near future. In this written version, the visual material is that presented at Saas-Fee but I have provided updates on progress from the literature up to August 2025. The material is aimed at early career researchers and should not be considered as a scholarly review of the entire JWST literature on high redshift galaxies.}

\end{abstract}

\medskip

\noindent{\bf Preface}

\medskip

It was a pleasure to give these lectures on the progress being made by the James Webb Space Telescope (JWST) in understanding the formation and early evolution of galaxies at the beautiful location of Saas-Fee in the Swiss Alps in January 2025. The timing of the meeting was perfect. Two and a half years after science operations began,  Early Release Observations, Cycles 1 and 2 General Observer and most of the Guaranteed Time programmes had been completed. The JWST archives were full of rich imaging and spectroscopic datasets and there were over 500 scientific articles to digest. The participants were at the beginning of their graduate careers and full of enthusiasm! 

The pace of discovery with JWST is, of course, rapid and I was mindful how quickly my lectures and their written record might become out of date. Therefore, in addition to summarising the state-of-the-art in our understanding of the JWST observations, I decided to focus also on the long-term goals, the analysis methods, their assumptions and limitations, hoping that the underlying material will retain some value in a few years' time. However, there is no way to avoid the fact that much of the material presented herein has a "time stamp" of January 2025. In this write up, I decided to keep the visual material I presented at Saas-Fee but have provided brief updates and references in the text of relevant articles up to August 2025.

The Saas-Fee series has provided an invaluable record of the progress in astrophysics for over 50 years. It was amusing to revisit my earlier lectures on the high redshift universe given at Saas-Fee in 2006!  I'm hoping the present notes will continue this tradition. They are intended primarily for early career researchers and should not be considered as some form of scholarly review of the entire JWST literature.

I thank the organisers, Pascal Oesch, Romain Meyer and Michaela Hirschmann for inviting me once again to Saas-Fee. Their support, guidance and hospitality has been wonderful. I also thank my many former students and postdocs who have kept me on the straight and narrow in this topic over many years. I particularly thank Yuichi Harikane, Koki Kakiichi, Romain Meyer, Kimihiko Nakajima, Tucker Jones, Guido Roberts-Borsani, Aayush Saxena, Dan Stark and Mengtao Tang who answered many of my questions on the latest results. 

Finally, and above all, it was inspiring to meet such an enthusiastic group of students at Saas-Fee. I wish them all the very best in their careers. We can certainly agree that, with JWST, the future is very bright!

\newpage
\begin{center}
{\bf \Large  Lecture 1: Cosmic Reionisation}
\end{center}

\section{Introduction}
\label{sec:1}
It is an exciting time for those of us studying early galaxies given the unprecedented opportunities now possible using NASA's James Webb Space Telescope (JWST). In these lectures, I will attempt to digest the astonishing progress made in the space of only two and half years since science operations began in July 2022. Although it is inevitable that many of the details discussed will soon become outdated, I hope by covering both the progress made and the challenges remaining, my lecture notes will still be of value, particularly to the graduate students attending this Saas-Fee course who are very fortunate to be witnessing this unique period in the history of extragalactic astronomy.

\begin{figure}[hbt!]
\center
\includegraphics[width=\textwidth]{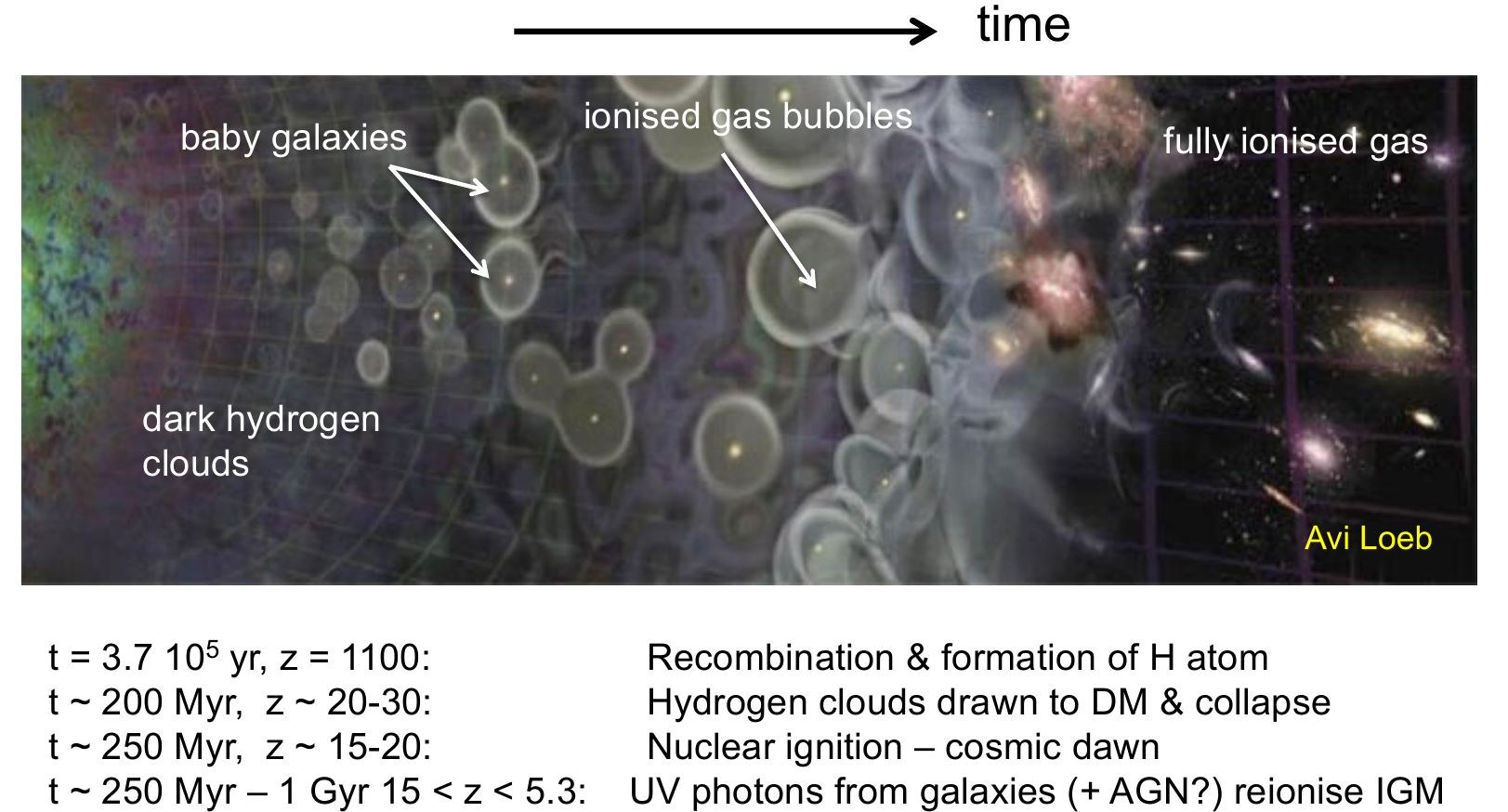}
\caption{\it Illustrative timeline of the first billion years of cosmic history. From left to right: the thermal glow from the Big Bang and last scattering surface when the hydrogen atom first formed (370,000 yrs, z=1100). Dark hydrogen clouds clump under gravity aided by dark matter becoming Jean unstable and collapse. The gas temperature rises igniting nuclear fusion and the universe is bathed in starlight (150-200 Myr, z$\simeq$20-30). Hot massive stars, free from metals, emit copious amounts of UV photons that reionise the local surroundings. Such ionised bubbles expand and become more numerous, eventually overlapping to fully ionise the intergalactic medium (1.1 Gyr, z$\simeq$5.3) (Courtesy: Avi Loeb).}
\label{fig:loeb_cartoon}
\end{figure}

In my first lecture I will cover the topic of cosmic reionisation: the period during which hydrogen in the intergalactic medium (IGM) transitioned from a neutral to ionised state. The cartoon in Figure~\ref{fig:loeb_cartoon} shows a possible timeline. At recombination, corresponding to z=1100 some 370,000 years after the Big Bang, the primordial plasma has cooled so that the hydrogen atom formed for the first time. At this point, free electrons were no longer available to scatter the thermal glow and we see this moment as the last scattering surface in the cosmic microwave background. Dark hydrogen clouds coalesced under gravity aided by non-baryonic dark matter halos which fractionated from the cosmic expansion at an earlier era. The clouds became Jeans unstable and collapsed raising the temperature sufficient to ignite nuclear burning from the first stars, possibly by z$\simeq$15-30 some 150-200 Myr later - an epoch often romantically referred to as `cosmic dawn.' Devoid of heavy elements at this stage, these stars are likely to be more massive than those in the present universe due to the absence of metal cooling at the time of collapse. Additionally there will be no metal line blanketing in their stellar atmospheres. For both reasons these early stars emitted copious amounts of ultraviolet photons which created local ionised gas bubbles. As time progressed the bubbles expanded and became more numerous, eventually overlapping until the entire IGM was fully ionised by z$\simeq$5.3, just over 1 Gyr after the Big Bang. Of course the cartoon is a possible scenario. It assumes cosmic reionisation was driven mostly by star-forming galaxies. But it does emphasise the importance of this reionisation era by positing an association with the birth of galaxies and the commencement of nuclear processing in stellar nuclei which, of course, eventually led to our own existence. 

The big questions we aim to address in studying this era with JWST include: when did reionisation begin and end; which sources were responsible and can we recognise a primordial stellar system first emerging from darkness. The last topic is reserved for Lecture 3.

\section{Quasar Absorption and the End of Reionisation}
\label{sec:2}
Let's begin with some important astronomical history. Astronomers have known the intergalactic medium is fully ionised today since 1965. When luminous quasars were discovered in the early 1960s, redshift records were rapidly broken. By 1965, using the Palomar telescope, Martin Schmidt was able to secure a spectrum of the radio source 3C9 at z=2.0 providing, for the first time, a glimpse at the rest-frame ultraviolet continuum and the prominent Lyman $\alpha$ emission line at 1216  \AA\ . The absence of any foreground absorption blueward of this emission was interpreted by \citet{Gunn1965} as evidence for a low value for the density of neutral hydrogen. They speculated that at some point in earlier cosmic history the IGM had been fully ionised. It wasn't until 2001 that astronomers secured their first indication of when this earlier period of cosmic reionisation occurred. The Sloan Digital Sky Survey (SDSS) was used to locate the first meaningful sample of quasars at a redshift z$\simeq$6 (Becker et al 2001). Contrary to the situation at $z\simeq$2, the spectra showed significant absorption blueward of Lyman $\alpha$ emission indicative of an increased opacity from neutral hydrogen. This was an early indication that we had reached a look-back time sufficient to detect neutral gas in the IGM and could begin probing the so-called reionisation era.

\begin{figure}
\center
\includegraphics[width=0.38\textwidth]{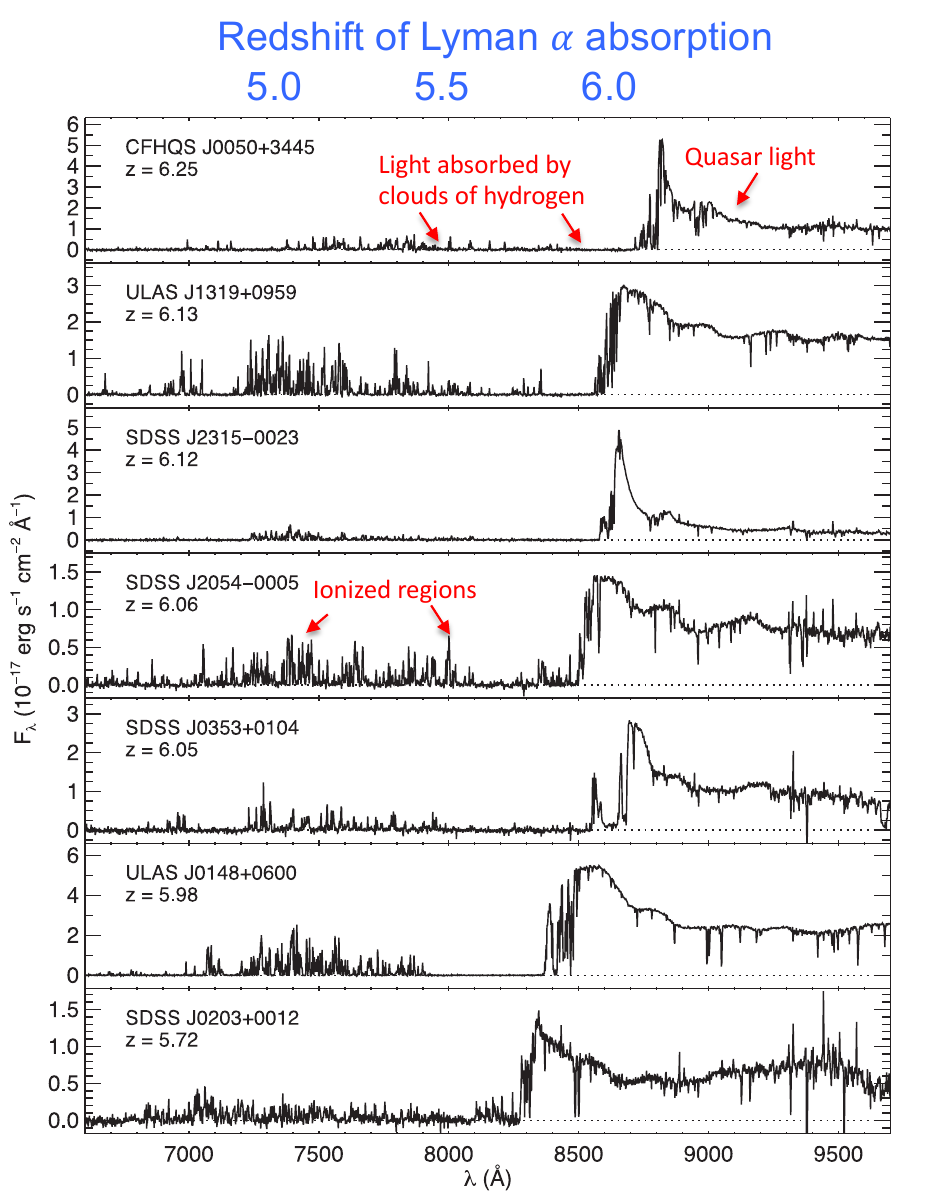}
\includegraphics[width=0.6\textwidth]{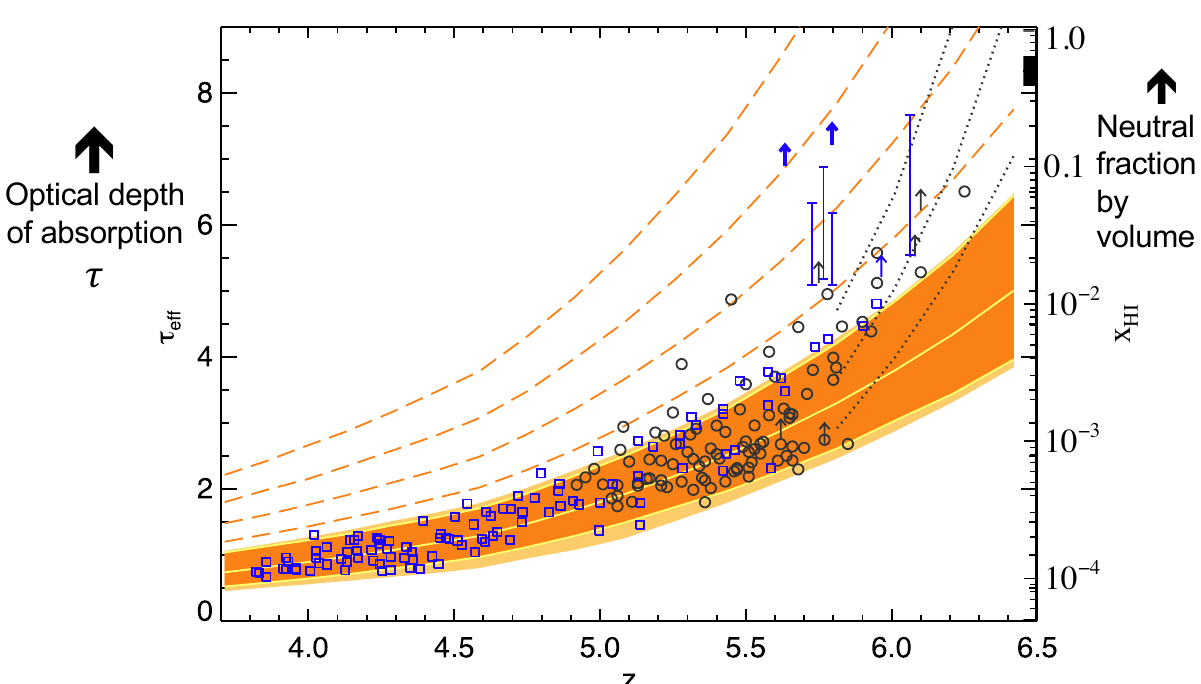}
\caption{\it Quasar absorption spectroscopy determines the end of cosmic reionisation: (Left) Selection of spectra of z$\geq$6 quasars. The depth of absorption shortward of Lyman $\alpha$ emission line traces the line of sight, redshift-dependent, opacity of neutral hydrogen (HI). Occasional tranmission spikes indicate localised ionised regions. (Right) The fractional volume density of neutral hydrogen $x_{HI}$ derived from the optical depth $\tau$ of Lyman $\alpha$ absorption for a large sample of quasar spectra. The noticeable upturn at redshifts $z>$5.5 corresponds to entering the partially neutral reionisation era. The scatter at a given redshift is an important indication of the patchiness of the reionisation process \citep{Bosman2018}}.
\label{fig:qso_absorption}
\end{figure}

Absorption line spectroscopy has given us the most precise indication of when reionisation ended. The left panel of Figure~\ref{fig:qso_absorption} shows some recent spectra from a sample of over 70 $z\simeq$6 quasars. We can quantify the Lyman $\alpha$ opacity via an optical depth $\tau$ using the observed continuum flux $I(\lambda)$ viz. 

\begin{equation}
I(\lambda) = I_0 \, exp(-\tau)
\end{equation}

where the unattenuated continuum flux $I_0$ is determined from an average rest-frame quasar spectrum at lower redshift when there is minimal hydrogen absorption. For low column densities of neutral hydrogen $n_{HI}$, the opacity is related via

\begin{equation}
    \tau = \frac{\pi e^2}{m_e c} f_{\alpha}\lambda_{\alpha}H(z)^{-1}n_{HI}
\end{equation}

where $f$ is the oscillator strength of Lyman $\alpha$ and H(z) is the Hubble constant.
Each quasar spectrum enables a measurement of $\tau(z)$ over a range in redshift, corresponding to the fluctuating signal in the Lyman $\alpha$ forest. Measures along different quasar sightlines then trace the patchiness of the signal. To trace the reionisation history, it is usual to quote the neutral gas fraction by volume $x_{HI} = n_{HI}/n_H$. The right panel of Figure~\ref{fig:qso_absorption} shows a recent compilation of opacities and neutral fractions from \cite{Bosman2018} where the scatter among different quasars at a given absorption redshift is evident. The most important feature, however, is a distinct upturn in the opacity at $z >5.5$ consistent with entering the partially neutral era.

Such spectroscopic analyses have been very productive in charting the end of reionisation because, over the redshift range $5 < z <6$, they provide direct measures of the neutral fraction $x_{HI}$. However, the Lyman $\alpha$ opacity method cannot easily be applied at higher redshift since only a modest neutral fraction ($x_{HI}\sim 10^{-3}$) saturates the absorption and eliminates any measured flux. Other methods, such as exploiting the effect of damped absorption on the Lyman $\alpha$ emission profile, are discussed in the reviews by \citet{Fan2006} and \cite{Fan2023}. Regardless of the methods used, a basic limitation however is the scarcity of quasars beyond a redshift $z\simeq$7.

Recently, through the remarkable efficiency of JWST's spectrographs, it has become observationally feasible to study the Lyman $\alpha$ forest at z$\simeq$5-6 in the spectra of {\it background galaxies} \citep{Meyer2025, Umeda2025}. This important development will enable many more IGM sightlines to be explored, and to higher redshift given the increased statistics will also enable studies of the Lyman $\beta$ absorption which saturates earlier in cosmic time than Lyman $\alpha$.

\section{Cosmic Microwave Background Constraints}
\label{sec:3}

NASA's WMAP and ESA's Planck satellites have provided valuable constraints on the duration and median redshift of reionisation through polarisation measures of the cosmic microwave background (CMB). Readers will be familiar with the angular power spectrum of temperature fluctuations $TT(l)$ where $l$ is the angular multipole moment at the surface of last scattering. This anisotropic radiation field suffers electron (Thomson) scattering by the integrated column of electrons along the line of sight starting from the commencement of reionisation down to the present day. The polarisation signal is commonly referred to as the E- or gradient mode and its angular power spectrum is $EE(l)$.

\begin{figure}
\center
\includegraphics[width=\textwidth]{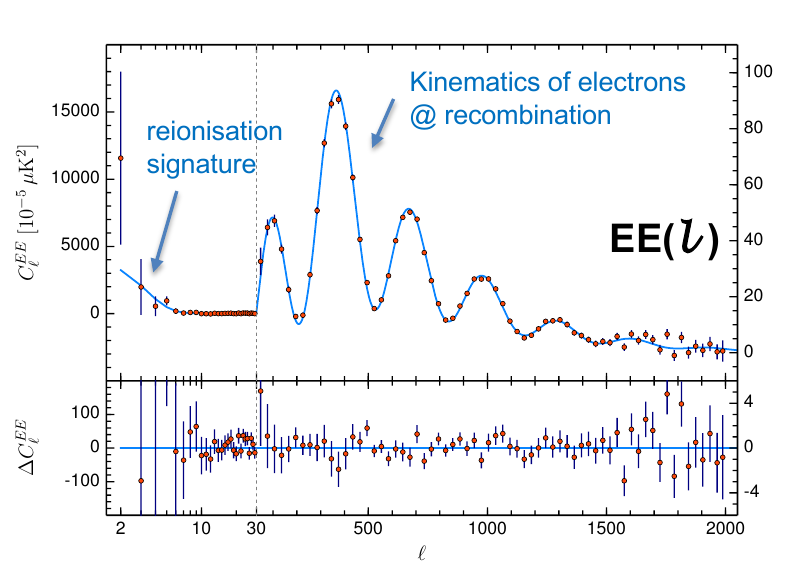}
\caption{\it Angular power spectrum of the polarisation or E-mode of the cosmic microwave background. The signal arises from Thomson (electron) scattering along the partial and fully-ionised intergalactic medium down to the present epoch. The optical depth $\tau$ is determined primarily from the upturn at large angular scales or low multipoles \citep{Planck2020}.}
\label{fig:cmb_eel}
\end{figure}

Reionisation also damps $TT(l)$ by an amount that is degenerate with its intrinsic amplitude. Fortunately this degeneracy can be broken since reionisation also produces an unique $EE(l)$ signal on large angular scales ($>$30 arcmin, $l<$200, Figure~\ref{fig:cmb_eel}). In practice cosmologists use the combined TT, EE and TE correlations to constrain the {\it optical depth of Thomson scattering} $\tau$ which was first detected by WMAP and whose value was considerably refined by Planck. Formally 

\begin{equation}
   \tau = <n_{H}>c \sigma_T \int_{0}^{z} x_{e}(z) (1+z)^2 H(z)^{-1}dz 
\end{equation}

so a larger $\tau$ implies an earlier period of reionisation. As an integrated measure $\tau$ constrains the median redshift $z_{mid}$ and duration of reionisation, but not the redshift-dependent neutral fraction $x_{HI}(z)$.

The first WMAP results in 2003 gave $\tau$=0.117$\pm$0.055 suggesting the start of reionisation might even be beyond reach of JWST. However, these results were significantly affected by Galactic foreground emission and over the next decade both the WMAP mission (2003-2013) and Planck (2009-2013) consistently lowered their estimates of $\tau$. The final Planck value $\tau$=0.057$\pm$0.007 \citep{Planck2020} implies a mid point of the reionisation era at $z_{mid}$=7.7$\pm$0.7, with parametric fits to the EE, TT and TE data that suggest reionisation began at $z\simeq$10-12 and ended at $z\simeq$6.  Subsequent reconsiderations of the polarisation calibrations indicate a modest {\it increase} to $\tau$=0.063 $\pm$0.005 \citep{deBelsunce2021}.

Regarding the beginning of cosmic reionisation and, by possible implication, the timing of ``cosmic dawn", the Planck team were fairly confident ``disfavouring any major contribution to the ionised fraction that could form as early as $z>$15". However, as mentioned, these conclusions were based on parametric models. Subsequent analyses in the context of $\Lambda\,CDM$ cosmology taking into account the Planck $\tau$ plus secondary anisotropies induced by bulk motions of electrons (the so-called "kinetic Sunyaev-Zeldovich effect") indicated that earlier ionised regions could exist to redshifts $z\simeq$16 and beyond \citep{Greig2017}. Although we are unlikely to learn substantially more on the timing of reionisation from CMB studies until the launch of the Japanese Litebird mission (currently $\simeq$2032), Planck clearly played a very significant role in demonstrating that the reionisation era can be explored in considerable detail by JWST.

\section{Probes of Lyman alpha emission}
\label{sec:4}

Returning to constraints relating to the end of reionisation, while recognising the limitations arising from saturation of Lyman $\alpha$ absorption in the spectra of quasars, in the late 2000s ground-based astronomers turned to probes based on {\it Lyman $\alpha$ emission}. Hot gas in H II regions heated by young stars shines via recombination in Lyman $\alpha$ emission \citep{Dijkstra2014}. As a resonant transition of hydrogen from $n$=2 to 1, the line is readily scattered by neutral hydrogen. Consequently if a galaxy lies in a neutral IGM, the line intensity will be significantly attenuated. Alternatively, if the galaxies lies in an ionised bubble, Lyman $\alpha$ photons can travel freely without hindrance through the bubble and will be redshifted out of resonance by the cosmic expansion at the bubble surface. A straightforward observational probe is to measure the fraction of galaxies at a given redshift that reveal Lyman $\alpha$ emission, $X_{Ly\alpha}(z)$. As we approach the neutral-dominated era, even though the intrinsic line emission will be present, $X_{Ly\alpha}(z)$ should decrease to zero. 

To apply this method, galaxies were selected in photometric redshift slices using the Lyman break technique \citep{Steidel1995} and the fraction $X_{Ly\alpha}$ showing line emission above a limiting equivalent width determined. Keck \citep{Stark2011, Schenker2014} and VLT \citep{Pentericci2011,Pentericci2014} surveys indicated an increase in the fraction from $z\simeq$4 to 6 (due to increased star formation and/or reduced dust attenuation) but revealed a marked decline in a short time interval earlier at $z>6$. With much uncertainty, the data suggested neutral fractions as high as $X_{HI}\simeq$0.4 at $z$=7 and $X_{HI}\simeq$0.65 at z=8. 

However, this Lyman $\alpha$ fraction method has several uncertainties. Using ground-based telescopes, the spectroscopic redshift of a galaxy often cannot be determined if there is no Lyman $\alpha$ emission line. Such non-emitting sources could therefore be lower redshift interlopers. Their inclusion in sample would act to incorrectly reduce the inferred value of $X_{Ly\alpha}$ mimicking a more neutral IGM. Moreover if a galaxy has outflowing hot gas its relative velocity will reduce the effect of resonant scattering making Lyman $\alpha$ visible even in a neutral IGM. To allow for such outflows would require a systemic redshift derived from another (usually weaker) emission line usually beyond reach of ground-based telescopes. Finally, early results were based on observations in a few independent sightlines and thus suffered from cosmic variance (see \citet{Treu2013}).

\begin{figure}
\center
\includegraphics[width=0.47\textwidth]{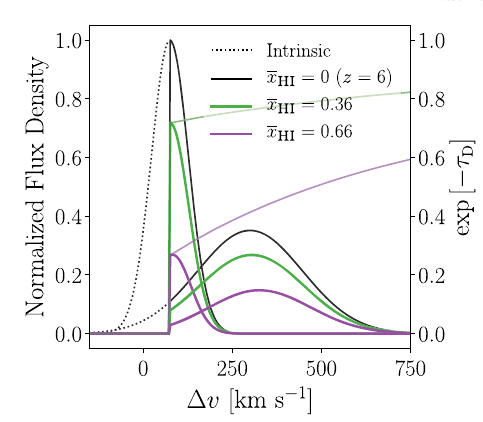}
\includegraphics[width=0.43\textwidth]{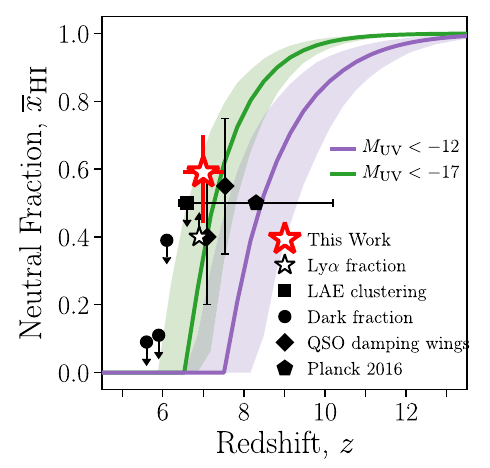}
\caption{\it The complexity of interpreting Lyman $\alpha$ emission. (Left) The intrinsic Gaussian line profile can be affected by HI scattering in any of the ISM, CSM and IGM such that the blueward wing is eliminated and the overall line flux diminished. However, if the emitting gas has a velocity offset $\Delta\,v$ , this can reduce the effect and render an otherwise extinguished line observable. (Right) By assuming the correlation between Ly$\alpha$ velocity and luminosity observed at lower redshift, pre-JWST data on 68 $z\simeq$7 galaxies was used to infer a neutral fraction at z=7 (red star), well into the neutral era \citep{Mason2018}.}
\label{fig:mason}
\end{figure}

To understand the complexity of deriving the IGM neutral fraction from the visibility of Lyman $\alpha$ emission, the various processes that affect the intensity and profile of Lyman $\alpha$ emission are shown in Figure~\ref{fig:mason} \citep{Mason2018}. Even in a fully ionised IGM, Lyman $\alpha$ photons can be scattered and attenuated by neutral gas in the interstellar medium (ISM) or circumgalactic medium (CGM) of the host galaxy. Photons propagating directly to the observer may lead to the blueward wing of the intrinsic line being eliminated. Likewise the peak intensity would be reduced in proportion to the combined ISM+CGM+IGM neutral components. In contrast, IGM attentuation will be {\it diminished} if the Lyman $\alpha$ emission has a significant outflowing velocity $\Delta\,v$ relative to the local IGM. Finally, line emission will be affected by any dust present within the galaxy. 

\citet{Mason2018} attempted to model these effects by evaluating the change in the equivalent width distribution of Lyman $\alpha$ emission across the transition in the ionisation state between $z\simeq$7 and z$\simeq$6. The authors also adopted an empirical correlation between the UV luminosity of a galaxy and its Lyman $\alpha$ velocity offset $\Delta\,v$, using systemic redshifts from other lines in the spectra of a large sample at $z\simeq$2-3. Finally, to estimate the ISM neutral hydrogen content $n_{HI}$, they adopted a correlation with halo mass from simulations. As the state of the art calculation prior to JWST, they applied the model to 68 $z\simeq$7 galaxies of which 12 had ground-based spectroscopic detections of Lyman $\alpha$, and derived $x_{HI}$ (z=7) = 0.59 $\pm$ 0.13 (Figure~\ref{fig:mason}). The result was the most convincing estimate of $x_{HI}$ in the reionisation era using Ly$\alpha$ emission prior to the arrival of JWST.

JWST has greatly improved the prospects of tracing neutral and ionised gas in the reionisation era via its extended infrared wavelength coverage compared to both ground-based telescopes and HST. The most significant feature is spectroscopic access with NIRSpec to other hydrogen lines such as H$\alpha$ (to $z\simeq$6.6) and H$\beta$ ($z\simeq$9.2). As the Balmer series lines are unaffected by resonant scattering, it becomes possible to estimate the intrinsic Lyman $\alpha$ flux. Assuming Case B recombination for a star-forming region with a typical electron density $n_e$= 100 cm$^{-3}$ and temperature $T_e=10^4$ K, Ly$\alpha$/H$\alpha$ = 8.2 thus the escape fraction of Lyman alpha $f_{esc}$(Ly$\alpha$) = L(Ly$\alpha$)/8.2*L(H$\alpha$). Similar calculations can be made at higher redshift using the combination of H$\beta$ and Lyman $\alpha$. 

\begin{figure}
\center
\includegraphics[width=\textwidth]{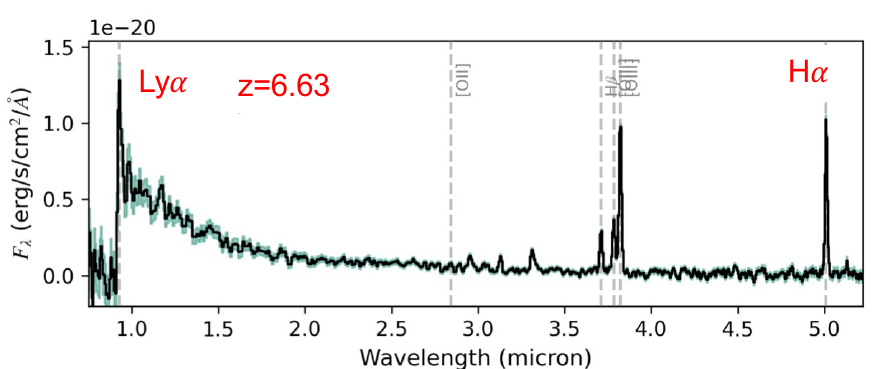}
\smallskip
\includegraphics[width=\textwidth]{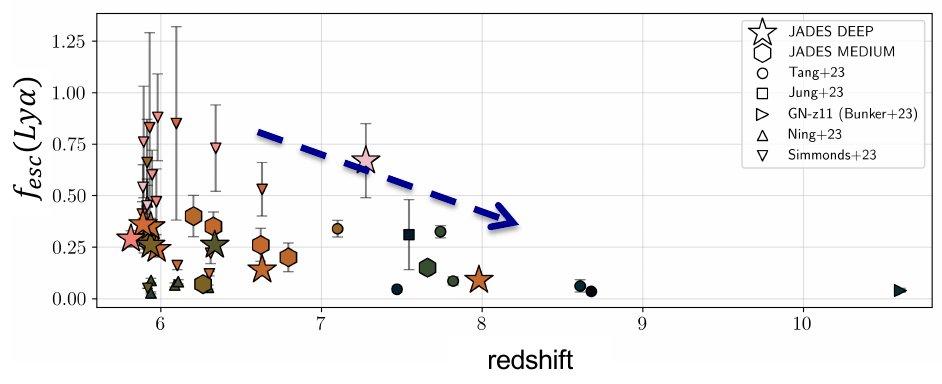}
\caption{\it JWST has significantly improved the prospects for constraining the IGM neutrality in the reionisation era using line emission measures. (Top) Spectrum of a z=6.63 galaxy revealing both Ly$\alpha$ and H$\alpha$ which, assuming recombination physics, permits determining the escape fraction of Ly$\alpha$. (Bottom) Early confirmation of the declining visibility of Ly$\alpha$ with redshift well into the reionisation era \citep{Saxena2024}.}
\label{fig:saxena}
\end{figure}

The two panels in Figure~\ref{fig:saxena} show an early derivation of the declining escape fraction of Lyman $\alpha$ with redshift from \citet{Saxena2024} (see also \citet{Napolitano2023}). While these trends give much credence to the tentative deductions made prior to JWST, redshift-dependent selection effects may still be present. Clearly only Lyman $\alpha$ emitting galaxies can be used in such analyses and these may lie in regions of lower than average $x_{HI}$.

The more recent JWST analysis undertaken by \citet{Tang2024} (see a similar analysis by \citet{Kageura2025}) follows the initiative taken by Mason et al discussed earlier by making reference to a comparison sample of Lyman $\alpha$ demographics in the fully-ionised era. The comparison sample utilises data from the JADES, FRESCO and Keck/VLT samples and comprises 79 spectra that reveal both H$\alpha$ and Ly$\alpha$. Probability distributions of equivalent widths, escape fractions and velocity offsets are constructed as a function of luminosity and UV continuum slope. The reionisation sample then comprises 210 6.5$<z<13$ spectra from taken JADES, the Early Release Science (ERS) campaigns and various General Observer programmes. 33 of these reveal Lyman alpha emission over $6.5 < z< 10.5$

\begin{figure}
\center
\includegraphics[width=0.5\textwidth]{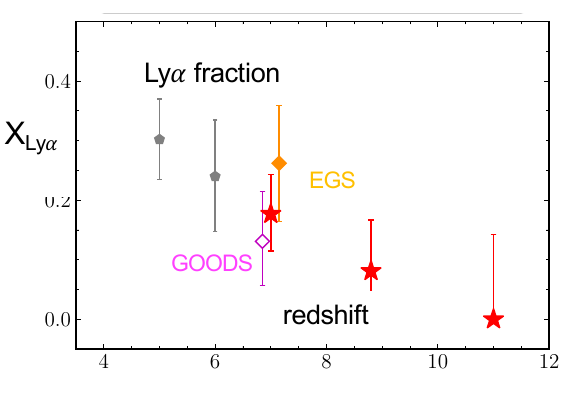}
\includegraphics[width=0.43\textwidth]{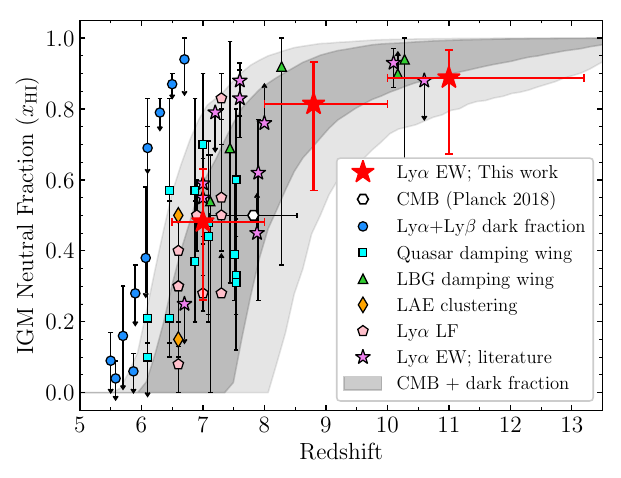}
\caption{\it (Left) Declining visibility of Ly$\alpha$ with redshift from over 200 $z>6.5$ JWST spectra and a controlled comparison sample in the fully-ionised era. (Right) Inferred history of the neutral fraction from Ly$\alpha$ is shown by the red points and represents the first self-consistent determination of $x_{HI}$ using one method through the bulk of the reionisation era \citep{Tang2024}.}
\label{fig:tang}
\end{figure}

This approach enables comparisons of the Lyman $\alpha$ demographics over a range of redshifts, taking into account biases that would otherwise occur due to observational limitations. The analysis reveals a convincing decline in the Lyman $\alpha$ emission fraction to $z\simeq10$ and the first meaningful estimate of the evolving neutral fraction $x_{HI}(z)$ using one method through most of the reionisation era (Figure~\ref{fig:tang}).

\begin{figure}
\center
\includegraphics[width=\textwidth]{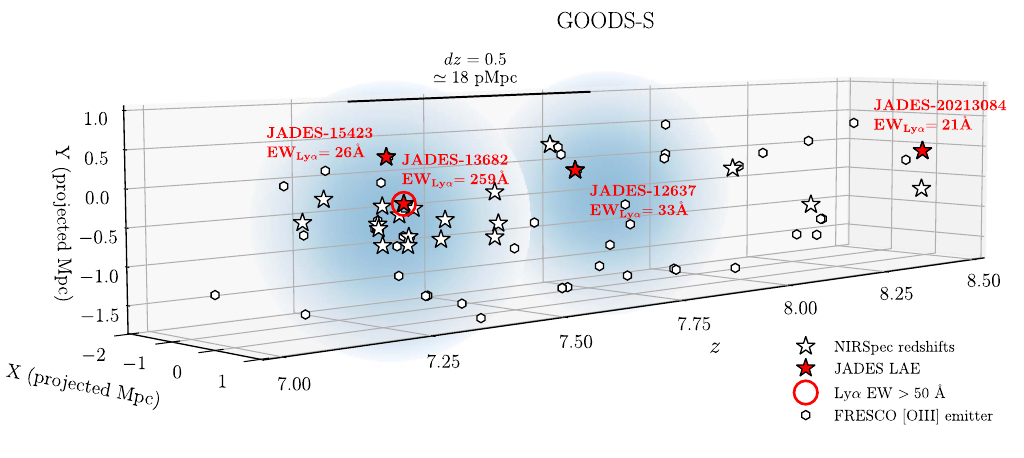}
\caption{\it Spatial distribution of Ly$\alpha$ emitters (red) with respect to non-emitters in GOODS-S from the analysis of \citet{Tang2024}. Whereas many emitters lie in clustered regions suggesting such associations may be responsible for the ionised regions (indicated in blue), there are also isolated examples.}
\label{fig:tang_3d}
\end{figure}

Tang et al also provide a glimpse of the future possibilities of IGM tomography. It is reasonable to assume that the escape of Lyman $\alpha$ can be used as a proxy for escaping radiation capable of ionising the IGM, i.e. the Lyman continuum (see next section) and hence for tracing ionised bubbles. They use the spatial distribution of [O III] emitters from FRESCO's NIRCam grism survey to trace overdensities in both GOODS fields. In the resulting 3-D map of the sample (Figure~\ref{fig:tang_3d}) some Lyman $\alpha$ emitters are found in clustered regions whose members may have collectively ionised the local region. Others are isolated examples which may be solely capable of ionising their surroundings. This work highlights by direct 3-D mapping, the role that galaxies play in cosmic reionisation (see \citet{Chen2025} for an update of this tomographic approach).

In summary, both Lyman $\alpha$ absorption and emission techniques have been useful in delineating the history of the redshift-dependent neutral fraction $x_{HI}(z)$ over 5$ < z < $10 consistent with the coarser constraints from the CMB. We can also see evidence for the patchiness of the process and ultimately we can hope to be able to connect ionised bubbles with the sources responsible - a question we will return to later. Perhaps the most exciting development from JWST is the emerging evidence from Lyman $\alpha$ tomography \citep{Tang2024, Chen2025} that galaxies are likely responsible for creating local ionised regions and thus play a dominant role in the overall process.

\section{The Reionisation Budget}
\label{sec:6}

In order to determine which sources governed the reionisation process we first need to understand the balance between the rate of ionising photons $\dot{n}_{ion}$ and the recombination time for neutral hydrogen $t_{rec}$. If the volume-averaged fraction of ionised hydrogen at a given time is $Q_{HII}$, then its evolutionary change with time is given by e.g. \citep{Madau2000, Miralda-Escude2000} :

\begin{equation}
    \dot{Q}_{HII} = \frac{\dot{n}_{ion}}{<n_H>} - \frac{Q_{HII}}{t_{rec}}
\end{equation}

where the recombination time depends on the IGM temperature $T_{IGM}$ through a recombination coefficient $\alpha(T_{IGM})$ and a dimensionless clumping factor 

\begin{equation}
C = \frac{<n^2>}{<n>^2}
\end{equation}

For Case A recombination, corresponding to the optically thick case

\begin{equation}
    t_{rec} = [n_H \, \alpha(T_{IGM}) \, C]^{-1}
\end{equation}

Simulations of large scale structure generally indicate the clumping factor is in the range $C \simeq$ 1-6 and not a crucial ingredient in changing the transition from a neutral ($Q_{HII}=0$) IGM to one that is fully ionised ($Q_{HII}=1$).If so, the pace of reionisation and the dominant sources are determined by understanding the ionisation rate $\dot{n}_{ion}$.

In the absence of a detailed picture, it has become customary to break 
the problem into three manageable components \citep{Robertson2010} : 

\begin{equation}
    \dot{n}_{ion} = \rho_{UV} \xi_{ion} f_{esc}
\end{equation}

\begin{figure}
\center
\includegraphics[width=0.45\textwidth]{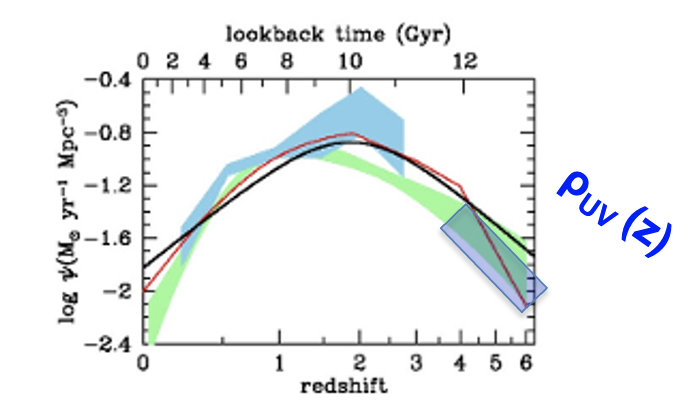}
\includegraphics[width=0.5\textwidth]{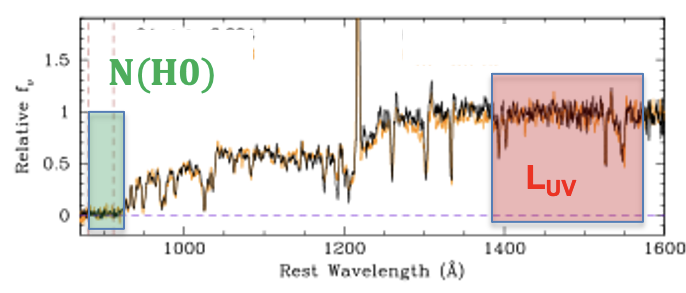}
\includegraphics[width=0.7\textwidth]{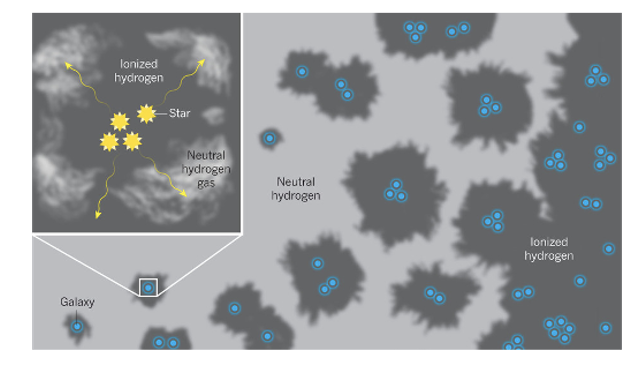}
\caption{\it A simple illustration of the three components of the reionisation rate $\dot{n}_{ion}$ (Left:) $\rho_{UV}$ is the redshift-dependent integrated UV luminosity density. (Right:) $\xi_{ion}$ is the intrinsic production rate of Lyman continuum photons $N(H^0)$ per UV luminosity $L_{UV}$. (Lower:) The escape fraction $f_{esc}$ is the fraction of Lyman continuum photons that can reach the IGM (Courtesy: Mark Dickinson, Chuck Steidel \& Dawn Erb).}
\label{fig:budget}
\end{figure}

\medskip
\noindent See Figure~\ref{fig:budget} for an illustration.

\medskip
\noindent Taking each component in turn:
\smallskip

\begin{itemize}
    
\item $\rho_{UV}$ is the integrated abundance (per comoving volume) of the chosen class of sources, across all luminosities. This is derived by integrating the associated luminosity function down to some fiducial absolute magnitude, often $M_{UV}$=-17.
\smallskip

\item $\xi_{ion}$ is the intrinsic production rate of photons capable of ionising hydrogen (i.e. with energies $ > $13.6 eV). It is defined as the number of such photons per second $N(H^0)$ divided by the UV luminosity in ergs at a longer wavelength, traditionally $\simeq$1500 \AA\ . It is usually presented as a logarithmic quantity in units of Hz erg$^{-1}$. This quantity can be estimated by diagnosing the stellar population from fits to the spectral energy distribution, or using nebular emission lines to gauge the hardness of the radiation field. However, several assumptions must be made regarding the nature of the stellar population e.g. whether stellar binaries make a major contribution to the radiation field and the nature of the stellar initial mass function.
\smallskip

\item $f_{esc}$ is the fraction of ionising photons that manage to escape the source into the IGM, often quoted as a percentage. This is very challenging to measure beyond a redshift $z\simeq4$ due to line of sight hydrogen absorption. Direct measures of the Lyman continuum ``leaking radiation" are only possible at redshifts $z \leq 3$ which has led to searches for ``analogues" of reionisation-era galaxies in order to make progress. 

\smallskip
Before discussing the latest JWST results, it's worth issuing some cautionary remarks on this simplistic view of the overall ionisation rate. Star-forming galaxies represent an enormously diverse population in terms of size, luminosity and stellar activity. It would seem naive to apply single values of $\xi_{ion}$ or $f_{esc}$ to an entire population, as was the standard practice prior to JWST. Moreover, simulations suggest that both parameters may increase with redshift. Lower metallicities will increase $\xi_{ion}$ and radiation pressure from energetic star formation in compact sources at higher redshift will create porous channels in the ISM increasing $f_{esc}$.
Finally, although rarer, AGN may contaminate the galaxy population and make a significant contribution to the ionisation process, particularly at later times.

\end{itemize}

\subsection{UV luminosity density}
\label{subsec: 3.1}

Let us discuss what is known about each of these 3 ingredients from both pre-JWST and JWST data.
\medskip

The UV luminosity density $\rho_{UV}$ is derived by integrating the redshift-dependent UV luminosity function. This is a simpler task than estimating the physically more important integrated star formation rate density $\rho_{SFR}$ commonly quoted \citep{Madau2014} because it avoids understanding the SFR-UV calibration which depends on metallicity, star formation histories, the initial mass function and contributions from dust-obscured sources.

The galaxy luminosity function (LF) is traditionally expressed in terms of the Schechter function \citep{Schechter1976}

\begin{equation}
    \Phi(L) \frac{dL}{L^*} = \Phi^*(\frac{L}{L^*})^{-\alpha}exp(-\frac{L}{L^*}) \frac{dL}{L^*}
\end{equation}

The seemingly complex function is simply a power law with a faint end slope of $\alpha$ multiplied by an exponential for the portion more luminous than a characteristic luminosity, $L^{\ast}$. The integrated luminosity density is then:

\begin{equation}
    \rho_L = \int \Phi(L)L dL = \Phi^* L^* \Gamma(\alpha + 2)
\end{equation}

For galaxies, the observed UV LF has a faint end slope that steepens from $\alpha\simeq$-2 at intermediate redshift to -2.4 in the reionisation era \citep{Bouwens2022}. Not only do intrinsically faint galaxies dominate the integral but, for $\alpha\leq$-2, the integral above diverges unless the faint end of the LF turns down at some point as theorists predict could be the case due to reionisation feedback effects. 

\begin{figure}
\center
\includegraphics[width=\textwidth]{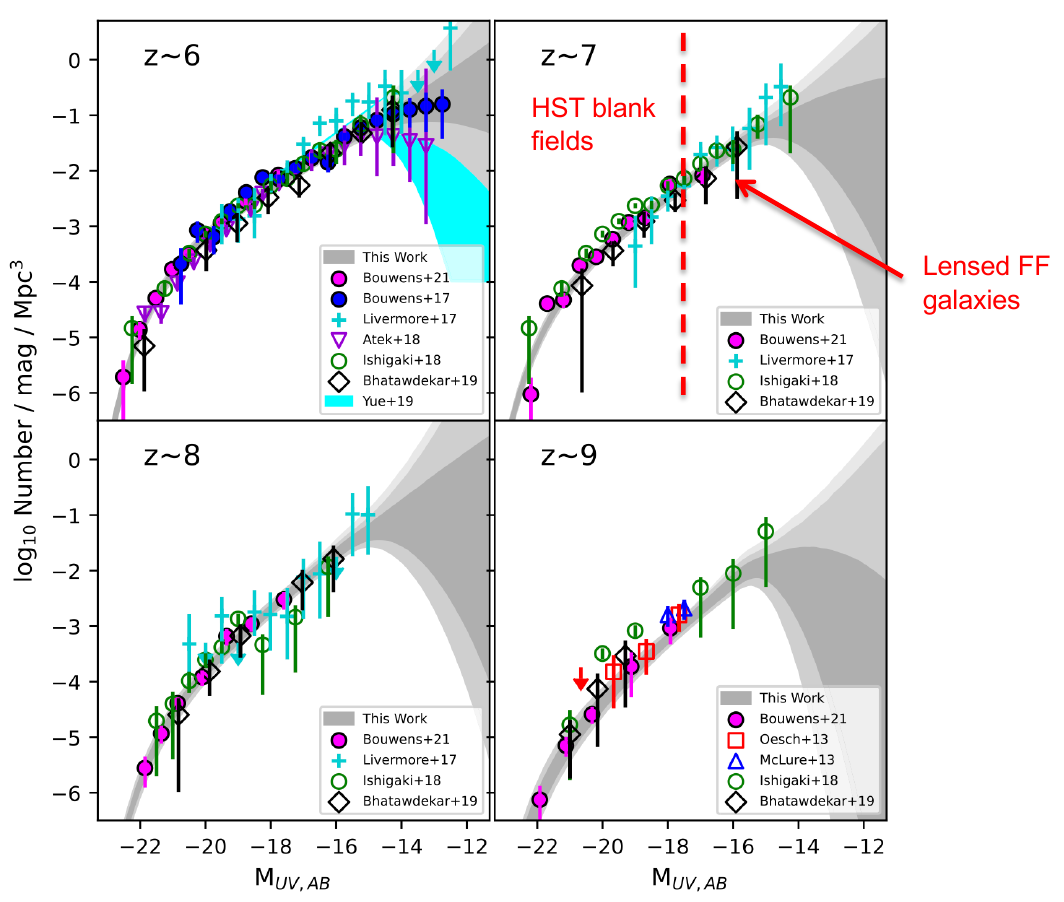}
\caption{\it Redshift-dependent UV luminosity function of galaxies combining HST's deep fields and the six lensing Frontier Field Clusters \citep{Bouwens2022}. The red dashed line indicates the luminosity limit of the deep fields, fainter than which the contribution analysed via imaging through lensing clusters is crucial. Since the Schechter faint end slope $\alpha$ [Eqn 8] is steeper than -2 in this redshift range, the integrated $\rho_{UV}$ would be divergent so a faint limit to the integration must be chosen. The grey shaded areas illustrates there is not yet any statistically-significant evidence of a decline to $M_{UV}=-13$.}
\label{fig:uvlf}
\end{figure}

Perhaps surprisingly, the current best estimate of $\rho_{UV}(z)$ still comes from HST observations, at least to redshifts $z\simeq$9. This is because HST has observed galaxies in several deep fields as well as through the six Frontier Field lensing clusters, thereby mitigating cosmic variance effects. However, JWST does probe the LF to higher redshift. Figure~\ref{fig:uvlf} shows the most recent LF measures over $6 < z< 9$ and highlights the important contribution from lensed sources as well as the constraints on any possible turn-down in the LF down to a luminosity of $M_{UV}$=-13 \citep{Bouwens2022}. To this limit the UV luminosity density is known to better than 20\% (making it the most accurately determined of the three components introduced above!)

\subsection{Ionising Productivity $\xi_{ion}$}

Ionising radiation could be thermally produced from hot, young stars, or non-thermal radiation from active galactic nuclei (AGN). In the past, $\xi_{ion}$ was mostly determined by using stellar population synthesis (SPS) models to fit to the spectral energy distribution (SED) compiled from broad-band photometry or even from more basic data such as the UV continuum slope calibrated by such models.

The advent of JWST has led to an increased awareness of using spectroscopic data and two techniques are now in common use.
\medskip

\noindent{\bf Balmer lines and recombination physics:} For an ionisation-bounded nebula, the ionisation and recombination rates are equal. The ionisation rate can therefore be derived from the dust-corrected luminosity of H$\alpha$ emission, via 

\begin{equation}
    L(H\alpha) = C(T,n_e) N(H^0)     
\end{equation}

Unfortunately, this leads to a degeneracy with the escape fraction $f_{esc}$, such that 

\begin{equation}
    \xi_{ion} = \frac{N(H^0)}{L_{UV}} (1 - f_{esc})
\end{equation}

An impressive example of this method is the analysis of 670 1.4$< z < 2.6$ Lyman break galaxies (LBGs) in the Keck MOSDEF survey \citep{Shivaei2018}. These authors determined $N(H^0)$ from Balmer line measures of H$\alpha$ and H$\beta$ and $L_{UV}$ from SED fits to 3D-HST photometry. Depending on the adopted UV extinction law, they find 
\smallskip

\begin{figure}
\center
\includegraphics[width=\textwidth]{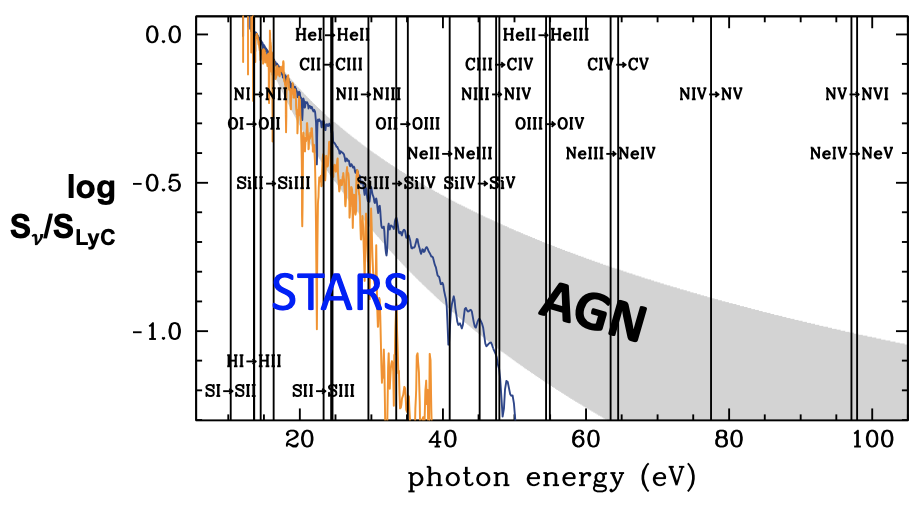}
\caption{\it Intrinsic spectral energy distributions $S_{\nu}$ (in terms of that at the Lyman limit $S_{LyC}$) versus photon energy (eV) for stars with 3\%  (orange) and 0.1\% (blue) solar metallicities, compared with AGN (the grey shaded area) with a range of likely power law indices. Vertical lines indicate the ionising energies of ions of different species. The visibility and ratios of such ionic spectrum lines can be used to gauge the hardness of the radiation \citep{Feltre2016}.}
\label{fig:feltre}
\end{figure}

\begin{center}
    {log $\xi_{ion} = 25.0-25.3$ Hz erg$^{-1}$}
\end{center}
\medskip
\noindent{\bf Rest-UV metal lines of high ionisation potential:} Photoionisation models of stellar populations of different metallicities or AGN can predict the expected line ratios for metals in various ionisation states. An illustrative figure from \citet{Feltre2016} in Figure~\ref{fig:feltre} shows how key emission lines such as CIII], CIV and NV with associate high ionisation potentials can be used to diagnose the hardness of the radiation field. 

An example of this method followed VLT rest-UV spectra of $\simeq$100 metal-poor, relatively dust-free Lyman $\alpha$ emitting galaxies (LAEs) at $z\simeq$3 considered to be better analogues of $z>7$ galaxies than Lyman break galaxies (LBGs) discussed above \citep{Nakajima2018}. These authors found: 
\smallskip

\begin{center}
{log $\xi_{ion}$ = 25.5-25.7 Hz erg$^{-1}$}
\end{center}

\begin{figure}
\center
\includegraphics[width=0.8\textwidth]{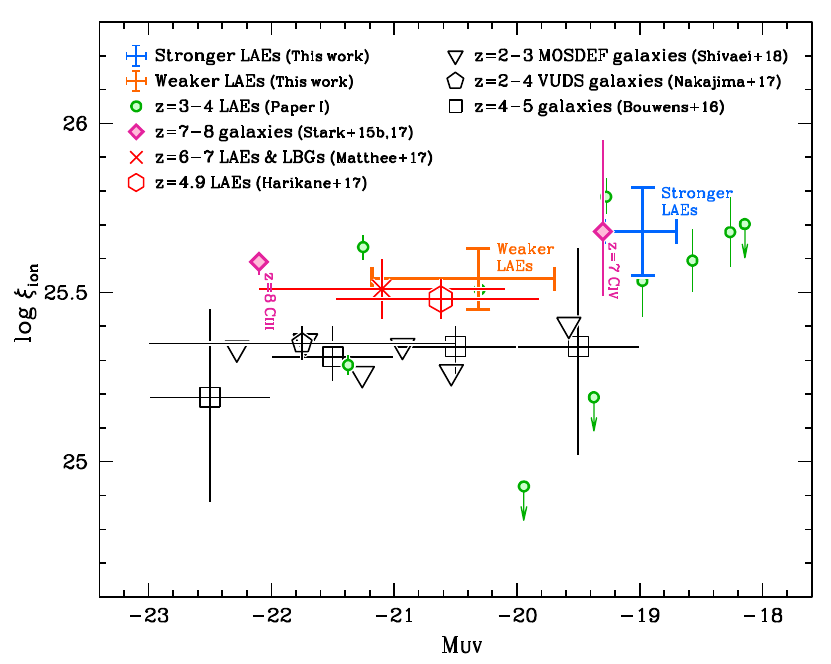}
\caption{\it Intrinsic ionising production rate $\xi_{ion}$ derived from photoionisation models for a sample of z$\simeq$3 Ly$\alpha$ emitters of  various UV luminosities (orange and blue points) compared to a coeval population of Lyman Break galaxies (open symbols). LAEs have a harder radiation field and are better analogues of galaxies in the reionisation era \citep{Nakajima2018}.}
\label{fig:nakajima}
\end{figure}

\medskip
i.e. a harder radiation field than for LBGs (see Figure~\ref{fig:nakajima}). Very limited pre-JWST samples of star-forming galaxies at $z>7$ \citep{Stark2017} revealed high ionisation lines consistent with those seen in intermediate redshift LAEs. 

The infrared capability of JWST means it is now possible to extend the first of the two methods introduced above into the reionisation era for the first time. There is a marked distinction between those adopting a {\it photometric approach} - using excess fluxes seen in key broad and medium band filts to infer Balmer and other line strengths - and those focusing on genuine spectroscopy. This is a recurring theme in JWST studies of the reionisation era. Photometric measures of galaxy properties offer the advantage of larger samples probing to higher redshift and perhaps permitting a selection method free from any bias. Spectroscopic measures are usually more precise but more limited in redshift range and, when applied to the merged database of many different JWST campaigns, there's naturally a greater likelihood of biases e.g. to systems with stronger line emission.

The JEMS imaging survey \citep{Simmonds2024} analyses 677 $4 < z < 9$ LBGs using NIRCam medium and broad-band photometry sensitive to the presence of H$\alpha$ and [O III] 5007 A (the latter used as a proxy for H$\alpha$ at higher redshift). They find a modest increase in $\xi_{ion}$ with redshift. By contrast, \citet{Saxena2024} using JADES spectra of 16 LAEs found no convincing increase with redshift with log $\xi_{ion}$=25.6 very similar to that found for LAEs at z$\simeq$3.  Differences have also been found between the photometric and spectroscopic analyses in claimed trends with absolute magnitude $M_{UV}$ and UV continuum slope. An enlarged spectroscopic survey of 167 galaxies to $z\simeq$6.7 from JADES and CEERS samples now finds weaker trends with redshift and only marginally harder radiation fields for luminous galaxies, possibly indicating the presence of AGN (\citet{Pahl2024}, see also \citet{Llerena2024}). Summarising, it seems likely that galaxies in the reionisation era may have radiation fields uniformly stronger than  Lyman break galaxies at intermediate redshift. Any trends with luminosity would be very important, particularly in understanding the relative contribution of intrinsically faint or luminous galaxies to reionisation. If star formation is stochastic, it is likely that $\xi_{ion}$ would be lower in sub-luminous sources observed during a down-turn in activity \citep{Stark2025}. 

\subsection{Escape Fraction $f_{esc}$}
\label{subsec: 2}

\begin{figure}
\center
\includegraphics[width=\textwidth]{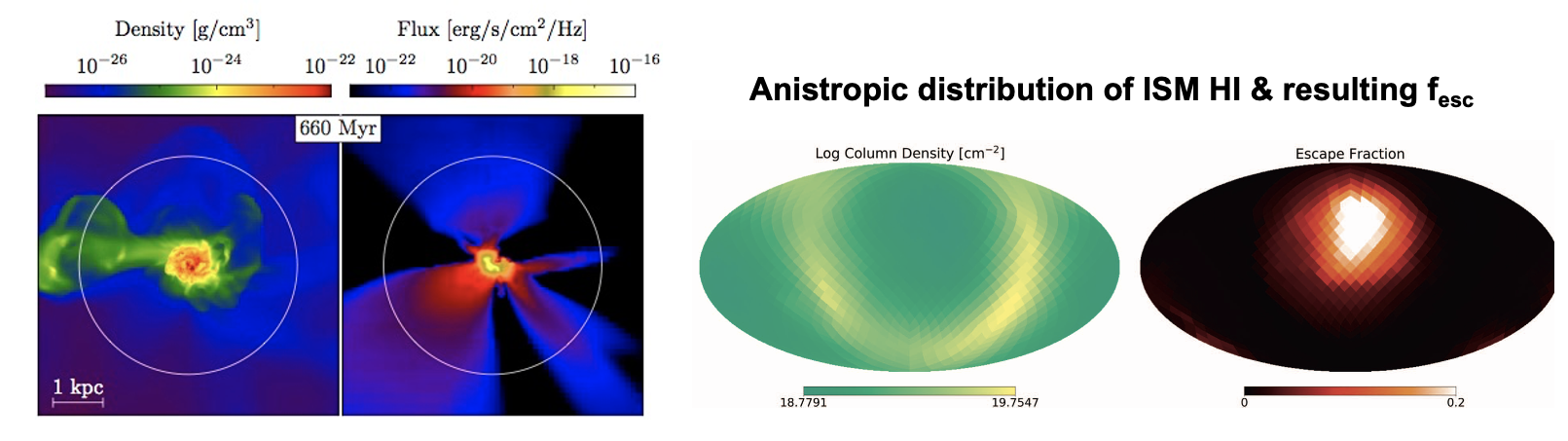}
\includegraphics[width=\textwidth]{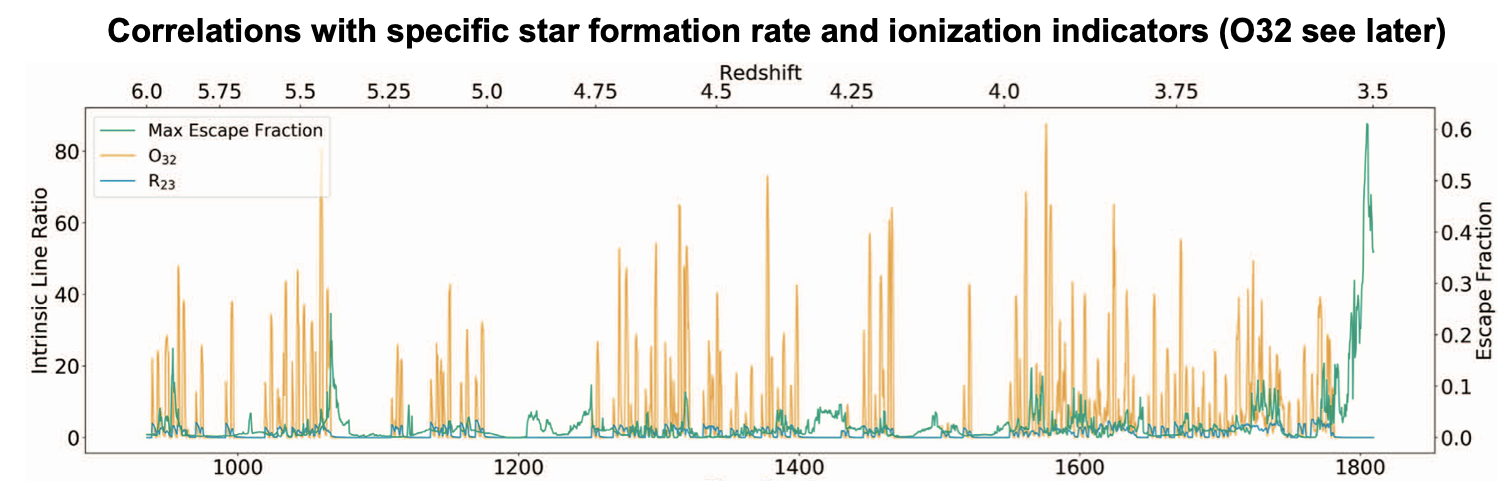}
\caption{\it Numerical simulations of the leaking Lyman continuum from star-forming galaxies \citep{Wise2014, Barrow2020}. (Top) Evidence for anisotropic leakage whereby radiation pressure creates specific channels through the inhomogenous neutral gas. (Bottom) Demonstration of how stochastic star formation influences the leakage and correlates with observable diagnostic line ratios (R23 and O32, see text)}.
\label{fig:fesc_theory}
\end{figure}

This final parameter presents the greatest challenge in understanding the process of reionisation for two very different reasons. Firstly, hydrodynamical simulations suggest that young star-forming galaxies are complex systems and that any emerging Lyman continuum radiation would be highly time-dependent due to bursty star formation and anisotropic due to the irregular distribution of porous regions in the ISM (\citet{Wise2014,Trebitsch2017,Barrow2020}, see Figure~\ref{fig:fesc_theory}).  A second problem is that the attenuation of the IGM increases beyond $z\simeq$4 \citep{Inoue2014} which means direct measures of $f_{esc}$ are restricted to redshifts $z\leq$3 where lower redshift "analogues" of reionisation era systems can be studied.

First some definitions. The ``ideal" definition of $f_{esc}$ would be the fraction of the intrinsic number of ionising photons that escape into the IGM without being absorbed by neutral hydrogen in the ISM/CGM. In practice however, we don't observe the intrinsic rate so the ideal value is sometimes referred to as the (unobservable) {\it absolute value} $f_{esc}^{abs}$. Observers usually refer instead to a {\it relative value} based on observations below and above the Lyman continuum limit at 912 \AA\ 

\begin{equation}
f_{esc}^{rel} = \frac{(f900/f1500)_{obs}}{(L900/L1500)_{int}}
\end{equation}

where the intrinsic (int) ratio is determined from a stellar synthesis model.

Direct methods at very low redshifts ($0.2<z<0.4$) have naturally provided the most detailed information but nonetheless some puzzles have arisen. So-called ``Green Pea" galaxies are compact low redshift galaxies with intense [O III] 5007 \AA\ emission. HST's UV-sensitive Cosmic Origins Spectrograph (COS) often reveals non-zero fluxes below the Lyman limit indicating, in some case, $f_{esc}$ values as high as 20-40\% \citep{Izotov2016, Izotov2022}. Various correlations have been presented between $f_{esc}$ and the strength of Lyman $\alpha$, Mg II and the line ratio O32 (= [O III] 5007/[O II]3727 \AA\ ). However, HST COS also recently undertook a systematic survey of 66 low metallicity galaxies chosen to sample a wider range of UV continuum slopes, O32 ratios and star formation rates \citep{Flury2022}. Although the mean escape fraction for this sample is $\simeq$10\% and a subset has $f_{esc} > $20\%, there is surprisingly little difference in the above physical properties between the Lyman continuum (LyC) leakers and non-leakers. A recent study of the Lyman Alpha and Continuum Origins Survey finds leakers have more compact Ly$\alpha$ cores suggesting both LyC and Ly$\alpha$ photons are emerging through sightlines cleared by central starbursts \citep{Saldana-Lopez2025}. This further suggests that whether a galaxy is a leaker or not depends not only on recent star formation but also on the viewing angle.

Direct LyC detections can be sought for large samples at redshift $z\simeq$3 which represents a ``sweet spot" in terms of redshift since the Lyman limit enters the optical range of ground-based multi-object spectrographs yet the IGM opacity is still modest. Even so, the LyC signal is mostly too faint to be detected in individual galaxies. Only 12 galaxies reveal individual LyC detections in the comprehensive Keck spectroscopic survey of 124 LBGs by \citet{Steidel2018}. That analysis is largely based on stacked spectra yielding average values of $f_{esc}\simeq$6-9\% depending on various modelling assumptions. Contamination of the LyC stacked signal from lower redshift sources along the sight line may inadvertently lead to an overestimated average $f_{esc}$. Such contaminants might be undetected in ground-based imaging and reveal no associated spectroscopic features.

\begin{figure}
\center
\includegraphics[width=\textwidth]{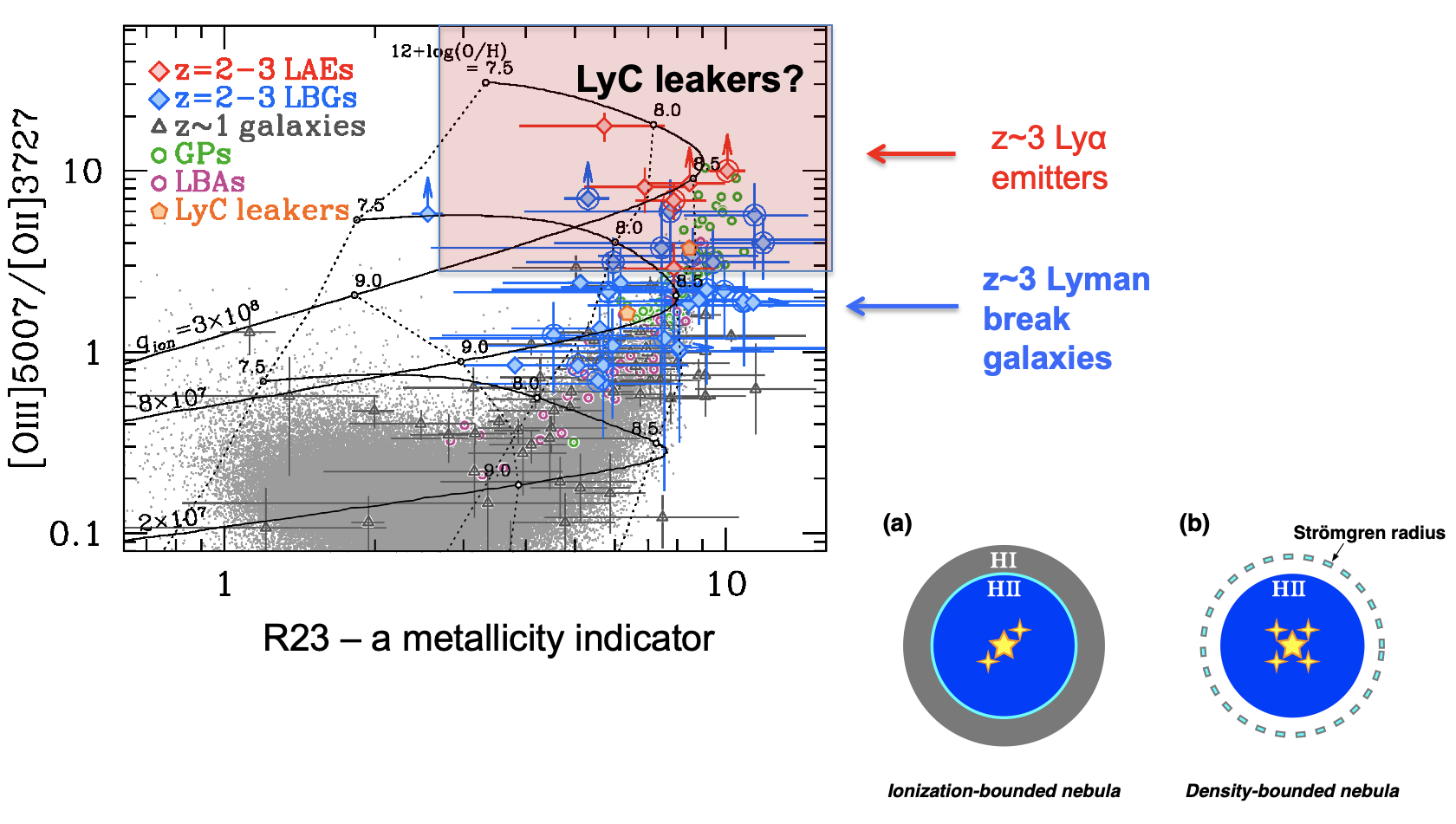}
\caption{\it Lyman $\alpha$ emitters at $z\simeq$3 (red lozenges) and local ``Green Pea" galaxies (green) have a larger O32 line ratio that Lyman break galaxies (blue). \citet{Nakajima2014} postulate such extreme line ratios may arise from density-bound (rather than ionisation-bound) HII regions (see lower right inset) which implying a higher escape fraction. }
\label{fig:O32}
\end{figure}

\begin{figure}
\center
\includegraphics[width=0.45\textwidth]{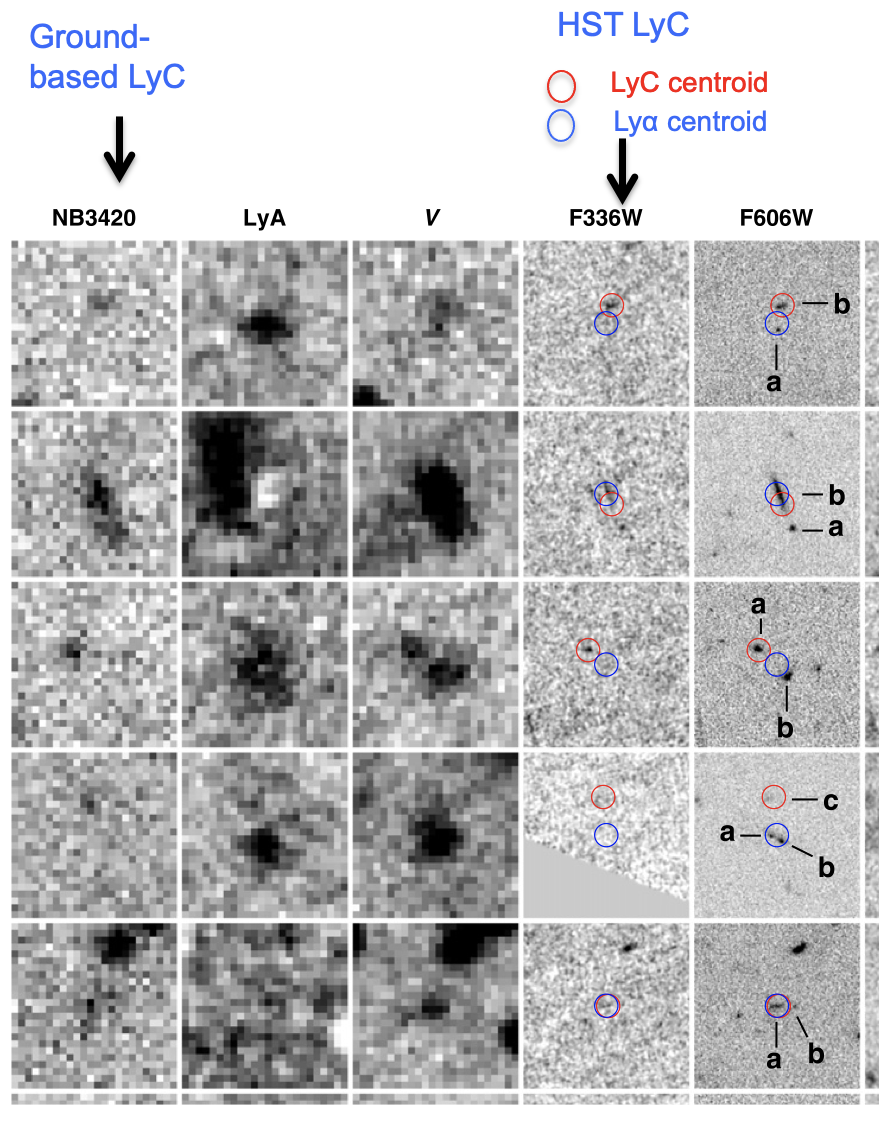}
\includegraphics[width=0.4\textwidth]{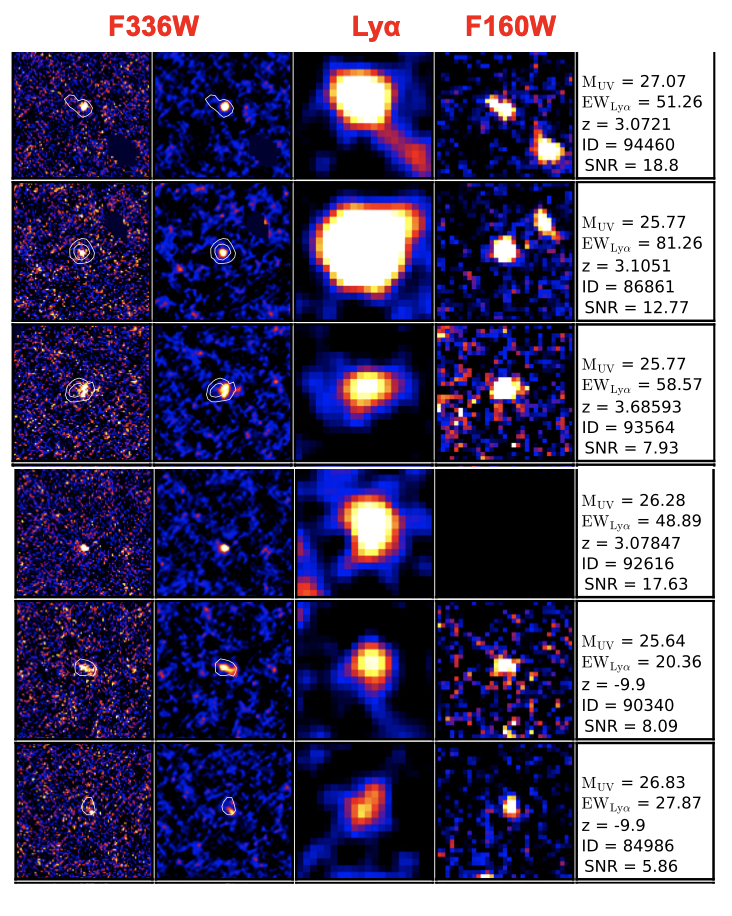}
\caption{\it Deep UV imaging of $z\simeq$3 Lyman $\alpha$ emitters as a probe of leaking radiation. (Left) While ground-based narrow-band 3420 \AA\ imaging revealed promising candidates, subsequent HST imaging in F336W demonstrates offsets between the putative LyC leakage (red circles) and Lyman $\alpha$ emission (blue circles) indicating likely foreground contamination \citep{Mostardi2015}. (Right) A selection of promising candidates from a similar HST-based F336W campaign \citep{Fletcher2019}. Over half this sample shows no discernible detections even when stacked, suggesting anisotropic leakage is the cause of such wide variations in $f_{esc}$.}
\label{fig:fesc_z3}
\end{figure}

\cite{Nakajima2014} proposed that high O32 line ratios are inconsistent with traditionally-assumed {\it ionisation-bound} nebulae (a.k.a. Str\"omgren spheres) and may reflect {\it density-bound} H II regions with larger escape fractions. They demonstrate (Figure~\ref{fig:O32}) that $z\simeq$3 LAEs have systematically higher O32 values than LBGs and thus may represent better analogues of reionisation-era galaxies. Following this suggestion, \citet{Fletcher2019} undertook deep (20 orbit) HST WFC3 F336W imaging of 61 $z\simeq$3 LAEs with associated HST/Subaru broad-band imaging and Keck/VLT spectroscopy. Fitting the SEDs enabled the authors to predict the intrinsic Lyman continuum flux in the F336W filter. Comparing with the observed flux and correcting for IGM absorption provides the likelihood distribution of $f_{esc}$. Although 20\% of the sample showed significant F336W detections with $f_{esc}$ value ranging from 15-50\% (right panel of Figure~\ref{fig:fesc_z3}), fully half the sample showed no detections at all implying individual values of $f_{esc}<$1.5\%.  As with the lower redshift HST COS samples, demographically the non-leaker population seems no different to the leaker population. Remarkably, even a F336W stacked image of the non-leakers reveals no signal suggesting an average $f_{esc}<$0.3\%. Once again, it seems Lyman continuum radiation is either observed or not at all.

Even with the superlative image quality of HST, foreground contamination remains a major concern for direct imaging measures of $f_{esc}$. Earlier claims of Lyman continuum detections based on ground-based narrow-band data were subsequently shown by improved HST imaging to arise from unassociated foreground sources (see left panel of Figure~\ref{fig:fesc_z3}, \citet{Mostardi2015}). JWST is even revisiting earlier claimed HST detections and apparently uncovering confusion between Lyman continuum signals and foreground galaxies (Schaerer, priv. comm.)

\begin{figure}
\center
\includegraphics[width=\textwidth]{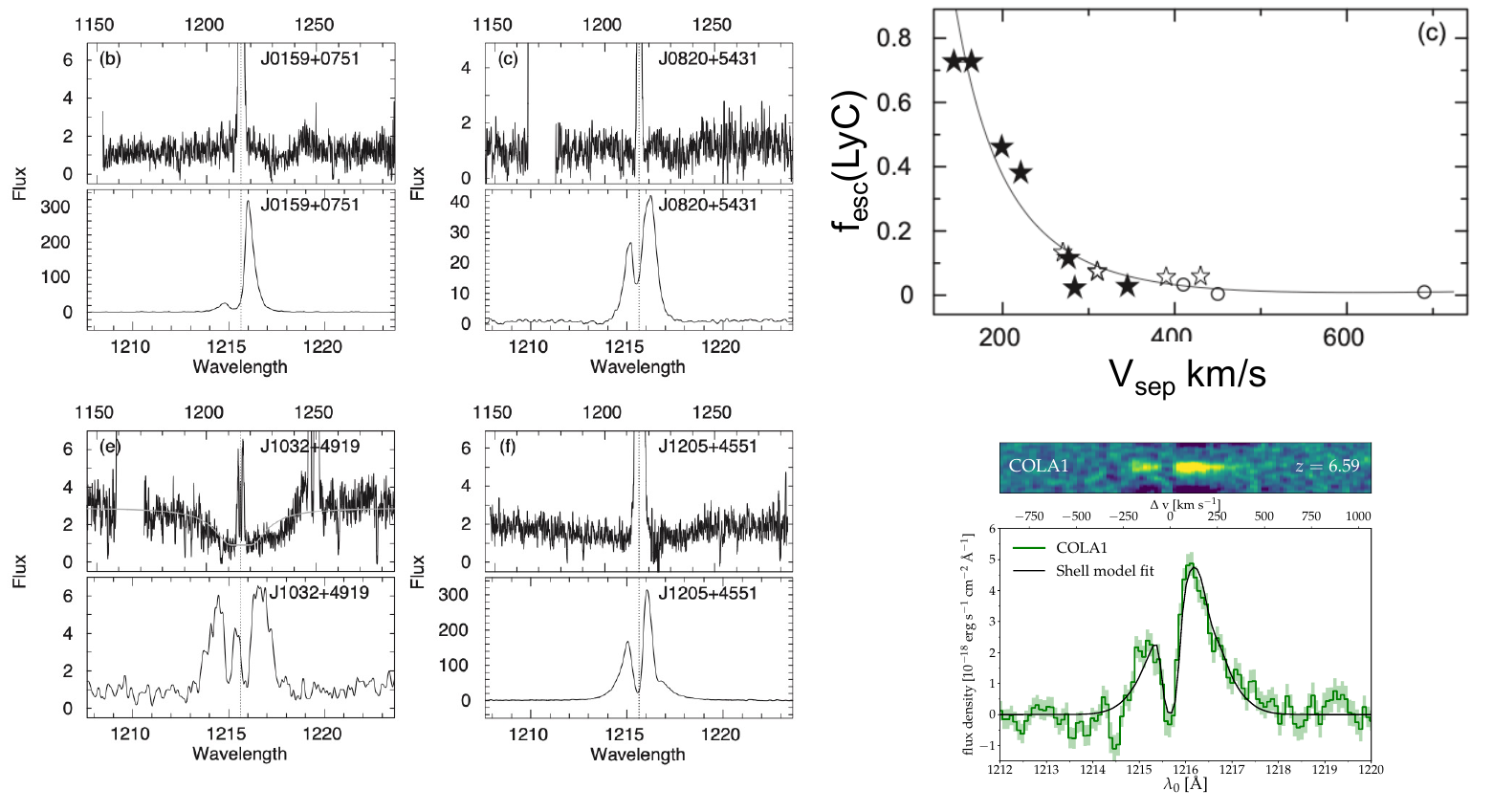}
\caption{\it Double-peak Lyman $\alpha$ emitters as a promising probe of $f_{esc}$. The main display shows high quality HST COS line profiles of selected $z\simeq0.03-0.06$ SDSS galaxies with high O32 \citep{Izotov2018} and a convincing anti-correlation between the derived $f_{esc}$ and the peak separation $V_{sep}$ which is the principal of the method (upper right). A convincing example of a double-peaked LAE in the reionisation era is shown in the lower right \citep{Matthee2018}.}
\label{fig:doublepeak}
\end{figure}

The situation regarding escape fractions prior to the launch of JWST was critically summarised by \citet{Naidu2020}. In addition to measures based on direct spectroscopy or UV imaging, less direct probes include measures of the covering fraction of neutral gas derived from high signal/noise absorption line spectra \citep{Jones2012, Jones2013, Leethochawalit2016} and studies of double-peaked Lyman alpha emitters where the velocity separation $V_{sep}$ between the peaks relates to the size of the ionised regions \citep{Verhamme2015,Michel-Dansac2020}. \citet{Izotov2018} show a convincing anti-correlation between $V_{sep}$ and $f_{esc}$ (Figure~\ref{fig:doublepeak}) but, to date, there are very few reionisation-era galaxies with such double-peaked Lyman $\alpha$ profiles (one is shown in the lower right of Figure~\ref{fig:doublepeak}, \citet{Matthee2018}).  As Naidu et al discuss, although the literature is rich in analyses of individual galaxies with high $f_{esc}$ there may be a natural bias against publishing negative results. Despite heroic observational efforts, prior to JWST there was little consensus on the average value for star-forming galaxies typical of those in the reionisation era. 

\begin{figure}
\center
\includegraphics[width=\textwidth]{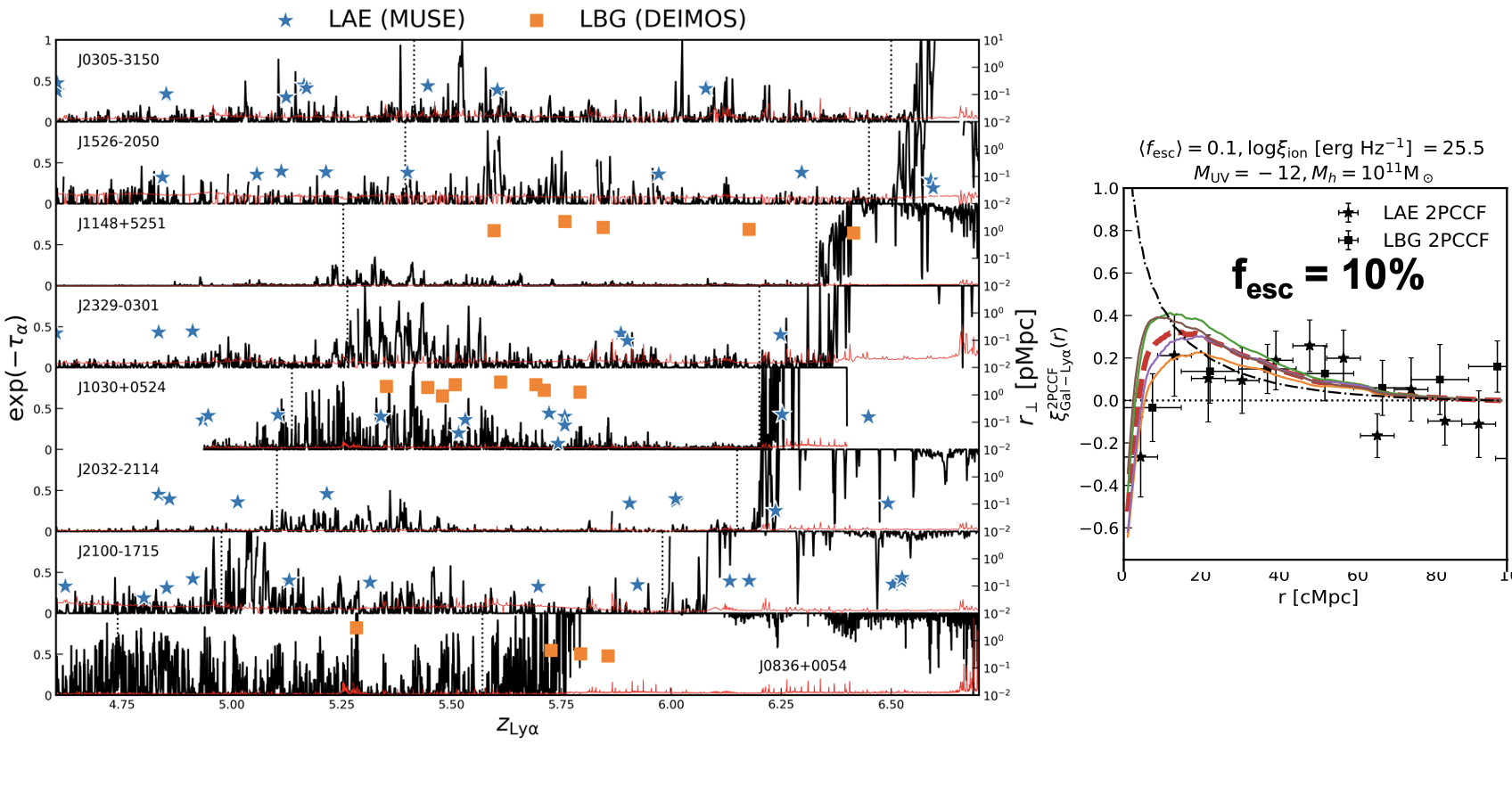}
\caption{\it Estimating $f_{esc}$ via direct spatial correlations between galaxies and the fluctuating transmission in the Lyman $\alpha$ forest as first proposed by \citet{Kakiichi2018}. Absorption spectra of 8 $z>$6.5 quasars sample the IGM over the redshift range 4.8$< z <$6.1. Keck and VLT redshift surveys locate galaxies and measure their distance to each sightline. The right panel shows the cross-correlation which is fit by a model which accounts for the likely clustering of sub-luminous galaxies beyond the spectroscopic limit of the galaxy survey. An average escape fraction of $\simeq$10\% is a reasonable fit to the data \citep{Meyer2020}.}
\label{fig:koki}
\end{figure}

Given the difficult of inferring $f_{esc}$ for sources in the reionisation era, theorists have tried to correlate the escape fraction seen in simulations with other properties observable at high redshift (e.g. \citet{Choustikov2024}). The results are somewhat inconclusive at the present time. Clearly $f_{esc}$ is a complex quantity that fluctuates according to irregular star formation histories and geometrical anistropies, and so a better approach is to statistically correlate the IGM opacity inferred from quasar absorption spectra of the Lyman $\alpha$ forest with the proximate presence of associated galaxies. First proposed as a proof-of-concept by \citet{Kakiichi2018}, \citet{Meyer2020} applied the method to the sightlines of eight $z\simeq$6.5 quasars securing Keck+VLT redshifts for 60 galaxies within the same cosmic volume as that probed by the Lyman $\alpha$ forest over $4.6 < z_{abs} < 6.6$ (Figure~\ref{fig:koki}). The positive cross-correlation over 10-50 comoving Mpc is consistent with an average $f_{esc}\simeq$10\% assuming the typical galaxy has an intrinsic production rate log $\xi_{ion}$=25.5 erg Hz$^{-1}$. One limitation of this promising method is that ground-based redshift surveys can only trace the most luminous galaxies at such high redshifts and so the associated population of intrinsically fainter sources which also contribute to the ionising flux must somehow be accounted for.

The JWST ASPIRE programme (PI: F. Wang) is adopting a similar strategy using NIRCam F356W slitless spectra to locate [O III] emitters associated with 25 $z>6.5$ quasar sightlines. A preliminary analysis based on the first 5 quasars and 50 [OIII] emitters detects a positive cross-correlation suggesting a similar result to that of Meyer et al \citep{Kakiichi2025}. Likewise the EIGER survey \citep{Kashino2025} has correlated [O III] emitters with the Ly$\alpha$ forest in 6 QSO sightlines and examines how the cross-correlation evolves over 5$<z<$7 suggesting that the influence of the background radiation field becomes important at higher redshift. 

Kakiichi and colleagues have proposed a further refinement of their method. As high redshift quasars are sparsely distributed on the sky and the above technique is limited to a number of 1-D sightlines, it is seriously affected by cosmic variance. Harnessing the unique wide field prime focus of the Subaru telescope, the new idea is to undertake deep narrow-band imaging to reveal photometrically the depth of Lyman $\alpha$ absorption at one selected redshift for all the background galaxies. The same narrow-band filter can also simultaneously locate Lyman $\alpha$ emitters at this redshift and thus undertake the cross-correlation. This provides a significasnt increase in the number of cross-correlated galaxy absorber pairs! The challenging requirements include ultra-deep Subaru narrow-band imaging which becomes progressively more demanding at higher redshifts and spectroscopic redshifts for all the background galaxies so there is no confusion regarding the origin of the absorption signal. This project is now being undertaken in the COSMOS-3D survey, a Cycle 3 Treasury Programme (PI: Kakiichi) where the redshifts are being determined using NIRCam F444W slitless spectroscopy.

Finally, two further intriguing ideas have emerged since the Lectures that bypass the inability to measure $f_{esc}$ directly beyond z$\simeq$4 due to IGM absorption. The first is to study galaxies lying in the ionised ``proximity zones" created by luminous quasars. \citet{Yue2025} apply this method to 15 z$\simeq$6-7 galaxies lying within 2500 km sec$^{-1}$ of a quasars and find significant flux blueward of Ly$\alpha$ compared to a control sample of galaxies not associated with any quasars. Their inferred escape fraction for Ly$\alpha$ emission is consistent with a LyC $f_{esc} \simeq$ 10\%. The second utilises topics I discuss in Lecture 2 that relate to the UV continuum slope and the contribution made by the nebular continuum emitted by young galaxies with hot stars (Lecture 2, Section 3.5). This nebular continuum is redder than its stellar counterpart and is naturally anti-correlated with $f_{esc}$; if $f_{esc}$=1, there is zero nebular contribution to the UV continuum. \citet{Giovinazzo2025} apply this idea to a large spectroscopic database and find a non-zero $f_{esc}$ for only 5\% of their sample. Nonetheless, they claim an average value overall of $\simeq$10\%.

Summarising, the escape fraction from typical star forming galaxies in the reionisation-era remains a fundamental challenge to confirming that cosmic reionisation is primarily driven by galaxies. I suspect the only reliable way forward is to secure statistical correlations between galaxies and the opacity of HI in local IGM, either via the methods being adopted by ASPIRE, EIGER and COSMOS-3D, or ultimately via matching JWST sources with 21cm tomography using the Square Kilometre Array (Lecture 3).

\section{Constraining the Reionisation History}
\label{sec: 4}

Let's now attempt to make sense of the observational constraints on the ionising budget noting again the simplistic assumption that the ionisation rate $\dot{n}$ can be broken into three constituent ingredients, which may themselves vary with redshift and galaxy luminosity. Of particular importance is the degeneracy between $\xi_{ion}$ and $f_{esc}$ since it is often only the product that is constrained observationally. 

\begin{figure}
\center
\includegraphics[width=0.6\textwidth]{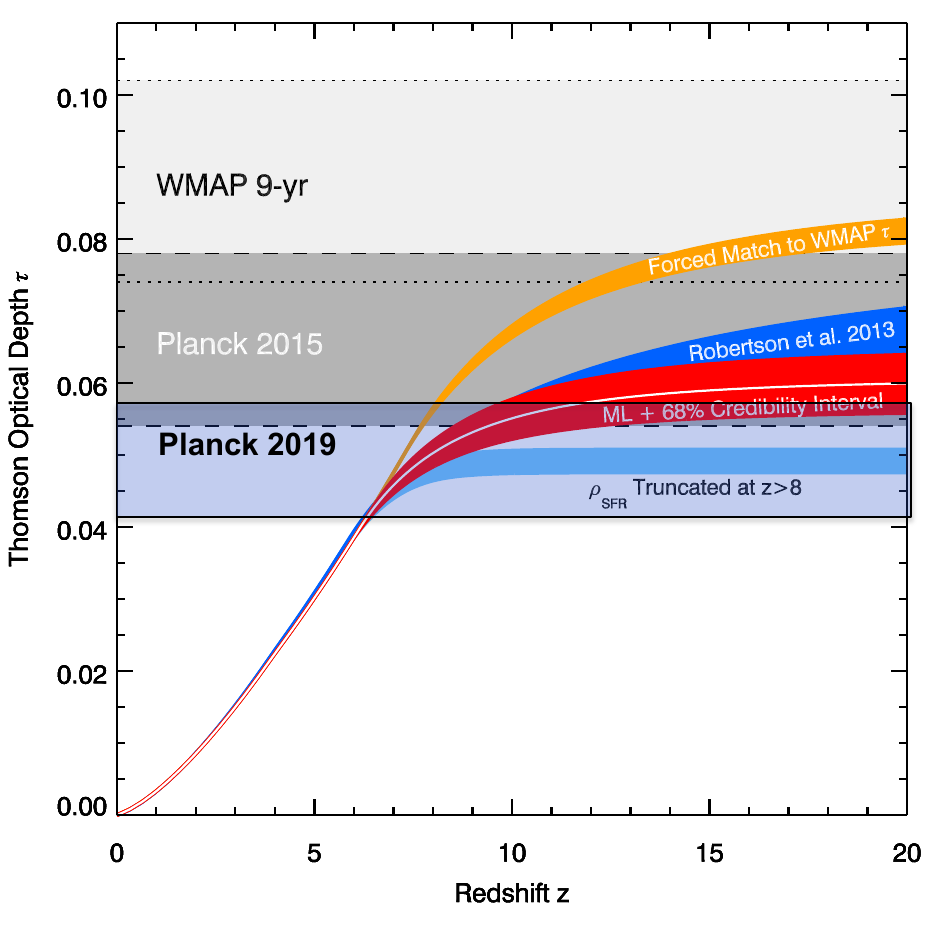}
\caption{\it Reconciling the demographics of galaxies determined with HST with the Planck optical depth $\tau$ to Thomson scattering \citep{Robertson2015}. The figure shows the integrated contribution of IGM electrons produced by leakage from star-forming galaxies to $\tau$ as redshift increases assuming log  $\xi_{ion}$=25.5 cgs and $f_{esc}$=10\% irrespective of luminosity and redshift as well as an option for a steeper rate of declining $\rho_{UV}$ beyond $z$=8 (blue curve). The unlabeled red curve is a reasonable match to $\tau$ assuming galaxies emerged in the redshift range 15$ < z < $20.}
\label{fig:tau_match}
\end{figure}

We begin with the popular view that galaxies dominate the process of reionisation. In this case, a simple question is whether the demographics of galaxies, as probed by the redshift-dependent UV luminosity density $\rho_{UV}(z)$ can be reconciled with the Planck optical depth $\tau$ when we adopt universal values for the other two ingredients $\xi_{ion}$ and $f_{esc}$. Following \citet{Robertson2015} who adopted $< log \ \xi_{ion} > $ = 25.5 Hz erg$^{-1}$ and $ < f_{esc} > $ = 10\% for all star-forming galaxies, the integrated contribution to $\tau$ is shown in Figure~\ref{fig:tau_match}. To the accuracy this is possible, with the exception of escape fraction adopted, the cumulative optical depth is satisfactorily matched with Planck's final value $\tau$= 0.063 by a redshift $z\simeq$11-15. 

But does this agreement exclude a contribution from AGN? In terms of pre-JWST constraints, \citet{Naidu2020} pointed out that Robertson et al's simplistic model fails to match the rapidly declining neutral fraction suggested by various quasar absorption probes and the fraction of Lyman $\alpha$ emitters (right panel of Figure~\ref{fig:tang}). At the time this article was written, there was conflicting information on whether $\xi_{ion}$ varies with luminosity. The rapid decline in $x_{HI}(z)$ for $z<8$ could be matched by assuming luminous sources played a more significant contribution, perhaps due to the onset of AGN activity.

In Lecture 2 I will discuss the growing body of evidence from JWST that low luminosity AGN are more numerous in the latter half of the reionisation era than suggested by earlier data. Noting the absence of persuasive evidence for a uniformly significant value ($\simeq$10-20\%) for the escape fraction from star-forming galaxies, this has led some to propose a dominant role by AGN. \cite{Madau2024} have presented a AGN-dominated model which can also satisfy the integrated Planck $\tau$ and updated constraints on the redshift-dependent neutral fraction $X_{HI}(z)$. However, the model has some very debatable assumptions. It assumes 15\% of galaxies (over all redshifts and luminosities) contain AGN whose escape fraction is greater than 80\%. This fraction is not yet justified given many AGN candidates claimed from JWST observations are hotly debated (see Lecture 2). A further dubious assertion is that, for those galaxies hosting AGN, at least 50\% of the UV luminosity arises from a non-thermal component. This cannot be justified across the entire UV luminosities at all epochs. Hopefully as further evidence for AGN activity in the reionisation era is gathered, the extreme views of "all done by galaxies" and "all done by AGN" will be compromised!

In concluding, \citet{Pahl2024} take their modest dependencies of $\xi_{ion}$ on redshift and luminosity and find a reionisation history very similar to that based on Robertson et al's simple model. A key issues is the question of whether there is an unacceptably rapid decline in $x_{HI}$ for $z < 8$ which might indicate an AGN contribution. Unfortunately, as Pahl et al discuss, the uncertainties remain quite large. In Lecture 3 we will briefly explore the promise of using the 21cm line as a tracer of the ionisation state of the IGM. This can be via emission line tomography revealing the location of ionised bubbles as well as in line of sight absorption in spectra of background radio galaxies. \citet{Sims2025} use current upper limits on the 21cm power spectrum from SKA pathfinders and incorporate it into a reionisation history that is quite similar to that of Pahl et al.

\section{Summary}
\label{sec: 6}

In this first lecture, we have addressed two fundamental questions:
\medskip
\begin{enumerate}
\item{} {\bf When} did reionisation begin and end and, ideally can we constrain the evolving neutral fraction $x_{HI}(z)$? A reminder we'll further address the beginning of reionisation in Lecture 3.
\smallskip
\item{}{\bf How} was it accomplished, i.e. by which sources, luminous or feeble galaxies and/or AGN?
    
\end{enumerate}

\medskip

In addressing {\bf when}: the quasar Lyman $\alpha$ forest shows a declining HI opacity down to $z=5.3$. As it was often assumed reionisation ended at $z=6$, the quasar observers are keen to point out their data requires neutral gas at this epoch. However, at some point defining an ``end" corresponding to a precise $x_{HI} \equiv$ 0 does seem somewhat arbitrary.

Regarding the beginning of reionisation, analyses of Planck and other data in the context of $\Lambda$CDM permits some level of ionisation in the IGM beyond $z=12$. A particularly significant development is the discovery of a Lyman $\alpha$ emitting galaxy at a redshift z = 13.0 \citep{Witstok2024} which implies the early existence of an ionised region. 

In addressing {\bf how}: the absence of strong luminosity trends in $\xi_{ion}$ would suggest that feeble galaxies dominate the process, although a fundamental limitation remains our understanding of the escape fraction. As we will see in Lecture 2, while JWST observations suggest a much-increased number of AGN, most calculations suggest they still only provide a minor contribution to the reionisation budget. While the rapid change in $x_{HI}$ below $z<$8 makes an appealing case for a late AGN contribution, the uncertainties in many of these $x_{HI}$ measures remain significant.

However, to end on a positive note, the continued increase in JWST spectroscopy will significantly improve the statistics of $\xi_{ion}$ and its dependence on luminosity as well as claims for AGN activity in the overall galaxy population. Likewise projects like COSMOS-3D, EIGER and ASPIRE promise much improved statistics on how galaxies contribute to the ionised state of their local IGM. Finally, as discussed in Lecture 3, the Square Kilometre Array will contribute further to the galaxy-IGM connection via tomography of the 21cm line of hydrogen.

\section{Recommended Reading on Cosmic Reionisation}
\label{sec: 8}

\noindent{\bf Major Review Articles:}

\begin{enumerate}
\item{} \citet{Loeb2001} - a classical theoretical review
\item{} \citet{Fan2006} - early quasar data and CMB discussion
\item{} \citet{Madau2014} - classic article on galaxy demographics
\item{} \citet{Dijkstra2014} - Lyman alpha emission as a probe of reionisation
\item{}  \citet{Stark2016} - galaxies in first billion years
\item{}  \citet{Ouchi2020} - analyses of Lyman alpha emitters
\item{}  \citet{Fan2023} - updated review of quasars and IGM
\end{enumerate}

\noindent{\bf Pre-JWST articles}

\begin{enumerate}
\item{} \citet{Robertson2015} - HST galaxy demographics in context of Planck CMB
\item{}  \citet{Planck2020} - final constraints on reionisation
\item{}  \citet{Greig2017} - fits to Planck in context of $\Lambda$CDM
\item{}   \citet{Naidu2020} - proposes special role of luminous sources
\item{}   \citet{Garaldi2022} - {\it THESAN} radiation-hydro simulations
\end{enumerate}

\noindent{\bf JWST review articles (so far)}

\begin{enumerate}
\item{} \citet{Bonn2024} - review written by selected attendants at ISSI Breakthrough Workshop held in Bonn, March 2024
\item{} \citet{Stark2025} - review by Stark, Topping, Endsley \& Tang submitted to Encyclopaedia of Astrophysics, November 2024
\end{enumerate}

\newpage
\begin{center}
{\bf \Large  Lecture 2: Galaxies in the Reionisation Era}
\end{center}

\medskip
\setcounter{section}{0}

\section{Introduction}
\label{sec:1}

Following my introduction in Lecture 1, it is worth understanding how we arrived at this remarkable period in observational astronomy where it has become possible to secure spectra of such quality that we can contemplate measuring the history of chemical enrichment in the universe back to the emergence of the first stellar systems. Over my own career (1974 to the present), I have witnessed the subject of galaxy evolution transition from simply counting faint galaxies on photographic plates without any statistically-sound information on their redshifts (late 1970s), through the first spectroscopic surveys probing barely to a redshift $z\simeq$1 (late 1980s), the launch and repair of Hubble Space Telescope (HST) (1990-93), the development of 8-10 metre class telescopes (late 1990s - early 2000s) to the promise of mid-infrared and sub-mm facilities such as the Spitzer Space Telescope, the James Clerk Maxwell telescope and ultimately the Atacama Large Millimetre Array (ALMA). JWST represents the final chapter in this remarkable revolution in observational capability and, some might argue, the largest single leap in performance over this dramatic half century of progress.

\begin{figure}[hbt!]
\center
\includegraphics[width=0.47\textwidth]{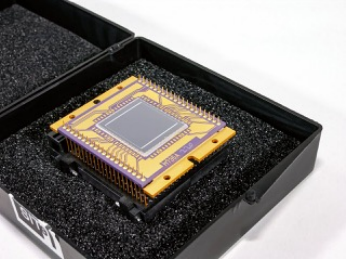}
\includegraphics[width=0.43\textwidth]{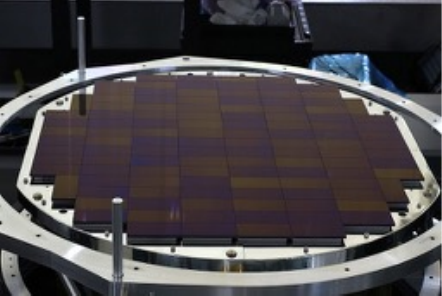}
\caption{\it The revolution in astronomical detectors. (Left) A Charge Coupled Device (CCD) of modest format from the early 1980s. Such digital detectors have quantum efficiencies 30-50 times those of the earlier hyper-sensitised photographic plates. (Right) The Hyper-Suprime Camera on the 8.2 metre Subaru telescope \citep{Miyazaki2018}. This remarkable 870 Megapixel camera comprises a mosaic of 116 CCDs and can capture the entire Andromeda spiral galaxy in a single exposure.}
\label{fig:ccds}
\end{figure}

Underpinning this story is one common factor - technological development. Focusing on a few of these technical revolutions that, in my opinion, transformed the field, I begin with the {\bf\it Charge Coupled Device (CCD)} which arrived on 4 metre class telescopes in the early 1980s. The initial examples were humble devices in a 256 $\times$ 256 pixel format but with sensitivities $\simeq$30-50 times that of hyper-sensitised photographic emulsions. This led to a major revolution in photometric measures of faint galaxies. One ultimate example of this development is the HyperSuprimeCam imaging system on the 8.2 metre Subaru telescope \citep{Miyazaki2018} comprising a mosaic of 116 2048 $\times$ 4096 CCDs which has completed an exquisite deep survey of 1500 deg$^2$ (Figure~\ref{fig:ccds}). The CCD era may soon be over as the number of manufacturers is limited and most commercial imaging systems now use lower noise Complementary Metal-Oxide-Semiconductor (CMOS) detectors.

\begin{figure}
\center
\includegraphics[width=0.4\textwidth]{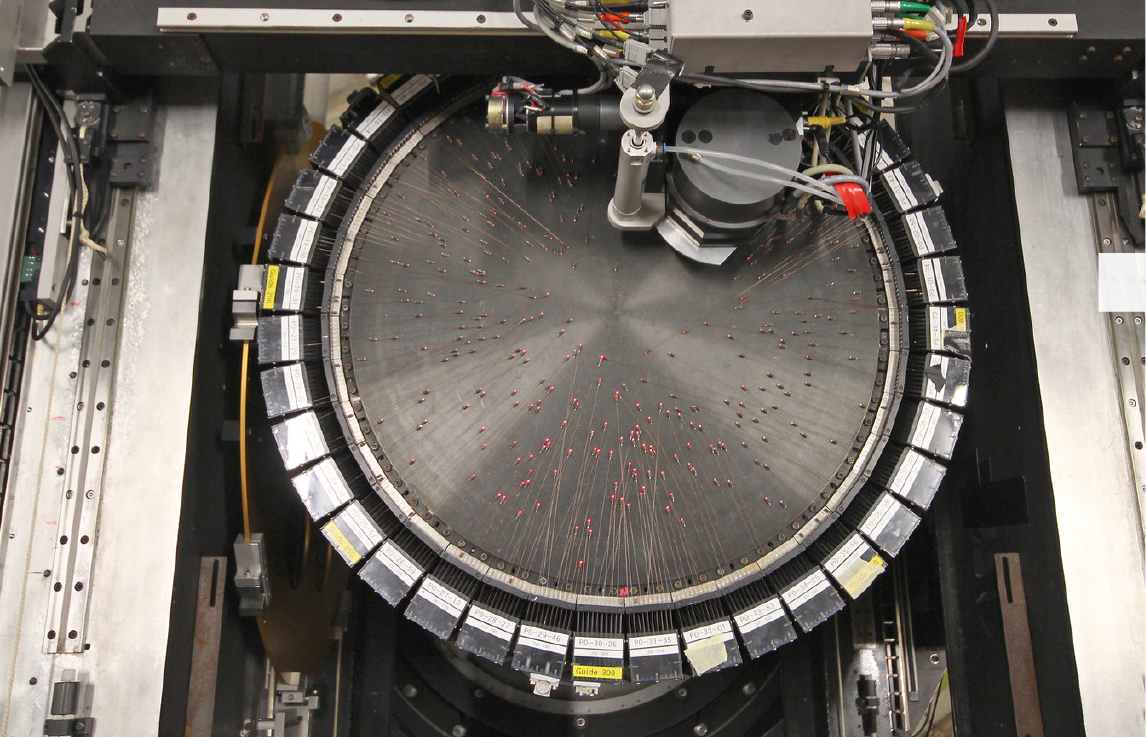}
\includegraphics[width=0.25\textwidth]{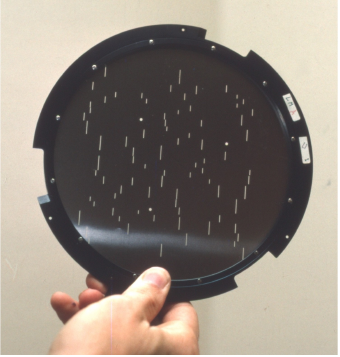}
\includegraphics[width=0.3\textwidth]{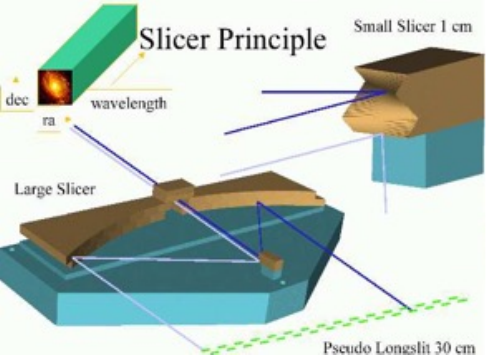}
\caption{\it Technologies for multi-object spectroscopy: (Left) The automated 400 fibre positioner for the 2 degree field facility at the prime focus of the Anglo-Australian Telescope. (Centre) A multi-slit mask for a ground-based imaging spectrograph. (Right) The image slicer concept for the MUSE integral field spectrograph on ESO's Very Large Telescope.}
\label{fig:mos}
\end{figure}

My second major development is {\bf\it multi-object spectroscopy} (see Figure~\ref{fig:mos}). The pioneering work to chase the cosmic expansion history using clusters of galaxies with the venerable 200-inch Hale telescope on Mount Palomar involved long slit spectroscopy on single targets, often with multiple night exposures. Multi-object spectroscopy using fibre optic couplers and machine-milled multi-slit masks in the telescope focal plane emerged in the mid-1980s. The former led to major discoveries in large scale structure via the Sloan Digital Sky Survey (SDSS) conducted on a dedicated 2.5 metre telescope and the 2 degree Field Redshift Survey on the Anglo-Australian Telescope; the latter demonstrated the power and efficiency of a robotic positioner with 400 fibres for the first time. Multi-slit masks can also be fabricated for imaging spectrographs providing resolved spectroscopy and improved sky subtraction for fainter sources. The final example of this revolution is the integral field unit (IFU) whereby a spectrum is produced for every point in a (typically) restricted field of view. This has been very effective in discovering line emitting galaxies without any prior imaging survey. A highly-readable account of the development of the IFU is given by \citet{Bacon2024} who pioneered the MUSE instrument on ESO's VLT. Unlike HST, JWST's instrumentation suite was developed to take advantage of the progress that followed the use of multi-object facilities on ground-based telescopes, e.g. in NIRSpec's Micro-Shutter Assembly (MSA) which represents an automated version of the multi-slit mask, and integral field capabilities on NIRSpec and MIRI.

The final major development that greatly contributed to progress was the construction of {\bf\it 8-10 metre class telescopes} in the 1990s. Alongside the development of large monolithic primary mirrors with active support (e.g. for the twin Gemini and four Unit Telescopes comprising ESO's Very Large Telescope) was the pioneering concept of a large segmented primary at the Keck observatory which demonstrated the way forward for ESO's Extremely Large Telescope, a 39 metre facility now nearing completion. The addition of laser guide star based adaptive optics led to angular resolutions better than HST which, when coupled with larger aperture and integral field spectrographs, led to resolved kinematics of $z>2$ galaxies.

Of course, powerful facilities are not, by themselves, sufficient to guarantee progress. New ideas and scientific intuition are essential for discoveries. In this category I have selected three developments that accelerated our understanding of galaxy evolution. The first is the demonstration and subsequent application of the {\bf\it Lyman break technique}. Although championed very effectively by Chuck Steidel and colleagues \citep{Steidel1995,Steidel1996}, the idea of using the Lyman continuum discontinuity to provide a redshift indicator can be traced to discussions of the ``faint blue galaxy problem" by Tyson and Guhathakurta \citep{Guhathakurta1990}. The photometric drop-out method has been extraordinarily effective in pre-selecting galaxies for spectroscopic study and providing redshift-dependent measures of the galaxy luminosity function and star-formation rate density. It spawned the entire field of photometric redshift determination and SED fitting from multi-colour broad band imaging which was not originally anticipated as a goal of HST imaging. 

\begin{figure}
\center
\includegraphics[width=0.9\textwidth]{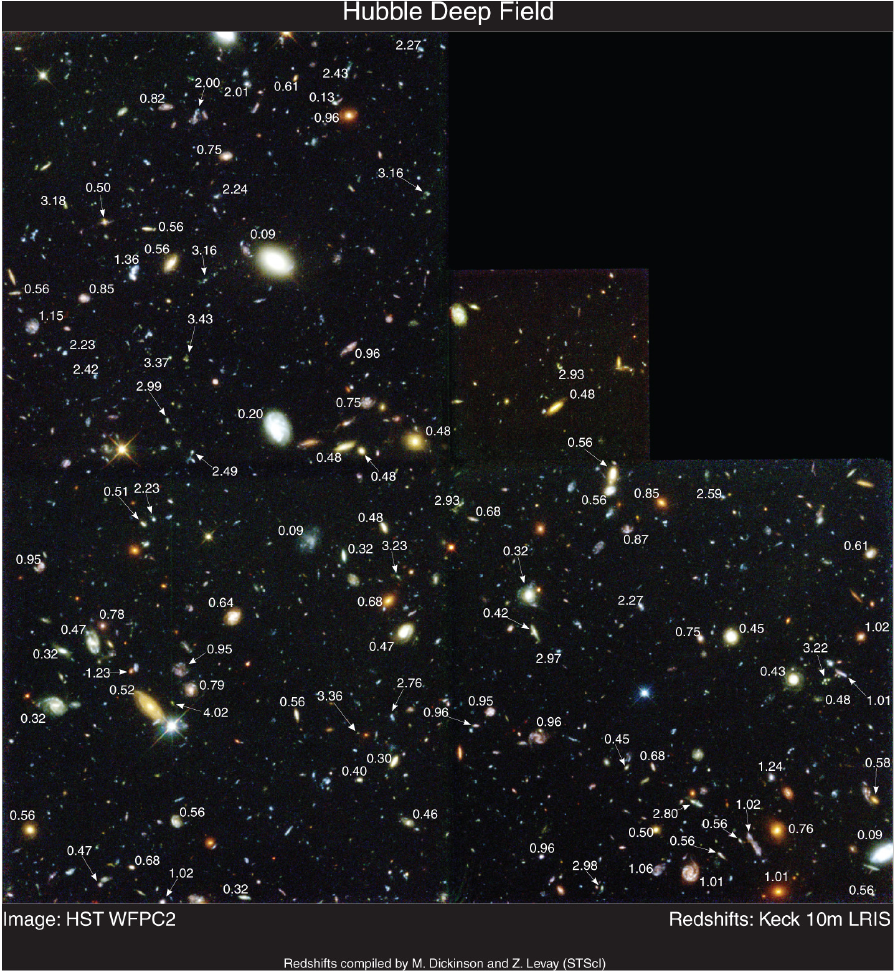}
\caption{\it The Hubble Deep Field (HDF) taken in 1996 following the initiative of Bob William, Director STScI at the time. The public release of this spectacular WFPC2 multi-band image led to cooperation amongst Keck observers in securing spectroscopic redshifts for a large sample of galaxies. HDF also led to a pattern of subsequent deep fields taken with HST imaging cameras. (Courtesy of Mark Dickinson \& Zolt Levay).}
\label{fig:hdf}
\end{figure}

A second development worthy of note was the first {\bf\it Hubble Deep Field (HDF)} proposed by Bob Williams, Director of the Space Telescope Science Institute, in 1996.  Williams had the ambitious vision of creating a very deep publicly-available multi-colour dataset that would galvanise and inspire the extragalactic community. Despite concerns from respected senior colleagues that such an image would waste a large amount of observing time so soon after the embarassment of HST's mis-shaped primary, it was a huge success and spawned an international effort with the Keck telescope to secured redshifts for hundreds of faint galaxies (Figure~\ref{fig:hdf}). This HDF led to a sequence of several further deep fields in the 2000s and similar programmes with JWST. The HDF story is described in detail in Williams' popular account which can be found online\footnote{https://iopscience.iop.org/book/mono/978-0-7503-1756-6.pdf}

\begin{figure}
\center
\includegraphics[width=0.55\textwidth]{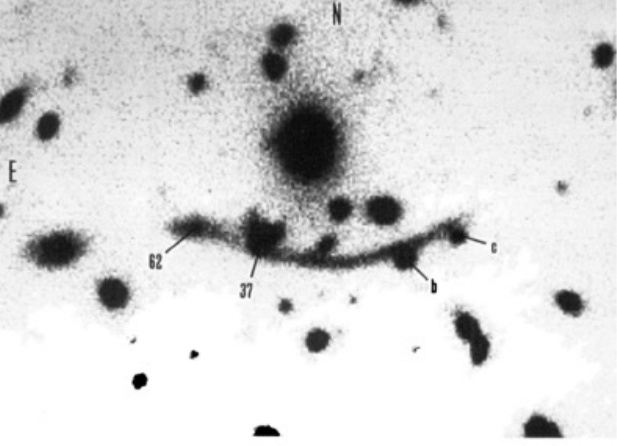}
\includegraphics[width=0.40\textwidth]{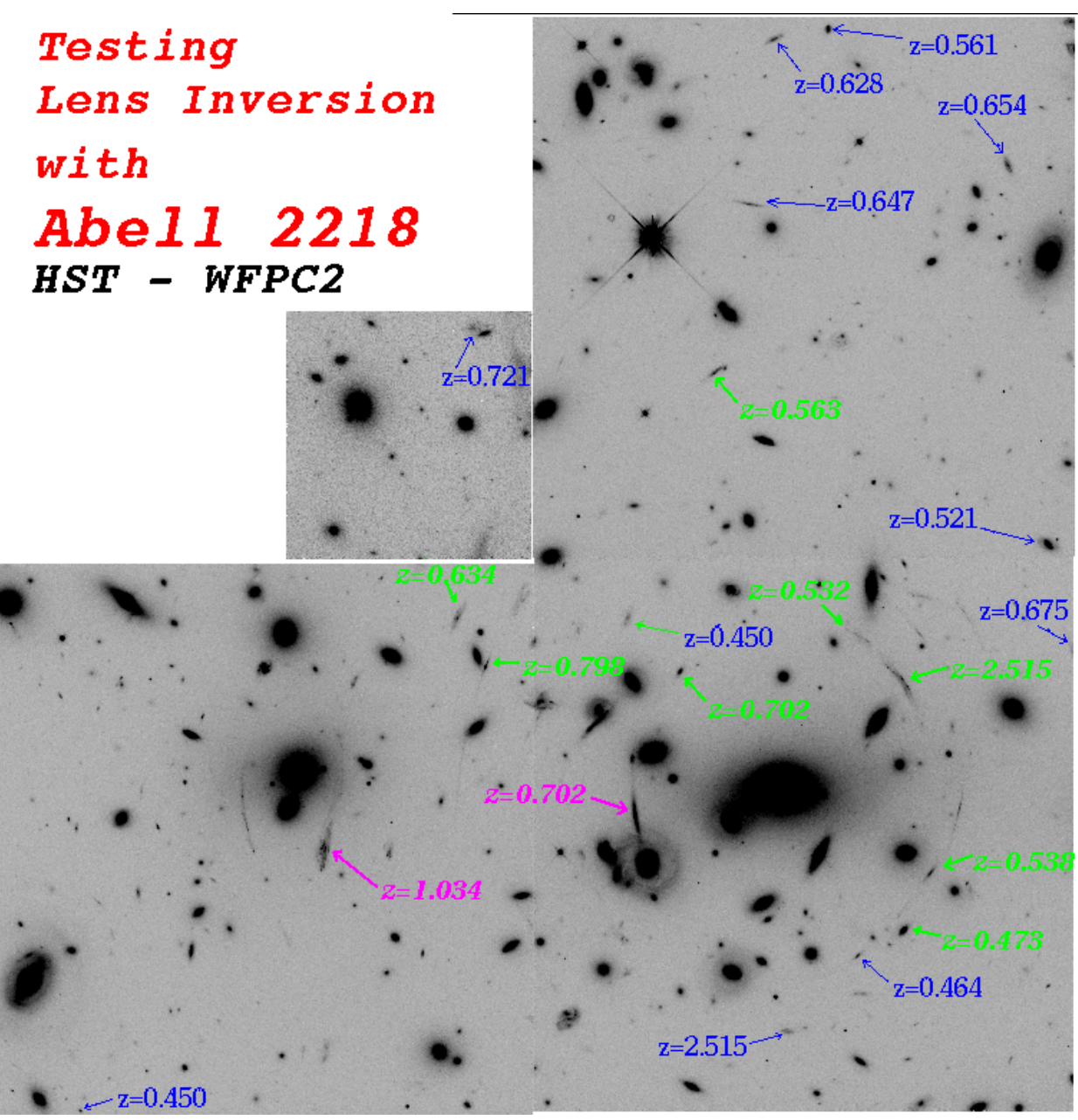}
\caption{\it The early promise of strong gravitational lensing. (Left) CFHT image of the giant arc in the cluster Abell 370 (z=0.37) which was spectroscopically confirmed to be a magnified and distorted image of a background z=0.72 galaxy by \citet{Soucail1988}. HST image of the lensing cluster Abell 2218. HST's improved resolution enabled the recognition of multiple images of the same background source improving mass models sufficiently that their redshifts could be estimated geometrically and later verified spectroscopically e.g. \citet{Ebbels1998}.}
\label{fig:lensing}
\end{figure}

The final scientific advance worth highlighting is when {\bf\it strong gravitational lensing} became a standard tool of the observational astronomer. Lensing was an obscure field of study until 1988 when Genevieve Soucail and colleagues in Toulouse used the Canada France Hawaii Telescope (CFHT) to demonstrate that the puzzling arc-like feature in the z=0.37 cluster Abell 370 was a galaxy at a redshift z=0.72 distorted and magnified by the foreground cluster \citep{Soucail1988}; see also a review of the early history by \citet{Treu2015}. HST's improved image quality revealed numerous examples of morphologically similar, multiply-imaged, lensed galaxies in other clusters and the development of mass models for the lensing clusters sufficiently precise they could even be used to estimate redshifts based purely on their geometrical positions (\citet{Kneib1996, Ebbels1998}, Figure~\ref{fig:lensing}). Strong lensing not only magnifies distant galaxies enabling probes of the galaxy luminosity function to fainter luminosities (Lecture 1) but also enlarges them enabling resolved kinematic studies with integral field spectrographs. As a result, the phenomenon has contributed significantly to our understanding of early galaxy formation and evolution. During 2013-16, HST conducted the Frontier Fields Survey \citep{Lotz2017}, an ambitious programme of multi-band imaging through six lensing cluster to exploit this phenomenon. Cluster lensing has also featured prominently in Early Release Science programmes with JWST, e.g. the GLASS survey \citep{Treu2022}.

\section{Pre-JWST results confirmed by JWST}
\label{sec: 2}

\begin{figure}
\center
\includegraphics[width=0.95\textwidth]{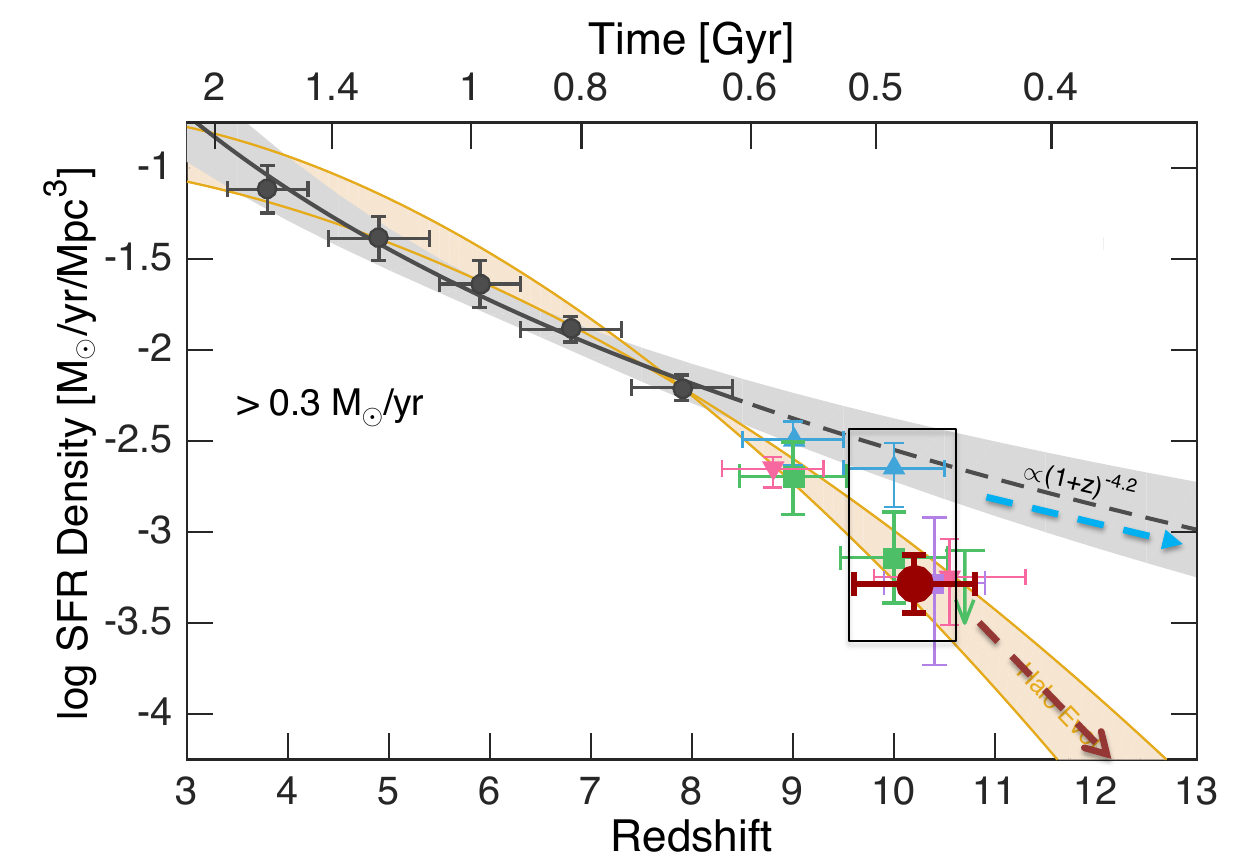}
\caption{\it A pre-JWST census of the redshift-dependent comoving star formation rate density of galaxies derived from rest-frame UV luminosities in all HST deep fields and the lensing Frontier Field clusters  \citep{Oesch2018}. The maroon arrow indicates the declining density proposed by \citet{Oesch2018} which matched a theoretical prediction based on a constant star formation efficiency (see text and Lecture 3). The blue arrow indicates a more gradual decline rate proposed by the analysis of \citet{McLeod2016}. Clearly the difference in the inferred decline at 8$<z<$10 made a considerable difference in predicting what JWST would find beyond a redshift z$\simeq$10.}
\label{fig:oesch}
\end{figure}

The most basic observable for probing the reionisation era is the redshift-dependent comoving abundance of star-forming galaxies. Through the succession of HST deep fields and the Frontier Field Survey discussed in Lecture 1, the evolving rest-frame UV luminosity function was derived using Lyman break selection to redshift z$\simeq$10 \citep{Bouwens2022}. Although the deep field and lensing searches, when considered independently, gave consistent results, there was debate in the community about the rate of decline in abundance beyond a redshift $z\simeq$8 (Figure~\ref{fig:oesch}). Models of galaxy assembly in standard $\Lambda$\,CDM cosmology together with a star formation rate (SFR) proportional to the baryon fraction in each dark matter halo predicted a steep decline in the SFR density over $8<z<10$. \citet{Oesch2018} argued this was consistent with the limited number of galaxies seen at redshifts z$\simeq$9-10 at the time. However, \citet{McLeod2016,McLeod2021} predicted a continuous decline rate $\propto (1+z)^{-4.2}$ at variance with most theoretical models. The difference was surprising given both teams were using more or less the same observational data, disagreeing only on individual photometric redshifts and how the counts should be binned. The implications were important for JWST given the significant difference in the number of galaxies expected beyond $z\simeq$10. 

\begin{figure}
\center
\includegraphics[width=0.95\textwidth]{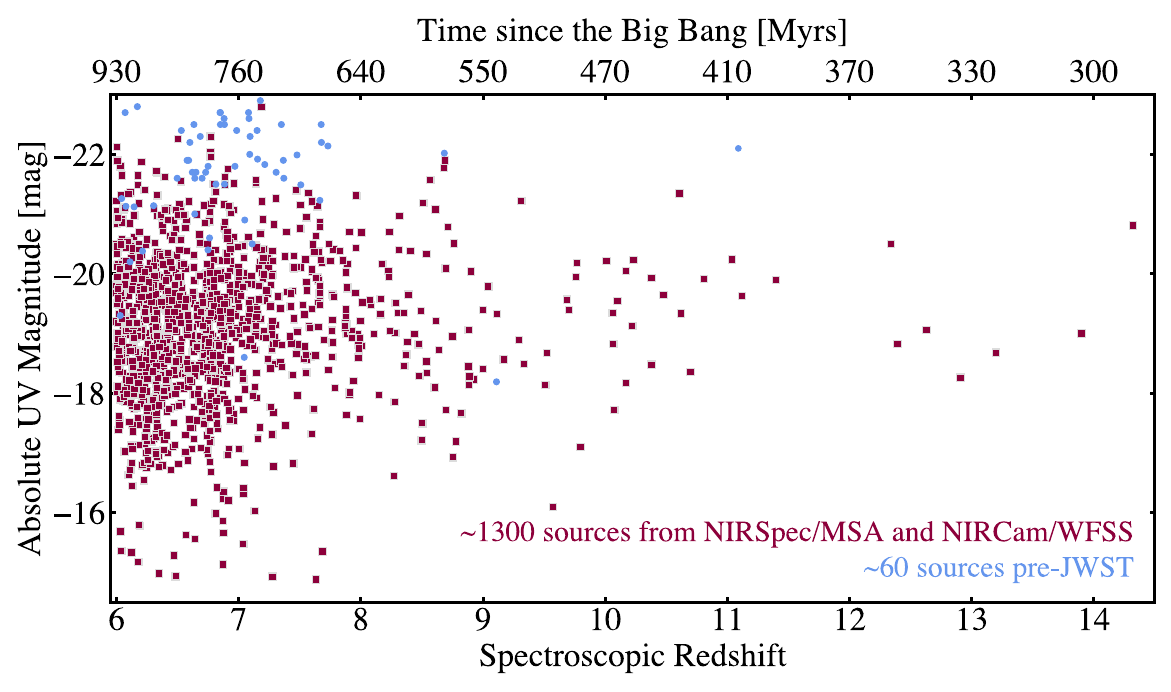}
\caption{\it The dramatic improvement in the $z>$6 spectroscopic redshift dataset from pre-JWST observations (blue dots) to JWST (maroon) as of January 2025. Note also the wider range in luminosity probed within the late reionisation era z$\simeq$6-8 (Courtesy of Guido Roberts-Borsani).}
\label{fig:nz}
\end{figure}

Of course spectroscopic data is crucial, even if only at the most basic level to confirm the photometric redshifts. After over a decade of heroic efforts using Keck's MOSFIRE, VLT's X-shooter, the WFC3 grism on HST and ALMA, a total of only $\simeq$60 spectroscopic redshifts was secured beyond a redshift z=6 collectively by the community. Such observations were usually taken at low spectral resolution and with low signal to noise in regions badly affected by intense airglow emission lines.  Figure~\ref{fig:nz} contrasts the situation pre-JWST with a recent compilation of 1300 JWST spectroscopic redshifts. As we will discuss in Lecture 3, JWST soon resolved the earlier debate on the question of the $z>10$ census by locating sources to $z=$14 and beyond.

\begin{figure}
\center
\includegraphics[width=0.99\textwidth]{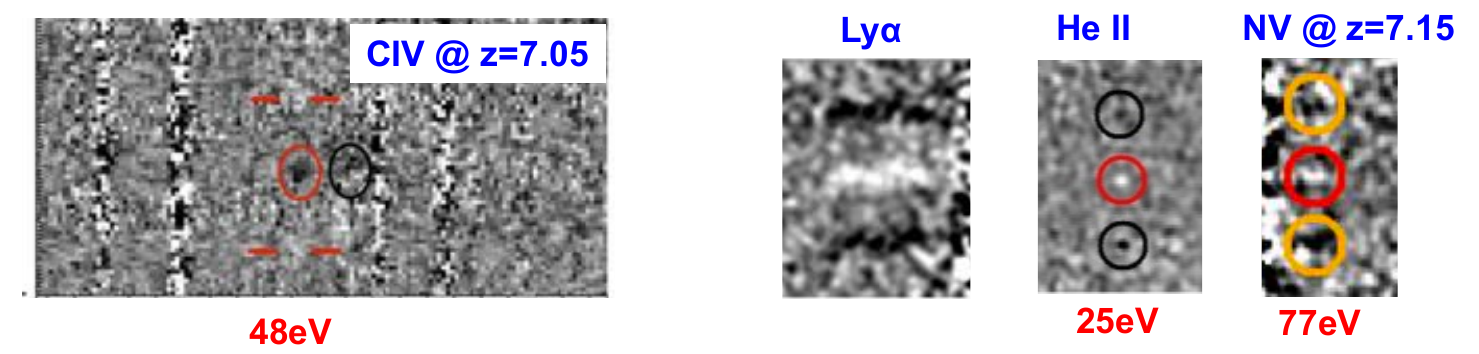}
\caption{\it Evidence for hard radiation fields from high ionisation emission lines and possible AGN from broad Ly$\alpha$ emission at $z>$7 in the reionisation era based on challenging observations with Keck prior to JWST \citep{Stark2015, Stark2017, Laporte2017}. Characterising the role of AGN in the reionisation ersa remains a major JWST activity (see Section 5).}
\label{fig:kecksp}
\end{figure}

Given the vastly improved quality of present-day JWST spectroscopy, it might be tempting to disregard all early pre-JWST articles as inconsequential but, in fact, important results emerged which have stood the test of time. Despite the observational challenges, ground-based spectra revealed emission lines indicative of hard radiation fields and the possible presence of active galactic nuclei at $z>7$. Figure~\ref{fig:kecksp} shows examples of emission in high ionisation lines from CIV (48 eV), NV (77eV) as well as broad Ly$\alpha$ \citep{Stark2015,Stark2017, Laporte2017} using ground-based telescopes. 

\begin{figure}
\center
\includegraphics[width=0.7\linewidth]{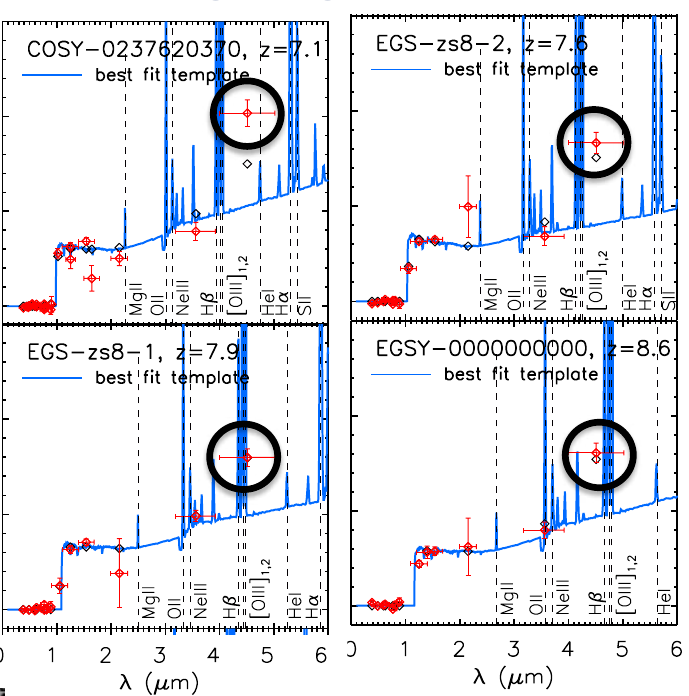}
\caption{\it The discovery of four intense [O III]-emitting galaxies at $7<z<$9 via an excess flux (black circles) seen in the Spitzer 4.5$\mu$m band in their spectral energy distributions \citep{Roberts-Borsani2016}. Subsequent Keck spectra revealed all four show Ly$\alpha$ emission indicating they have created ionised bubbles in an otherwise neutral IGM. This work paved the way for comprehensive surveys of charting ionised bubbles using Ly$\alpha$ emission seen in JWST spectra e.g. \citet{Tang2024}.}
\label{fig:rbs}
\end{figure}

A second important result was the early use of broad-band photometry to trace the presence of intense [O III] emission at redshifts $z\simeq$7-9 using Spitzer 4.5$\mu$m photometry. Four such candidate [O III] emitters were located in the CANDELS fields. Whilst [O III] was beyond spectroscopic reach with both HST and ground-based telescopes, spectroscopic redshifts were nonetheless secured for all four revealing Lyman $\alpha$ emission \citep{Roberts-Borsani2016} (Figure~\ref{fig:rbs}). This is surprising given the IGM should be substantially neutral at these redshifts and, as discussed in Lecture 1, Lyman $\alpha$ should be highly attenuated by resonant scattering. The implication is that each [O III] emitter has, either on its own or together with nearby clustered sources, created a local ionised bubble enabling Ly$\alpha$ photons to propagate. Both results that demonstrated hard radiation fields and ionised regions in the heart of the reionisation era have been confirmed with subsequent JWST data.

\begin{figure}
\center
\includegraphics[width=0.95\linewidth]{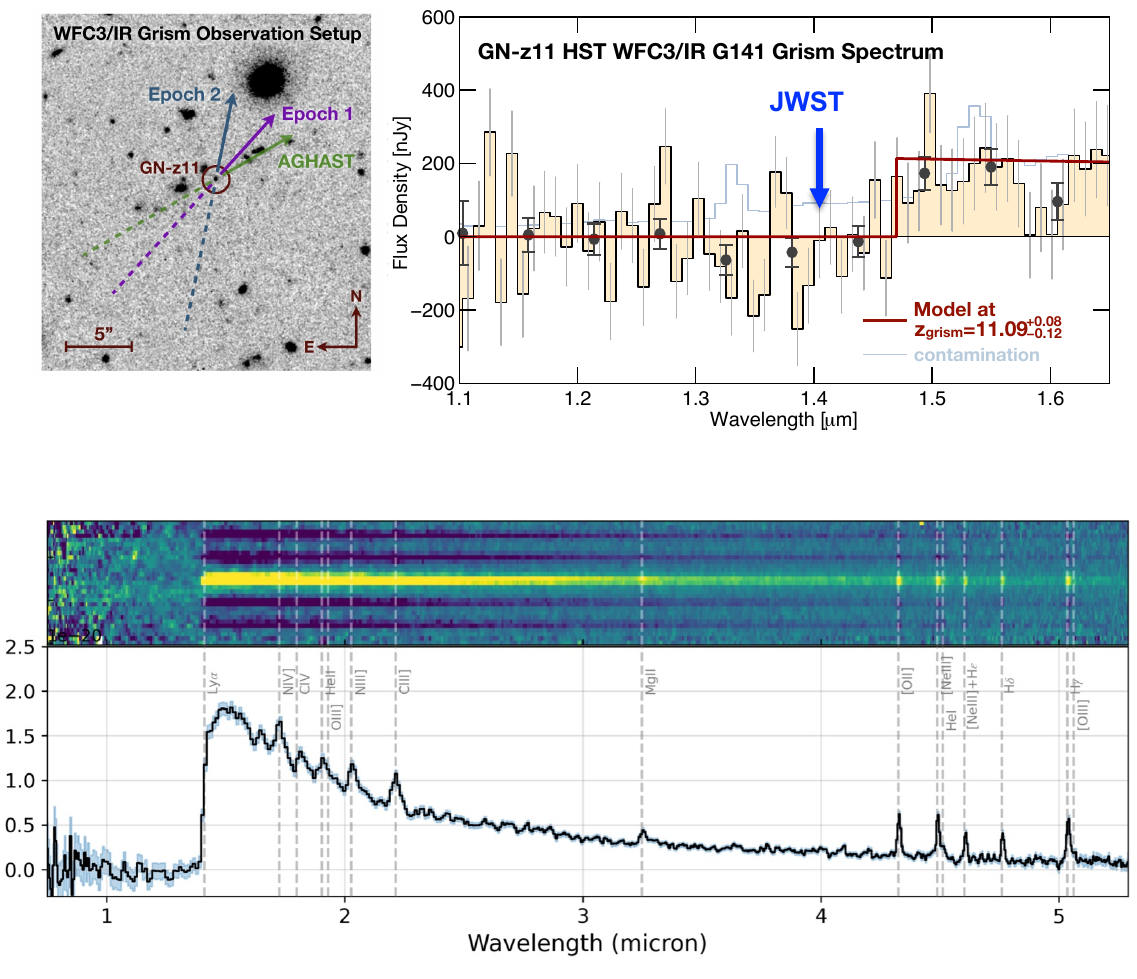}
\caption{\it A comparison of the HST grism and JWST NIRSpec spectrum of the z=10.603 galaxy GNz-11, dramatically illustrating the remarkable progress possible with JWST. The sole feature in the 12 hour HST grism spectrum is the Lyman break, which offers a crude redshift indicator (z=11.1$\pm$0.1) because of the uncertainty of a Lyman damping wing \citep{Oesch2016}. By contrast the 6.9 hour JWST spectrum locates an abundance of diagnostic emission lines yielding a precise redshift of z=10.603. \citep{Bunker2023}.}
\label{fig:gnz11}
\end{figure}

It is also encouraging how the most distant known sources based on challenging observations made with HST and ground-based telescopes have  been confirmed by JWST. Figure~\ref{fig:gnz11} compares the HST and JWST spectra of GNz-11, the brightest $z>10$ Lyman break source in the CANDELS field. Originally claimed to have a photometric redshift of z=10.2 $\pm$ 0.4, a heroic 12 hour HST WFC3 grism observation proposed a spectroscopic redshift z = 11.1 $\pm$0.1 \citep{Oesch2016}. A Keck spectrum suggested a similar redshift based on what transpired to be unlikely emission lines \citep{Jiang2021}. The 6.9 hour NIRSpec prism spectrum demonstrates dramatically the progress now possible with JWST revealing a prominent Lyman break at a redshift z=10.603 intermediate between Oesch et al's photometric and spectroscopic claims \citep{Bunker2023}. An even more remarkable confirmation has been achieved for the most distant claimed source in the Ultra Deep Field \citep{Ellis2013}. This source was only seen in the F160W filter and its absence in the F140W led to two possible explanations - either a Lyman break within the narrow portion of the broad F140W filter transmission that is not covered by the F160W filter, or an unusually strong emission line in the F160W filter e.g. H$\alpha$ at z$\sim$1. Naturally the claim for a photometric redshift of z=11.9 $\pm$ 0.3 based on a detection in a single filter was viewed with some skepticism! However, JWST has confirmed a spectroscopic redshift z=11.1, making this retrospectively the most distant source detected by HST.

\section{Physical Properties}
\label{sec: 3}

We will first introduce the methods in common use to derive the most important physical characteristic for galaxies in the reionisation era and then discuss the results in the context of the emerging picture of galaxy evolution.

\subsection{Stellar Population Synthesis Models}

At the heart of most analyses of galaxy properties is the {\bf\it stellar population synthesis (SPS) code} whose goal is to fit the spectral energy distribution and/or spectrum of a galaxy in the context of a model which incorporates theoretical tracks or ``isochrones" of how stars evolve on the Hertzsprung-Russell diagram according to an assumed initial mass function and metallicity for various star formation histories (SFH). The foundation of this method was established by Beatric Tinsley (1941-81) and her contribution can be found in a remarkably detailed article \citep{Tinsley1980} which, although in a rather obscure journal, has conveniently been resubmitted on the archive\footnote{https://arxiv.org/abs/2203.02041}. 

Popular variants of this basic method in common use include (NB: not an exhaustive list):

\begin{enumerate}
    \item{\it Bruzual \& Charlot - \quad \citep{Bruzual1993}}: A series of empirical models developed from 1980 with regularly updated stellar spectra predicting integrated colours and spectra for various SFHs including single bursts (a.k.a. simple stellar populations, SSP), constant SFR (so-called ``c-models"), exponentially-declining SFRs (so-called ``$\tau$ models" with SFR $\propto\, e^{-\tau}$). 
    \item{\it Conroy \& Gunn - \quad \citep{Conroy2010}}: The Flexible Stellar Population Synthesis (FSPS) code predicts SSPs for various stellar initial mass functions (IMF) and metallicities. The user can then combine these appropriately to create various SFHs. The allied code {\tt Prospector} derives physical parameters from fits to observational data.
    \item{\it BAGPIPES - \quad \citep{Carnall2018}}: Analyses spectral energy distributions using the Bruzual \& Charlot library of stellar spectra including nebular data from the {\tt Cloudy} code \citep{Ferland2017} with Bayesian nested sampling to fit parameters from observations.
\end{enumerate}

\begin{figure}
\center
\includegraphics[width=0.9\linewidth]{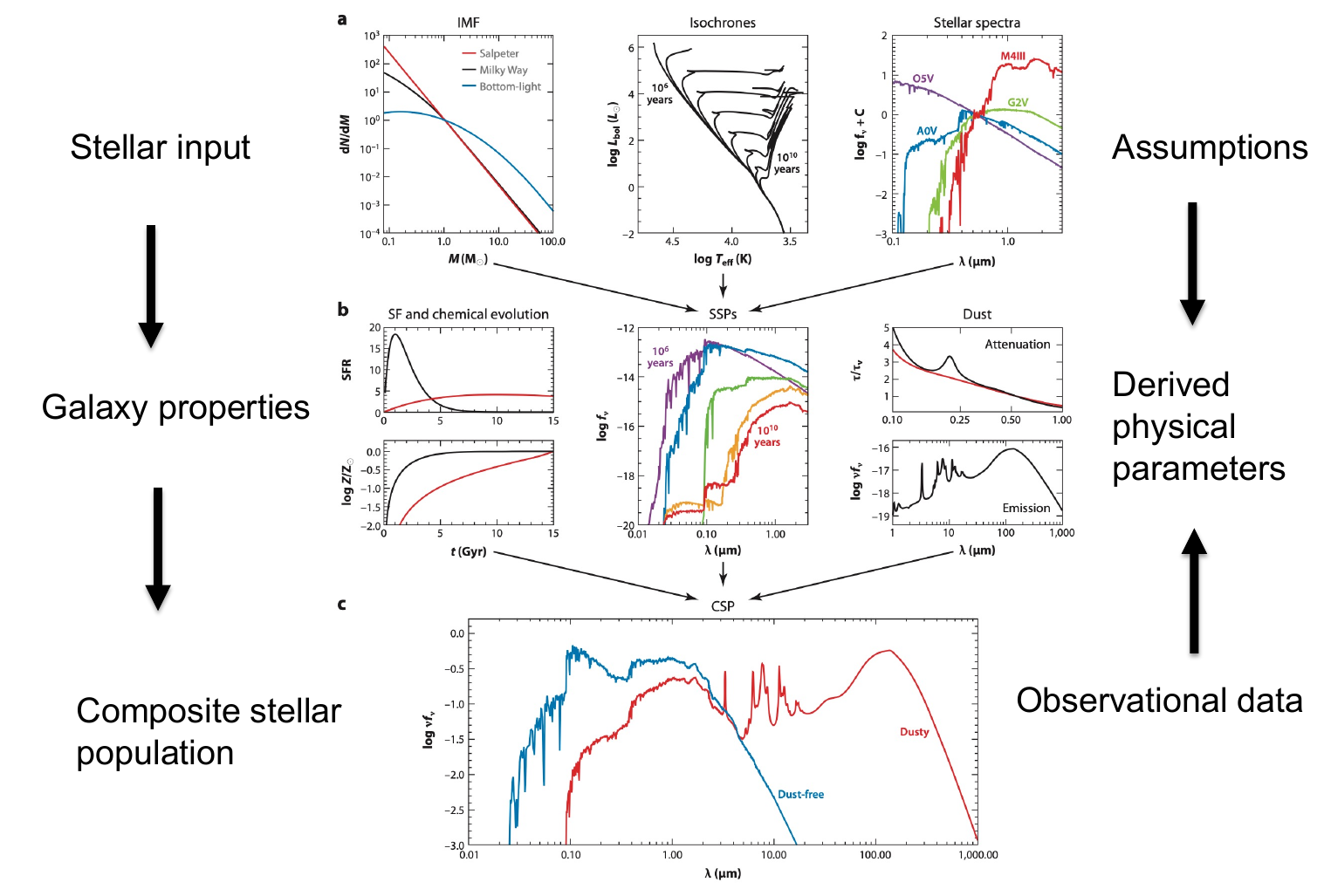}
\caption{\it A succinct summary of a process by which galaxy properties may be inferred from observations using stellar population synthesis models \citep{Conroy2013}. The input properties of stars include their initial mass function, isochrones which predict their luminosity and colour evolution for a given stellar mass and composition, plus observed spectra for the full range of masses, metallicities and ages. A range of star formation and chemical enrichment histories are then adopted that yield a large array of possible composite stellar populations and their spectra and SEDs. Associated codes optimally derive the various parameters by fitting the observations.}
\label{fig:conroy}
\end{figure}

\citet{Conroy2013} provides an excellent review of recent progress in SPS models. In Figure~\ref{fig:conroy} I reproduce a figure from his review that nicely summarises how the input parameters (IMF, isochrones, library of stellar spectra) can be used to deliver a range of composite stellar populations which are fitted to the observational data (spectral energy distributions and/or spectra) to yield physical parameters including star formation rates, stellar masses, metallicities, measures of dust extinction and potentially star formation and chemical enrichment histories.

Whilst these codes represent a considerable investment of effort following decades of work assembling stellar spectra for the full range of masses, ages and metallicities and, as a consequence, are now standard tools in the literature, new researchers may be tempted to use them as a ``black box" without fully appreciating the basic physics. So in what follows, I try to simplify the discussion so the key observational requirements and uncertainties are more apparent.

\subsection{Star Formation Rates}

Although in the local universe there are many independent probes of the star formation rate (SFR) of a galaxy including probes of radio emission from Type II supernova remnants and far-infrared measures arising from dust heated by young hot stars, in practice radio probes cannot yet be applied to meaningful samples at high redshift. The two main probes remaining are Balmer emission lines, notably H$\alpha$, and the rest-frame UV continuum \citep{Kennicutt1998, Madau2014}. H$\alpha$ emission in galaxies normally arises from gaseous HII regions heated by O stars whose main sequence lifetimes are less than 20 Myr. Accordingly it gives a near-instantaneous measure of the SFR. By contrast, the UV continuum luminosity (typically measured at 1500 \AA\ ) arises from a broader range of stars whose main sequence lifetimes extend to $\simeq$100 Myr. Both measures are affected by dust attenuation although that can be estimated using the Balmer decrement, H$\alpha$/H$\beta$. The UV continuum also depends on the stellar initial mass function and stellar metallicity which induces line blanketing at UV wavelengths. Comparisons of both H$\alpha$ and UV continuum-based measures in local samples reveal a scatter which can be used to infer the stochasticity of the star formation history \citep{Sullivan2001}.

\subsection{Stellar Masses}

For those of us studying high redshift star-forming galaxies, it is tempting to focus mostly on the rest-frame UV which, as we discussed above, has featured almost exclusively in constructing the census of early galaxies because detailed measures at longer wavelengths were beyond reach at high redshift with HST and ground-based observatories. However, low mass stars are far more abundant in all stellar populations and since they have long main sequence lifetimes, for extended star formation histories they accumulate over cosmic history thereby contributing significantly to the overall galaxy stellar mass. For a galaxy of a given stellar mass, the K-band (2 $\mu$m) brightness is largely independent of how that mass was assembled and so, to first order, is a reasonable proxy for the stellar mass. This is in marked contrast to the UV and optical colours which reflect only the younger stars \citep{Kauffmann1998}. To accurately determine the stellar mass of galaxy it is thus very important to have observational data at rest-frame infrared wavelengths. At very high redshifts being probed by JWST, the rest-frame K band shifts to 10-20 $\mu$m which is only accessible with MIRI.

\begin{figure}
\center
\includegraphics[width=0.99\linewidth]{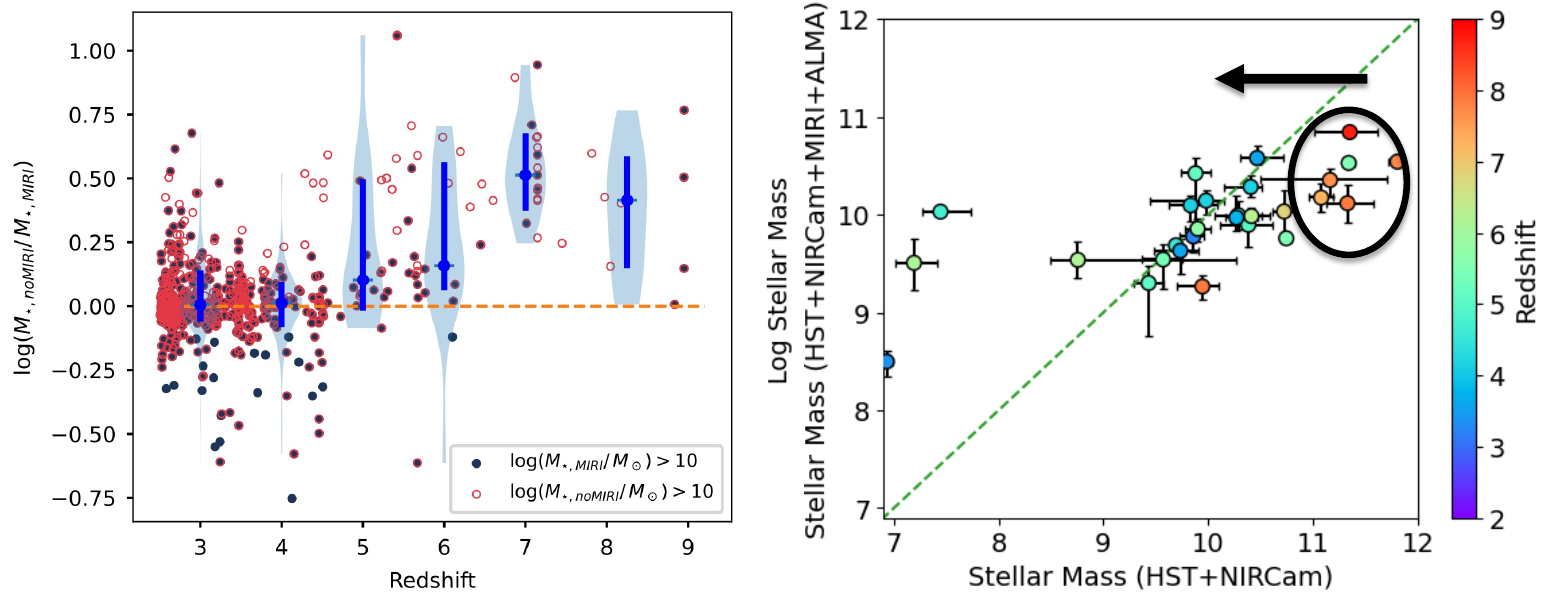}
\caption{\it The importance of MIRI photometry in estimating stellar masses in the reionisation era. (Left) The logarithmic difference between masses determined without and including MIRI as a function of redshift from \citet{TWang2024}. (Right) A direct comparison of MIRI+NIRCam and NIRCam-only masses from \citet{Williams2024}. Due to an inability to break the degeneracy of SED fitting between age and dust content, NIRCAM-only stellar masses at $z>7$ can be overestimated by factors of $\times$3-5.}
\label{fig:miri_mass}
\end{figure}

The technique for inferring stellar masses $M_{\ast}$ in galaxies follows the approach discussed in \citet{Brinchmann2000}. The spectral energy distribution (SED) is fit with SPS models to derive a mass/light ratio of the overall stellar populations e.g. for the K band this would be $(M_{\ast}/L)_{K}$. Given the redshift of the galaxy, the K-band luminosity $L_{K}$ is known and hence the stellar mass can be determined. Despite the well-established importance of ensuring the rest-frame infrared is incorporated in SED fits, much of the recent literature on JWST stellar masses is still primarily derived only from NIRCam photometry. At $z>7$ this corresponds to an upper rest-frame wavelength of only $\lambda <$ 6000 \AA\ . The dangers of not including MIRI data for galaxies beyond z$\simeq$5 have been illustrated by \citet{TWang2024, Williams2024}. Without MIRI data, masses are overestimated since, over the restricted rest-frame coverage available with NIRCam, there is a degeneracy between the effects of age and dust. As shown in Figure~\ref{fig:miri_mass}, this can amount to overestimates as large as a factor of $\times$3-5.

\subsection{Stellar Ages}

\begin{figure}
\center
\includegraphics[width=0.99\linewidth]{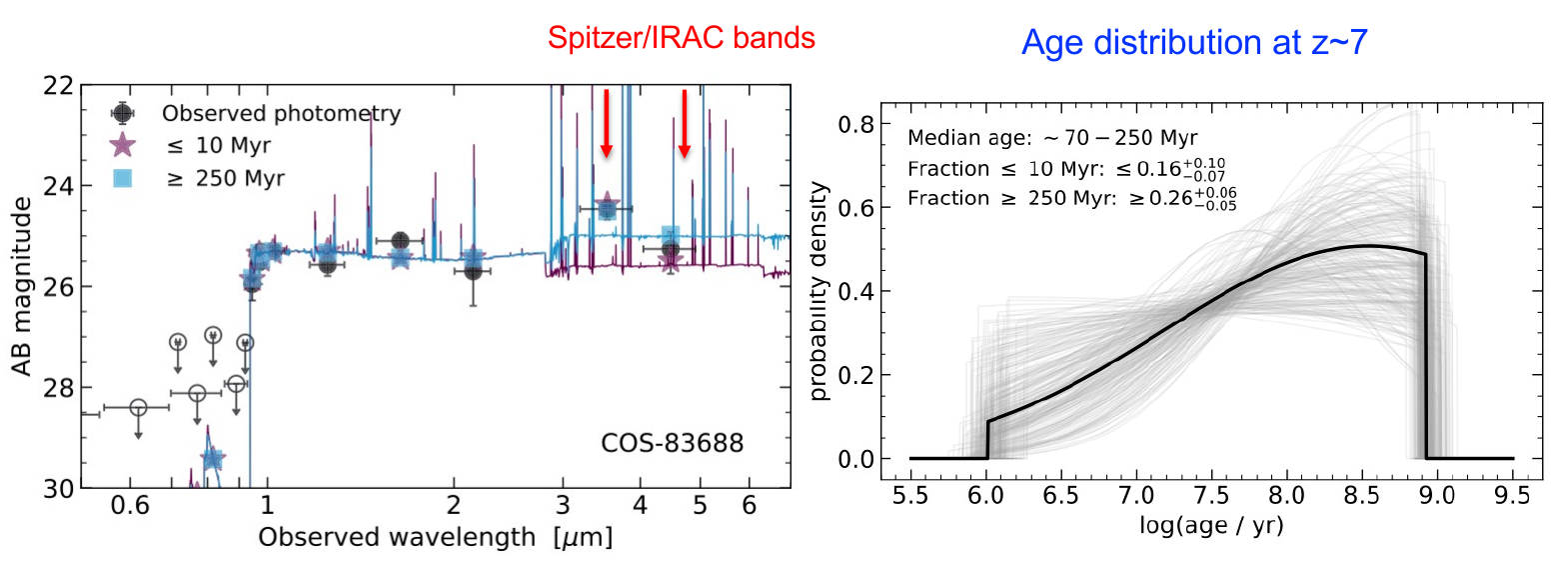}
\caption{\it Pre-JWST illustration of the important role of rest-frame infrared photometry in detecting old stars in actively star-forming galaxies at high redshift \citep{Whitler2023}. (Left) The SED of a z=6.6 galaxy where the Spitzer/IRAC 3.6 and 4.5$\mu$m photometry indicates an excess flux which cannot arise from known emission lines and thus represents the continuum radiation from a population of old stars. Ages as diverse as 10 and 250 Myr cannot be distinguished from SED fits to the shorter wavelength data. (Right) Incorporating the Spitzer data for a sample of 6.6$<z<$6.9 galaxies, the inferred star formation history indicates less than 16\% are truly young ($<$10 My) whereas at least 26\% have ages $>$250 Myr corresponding to redshifts of formation as high as $z\simeq$9.}
\label{fig:whitler}
\end{figure}

The age of a stellar population is naturally similarly affected if the underlying older, redder stars cannot be adequately probed. Numerous articles in the recent JWST literature have claimed $z\simeq$7-8 galaxies are ``young" with ages $<$10 Myr but this likely represents the ``outshining" effect whereby at rest-optical wavelengths the contribution to the SED is dominated by such young stars. Although it pre-dates the JWST era, this outshining effect was nicely illustrated for a sample of $z\simeq$6.6-6.9 Lyman break galaxies using Spitzer 4.5$\mu$m photometry \citep{Whitler2023}. Only by incorporating the Spitzer photometry is it possible to distinguish between a mean stellar age of 10 and 250 Myr. In this particular sample, 25\% of the population has stellar ages greater than 250 Myr consistent with formation at redshifts z$\simeq$9 and earlier (Figure~\ref{fig:whitler}). A comprehensive demonstration of the dangers of outshining in analysing JWST photometry in the case of bursty star formation is given by \cite{BWang2025}.

This is not to diminish the importance of age-related signatures in rest-frame optical spectra. The most significant feature of importance is the Balmer break which is most prominent for A-type stars whose ages are $<$500 Myr. The physical basis of this age-indicator is that hydrogen absorption in the outer atmospheres of stars is highly dependent on the star's mass. At low stellar masses with cooler temperatures, the principal source of opacity is the negative hydrogen ion $H^-$, whereas in the most massive hot stars, the hydrogen is fully ionised. Only in the intermediate mass stars with lifetimes of several 100 Myr is this Balmer break visible. Again, it is easily diluted if there has been residual star formation within the last few Myr and even if the break is located it only provides a lower age limit for the stellar population. 

\begin{figure}
\center
\includegraphics[width=0.99\linewidth]{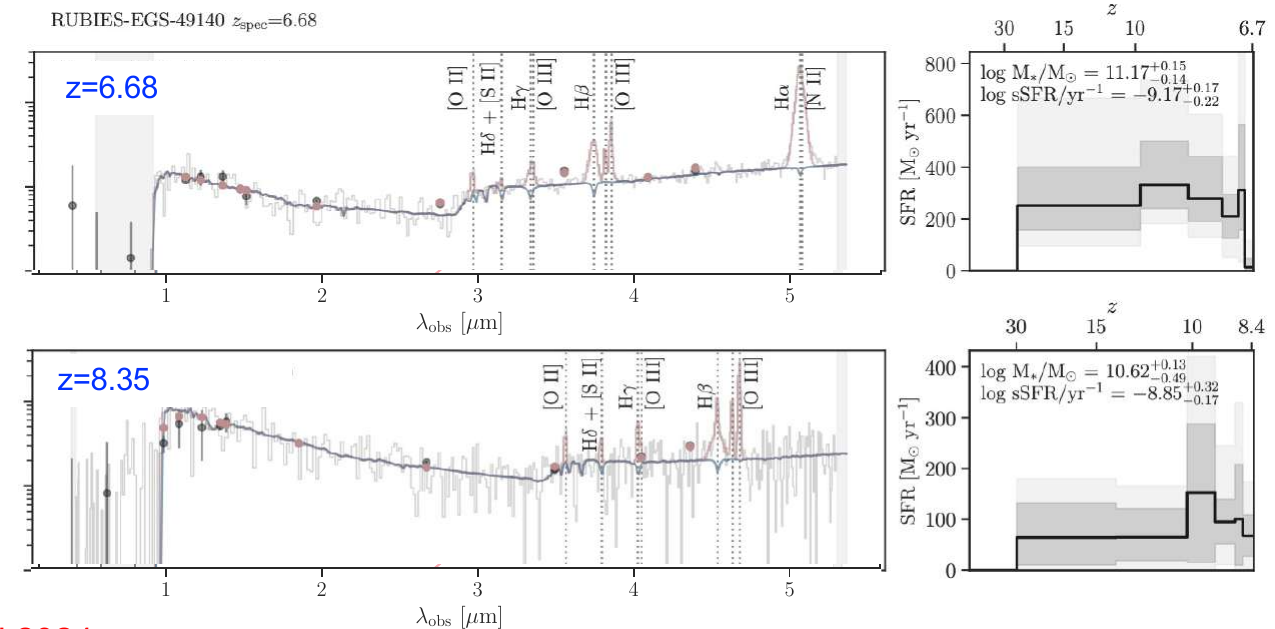}
\caption{\it The Balmer break as an age diagnostic \citep{BWAng2024}. (Left) Spectra and SED for two galaxies from the RUBIES programme which show clear evidence of both a photometric and spectroscopic Balmer break seen at 3$\mu$m. The feature arises only in stars with main sequence ages of several 100 Myr and thus can be used as a lower limit for the age of the stellar population. (Right) Inferred star formation histories for both galaxies indicating possible contributions beyond $z\simeq$10. (The sensitivity to activity at yet higher redshifts is limited.)}
\label{fig:balmer}
\end{figure}

High quality spectra are needed to locate and accurately measure this Balmer break. The interpretation of its strength from broad-band photometry alone is often challenging due to contamination from rest-frame optical emission lines e.g. [O II]. The RUBIES JWST programme has demonstrated the presence of Balmer breaks in several reionisation-era galaxies and derived past SFHs extending to $z\simeq$10 and earlier (Figure~\ref{fig:balmer} , \citet{BWAng2024}, see also \citet{Kuruvanthodi2024, Roberts-Borsani2024}).

\subsection{UV continuum slopes and nebular contributions}

The UV continuum slope is a further important diagnostic than can be derived either photometrically or spectroscopically. It is not only sensitive to the age of the actively star-forming population in high redshift galaxies, but also to the stellar IMF, dust content, metallicity and the ionising production rate ($\xi_{ion}$, see Lecture 1). Traditionally the slope $\beta$ is measured between 1250 and 2600 \AA\ according to the following relation:

\begin{equation}
f_{\lambda} \propto \lambda^{-\beta}
\end{equation}
such that steeper UV slopes correspond to more negative values of $\beta$.

\begin{figure}
\center
\includegraphics[width=0.9\linewidth]{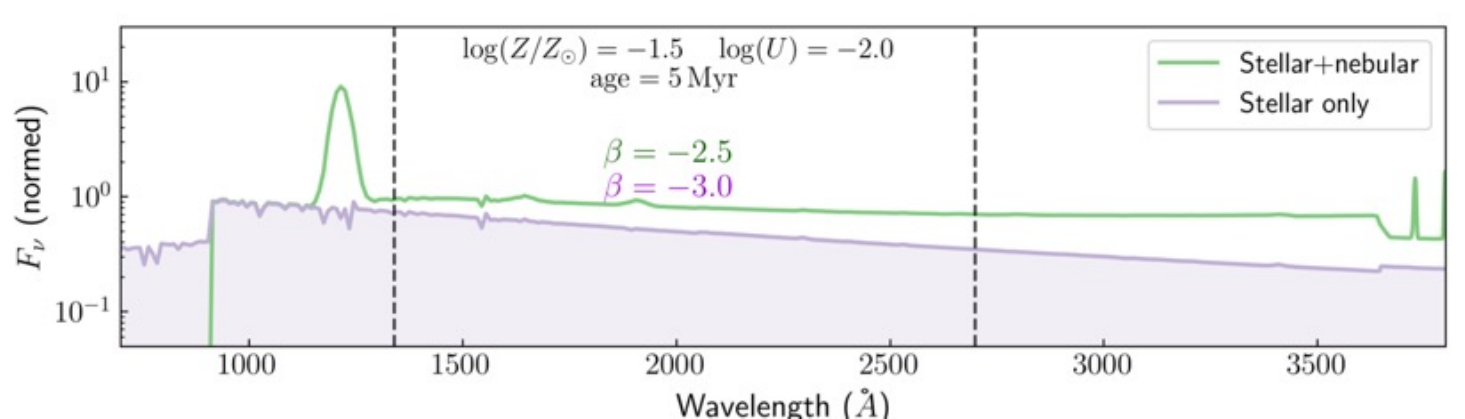} \\
\includegraphics[width=0.45\linewidth]{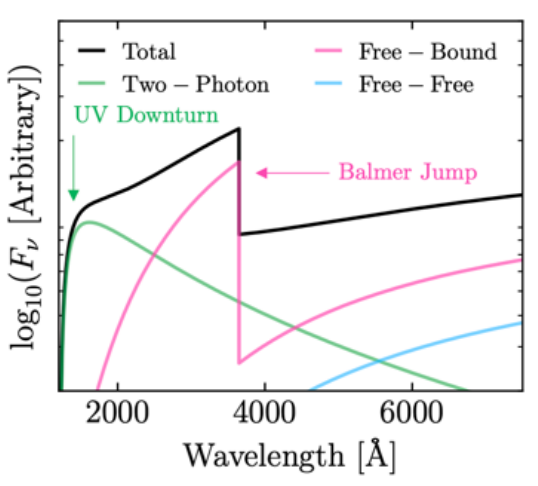}
\vspace{+1.5cm}
\includegraphics[width=0.50\linewidth]{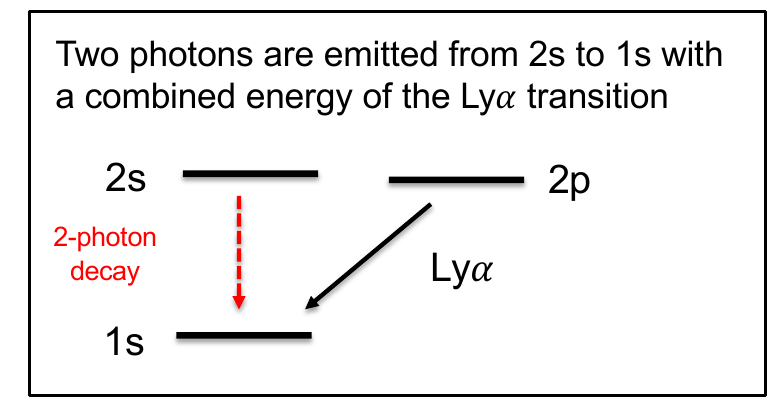} \\
\includegraphics[width=0.95\linewidth]{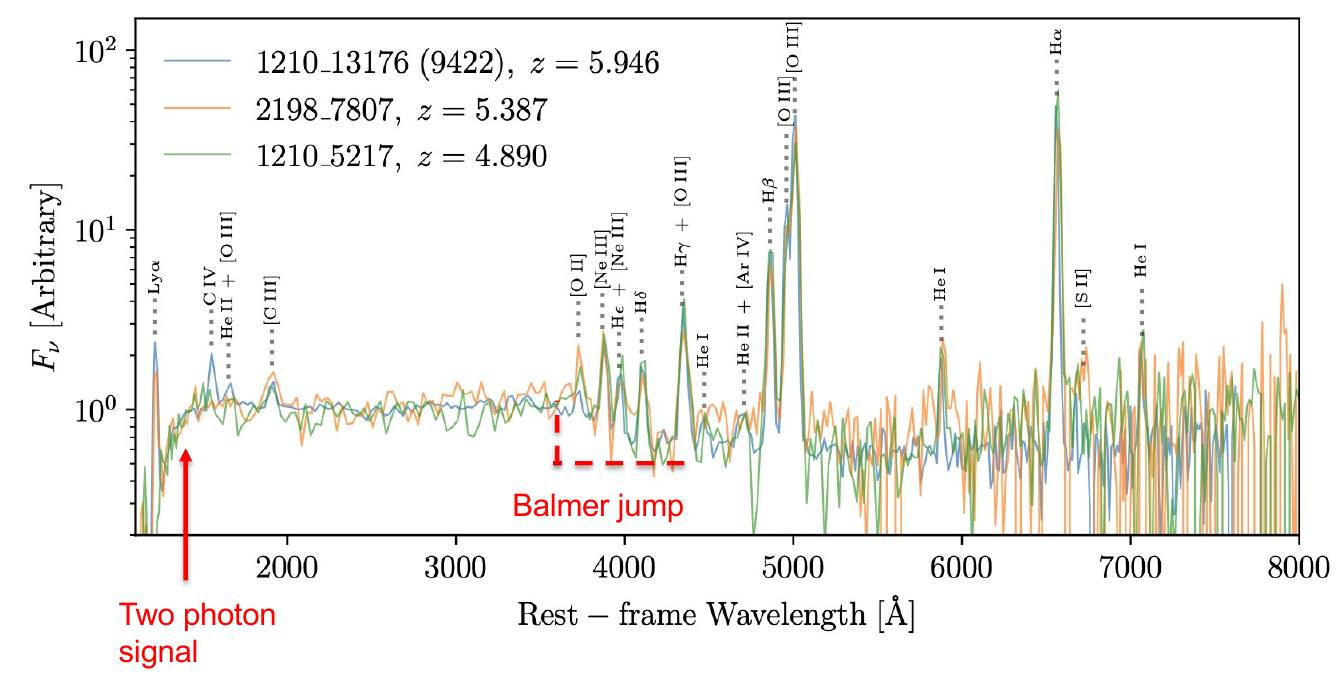}
\caption{\it Aspects of the UV continuum slope. (Top) Model UV spectrum for a instantaneous starburst illustrating the steepest UV slope $\beta$ possible for a stellar only and stellar + nebular contribution \citep{Saxena2024}. (Centre left) Components of the nebular continuum which arise from free-free, free-bound and two-photon emission. The key observational features are a strong negative Balmer jump arising from free-bound emission and a UV downturn longward of Ly$\alpha$ arising from two-photon emission \citep{Katz2024}. (Centre right) Physical origin of two-photon decay. Two photons are simultaneously emitted from the 2s to 1s state of hydrogen which can break the normal selection rule if their combined energies equals that of Lyman $\alpha$ (2p to 1s). (Bottom) Spectra for three galaxies showing the negative Balmer jump and two-photon UV downturn \citep{Katz2024}. }
\label{fig:beta}
\end{figure}

Stellar population synthesis models for an instantaneous burst of low metallicities predict $\beta >-3.0$ for a standard IMF. However, as young massive stars are capable of heating up the surrounding gas the slightly redder nebular continuum may also contribute, and when included the instantaneous limit becomes $\beta >-2.6$ (see Figure~\ref{fig:beta}). First encountered astrophysically in planetary nebulae, hot stars can induce free-bound, free-free and two-photon continuous emission in gaseous hydrogen. Its presence can be distinguished from starlight via a {\it negative Balmer jump} and a {\it UV downturn at 1400\AA\ } arising from two-photon emission \citep{Katz2024}. In the latter case two photons are emitted from the 2s to 1s data with a combined energy equivalent to the Lyman $\alpha$ transition (see centre right panel of Figure~\ref{fig:beta}). The importance of the nebular continuum follows from its sensitivity to a top-heavy IMF which some believe may be a signature of Population III stars (Lecture 3). Convincing examples of galaxy spectra revealing nebular continuum emission are shown in the lower panel of Figure~\ref{fig:beta}. Model fits to such spectra suggest the fractional contribution from nebular emission at 1500 \AA\ can be as high as 60\%. Although a dominant nebular continuum would temporarily increase the UV luminosity of early galaxies, it will be challenging to identify such signatures at redshifts $z>10$. Indeed, the downturn signature of two-photon emission is difficult to distinguish from damped Lyman $\alpha$ absorption.

The question of how the UV continuum slope $\beta$ evolves with redshift has been the focus of much attention in the early years of JWST observations. There have been two basic approaches, each motivated by attempts to chart the effects of decreasing stellar age, metallicity and dust content with increasing redshift. Initially photometric measures had the advantage of large well-controlled samples but individual slope measures can be affected by (unseen) emission lines. Spectroscopic measures are more accurate as any strong emission lines can be masked. However, the spectroscopic samples were initially more modest and could be biased. A common technique is to compile a redshift catalogue from diverse JWST spectroscopic campaigns, each with different selection criteria. Inclusion of a galaxy in such a catalogue may depend on the star-formation rate and/or the presence of emission lines often essential for a robust redshift, possibly rendering the overall sample unrepresentative of the underlying population.

\begin{figure}
\center
\includegraphics[width=0.9\linewidth]{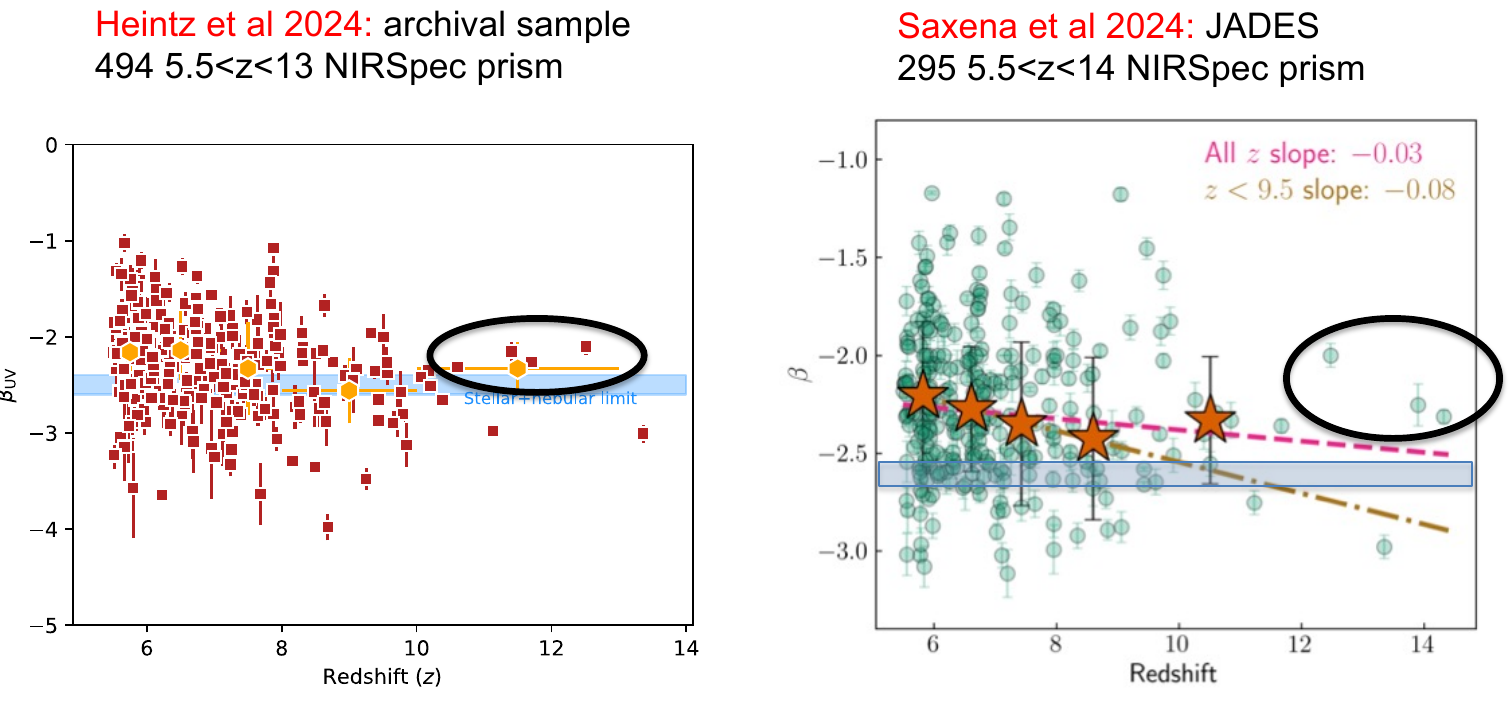}
\caption{\it Spectroscopic measures of the UV continuum slope $\beta$ from archival data \citep{Heintz2025} and the JADES GTO programme \citep{Saxena2024}. The stellar + nebular limit of $\beta<-2.6$ for an instantaneous burst of star formation is indicated by the horizonal blue shaded line. Redshift-binned averages are shown as star symbols. There is a reasonably significant trend for steeper $\beta$ with increasing redshift and a tentative upturn for $z > $10 which may arise from an increasing contribution from a redder nebular contributions (highlighted in black).}
\label{fig:beta_z}
\end{figure}

Early photometric studies \citep{Cullen2023, Topping2024} found declining values of $\beta$ over 8$< z <$14 and steeper values for lower luminosity galaxies as might be expected if they are more metal-poor. However, many high redshift galaxies were found to have unphysically steep values (e.g. $\beta\simeq$-4) raising questions of accuracy or possibly significant Lyman continuum leakage. If the escape fraction $f_{esc}$ (Lecture 1) is very high, clearly the nebular continuum will be weakened, permitting more extreme value of $\beta$. Generally speaking the trends are similar for the later spectroscopic studies \citep{Heintz2025, Roberts-Borsani2024, Dottorini2025}, but with the benefit of less scatter and fewer unphysical values (Figure~\ref{fig:beta_z}). $\beta(z)$ tends to asymptote to the required instantanous value of -2.6 but it's interesting to note a marginal upturn to redder values for the small sample of $ z > 10$ galaxies that may indicate an increased contribution from nebular continuum in young systems with very massive stars.

\section{Chemical Evolution}
\label{sec: 4}

Arguably the most important progress with JWST has been through spectroscopic studies of early galaxies. Although basic redshift measures were possible with the WFC3-IR grism on HST and some diagnostic indicators of $\xi_{ion}$ and potential AGN activity followed ambitious exposures on Keck and VLT, the absence of atmospheric airglow in space and the multiplex gain of JWST's spectrographs has provided a major revolution. Here we focus on what has been learned in tracing the gas-phase chemical composition with increasing redshift (see \citet{Curti2025} for a timely review). 

\begin{figure}
\center
\includegraphics[width=0.9\linewidth]{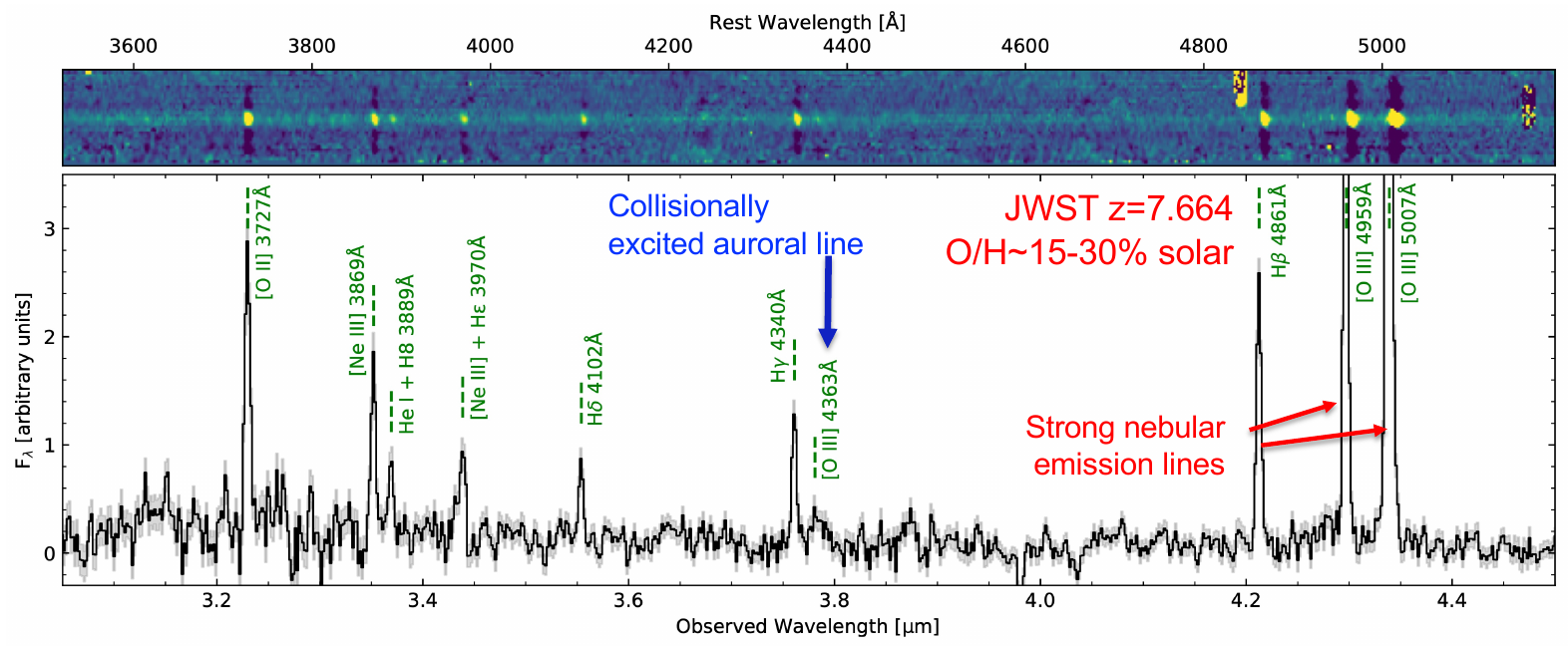}
\caption{\it The rich emission line JWST spectrum of a z=7.66 galaxy taken in the early period of operations \citep{Schaerer2022, Curti2023}. In addition to the familiar rest-frame optical lines of [O II], [O III] and Balmer H$\beta$ which have formed the basis of metallicity measurements from strong line ratios for decades, the detection of the $T_e$-sensitive collisionally-excited auroral line of [O III] 4363 \AA\ can be used to calibrate strong-line data in the reionisation era.}
\label{fig:z7spec}
\end{figure}

A good example is the spectrum of a z=7.66 galaxy released shortly after science operations commenced in July 2022 (Figure~\ref{fig:z7spec}, \citet{Schaerer2022, Curti2023}. For the first time in the reionisation era, JWST provides access to the familiar strong nebular emission lines of [O III] 4959 and 5007 \AA\ and Balmer H$\beta$, which have underpinned metallicity measures for almost a century. However, of greater significance is the detection of the weak auroral [O III] line at 4363 \AA\ which provides a crucial calibrating measure of the electron temperature $T_e$. Prior to JWST, observers struggled to calibrate gas phase metallicities derived from the strong [O III] lines and had to rely on local $T_e$ measures which were unlikely to be representative at high redshift. Similar progress with $T_e$-sensitive auroral lines has been accomplished for other atomic species. 

\begin{figure}
\center
\includegraphics[width=0.9\linewidth]{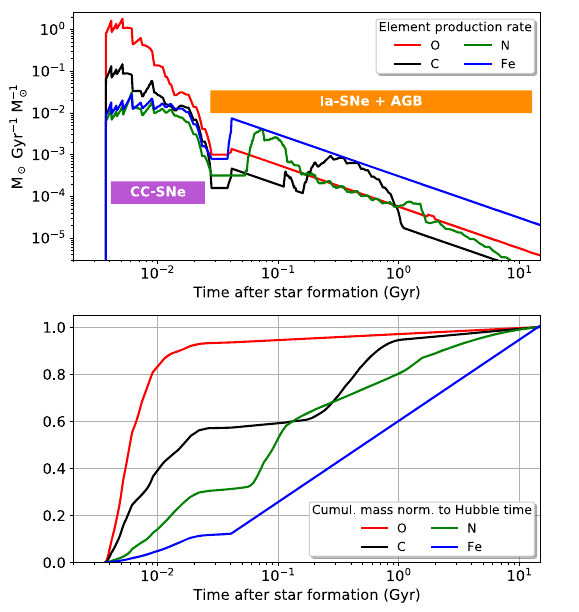}
\caption{\it Chemical enrichment history for various atomic species following a single burst of star formation \citep{Maiolino2019}. (Top) The mass production rate in solar masses per Gyr normalised to a solar mass of formed stars. (Bottom) The cumulative fraction produced normalised to that produced in a Hubble time. Oxygen forms rapidly from Type II supernovae whereas iron production is dominated by Type Ia supernovae which requires completion of stellar evolution for lower mass stars. Carbon and Nitrogen involve contributions from several processes including mass loss from intermediate age AGB stars.}
\label{fig:cnofe}
\end{figure}

Before discussing the recent JWST progress, let us recap on the likely timescale of enrichment for the various chemical elements we can detect in JWST spectra. Heavier nuclei are formed continuously via hydrostatic burning in the cores of all stars during their lifetimes. This includes hydrogen to helium while on the main sequence and progressively heavier elements up to iron during various stages of post-main sequence burning depending on the stellar mass. Although mass loss during the asymptotic giant branch (AGB) phase contributes to the enrichment of the interstellar medium, explosive nucleosynthesis that occurs during the final stages of supernovae is the most important mechanism. The so-called $\alpha$ elements (C, O, Mg, Ne, S, Si) are promptly produced in Type II supernovae following the death of short-lived massive stars. In contrast, iron-peak nuclei are produced more slowly in Type Ia supernovae which result from the destruction of a low mass white dwarf whose stellar lifetime is much longer. As we proceed towards stellar systems whose ages are less than a few hundred million years at redshifts $z>$10, we expect little enrichment from AGB mass loss and Type Ia supernovae \citep{Maiolino2019} (Figure~\ref{fig:cnofe}).

How then do we estimate the chemical abundances from an emission line spectrum? In general an emission line flux $F(\lambda)$ is the product of the ionic abundance $Z_{ion}$ and an emissivity function $\epsilon$ 

\begin{equation}
   F(\lambda) = Z_{ion} * \epsilon\,(T_e, n_e) 
\end{equation}

the latter of which depends on the physical conditions in the gas, particularly the electron temperature and density. The derived abundance of the element $Z$ follows corrections for the ionic fraction observed and is usually quoted with respect to the hydrogen abundance measured from recombination lines, i.e. $Z/H$. When quoted on a logarithmic scale with respect to the solar abundance ratio, it is placed in square brackets, e.g. $[O/H]$=-1.0 would indicate an oxygen abundance of 10\% solar. Another convention is to quote the abundance on a scale where H=12, e.g. 12 + log (O/H) = 7.6 would also represent an abundance of oxygen 10\% solar given the sun's oxygen abundance is 12 + log (O/H) $\simeq$8.6.

\begin{figure}
\center
\includegraphics[width=0.99\linewidth]{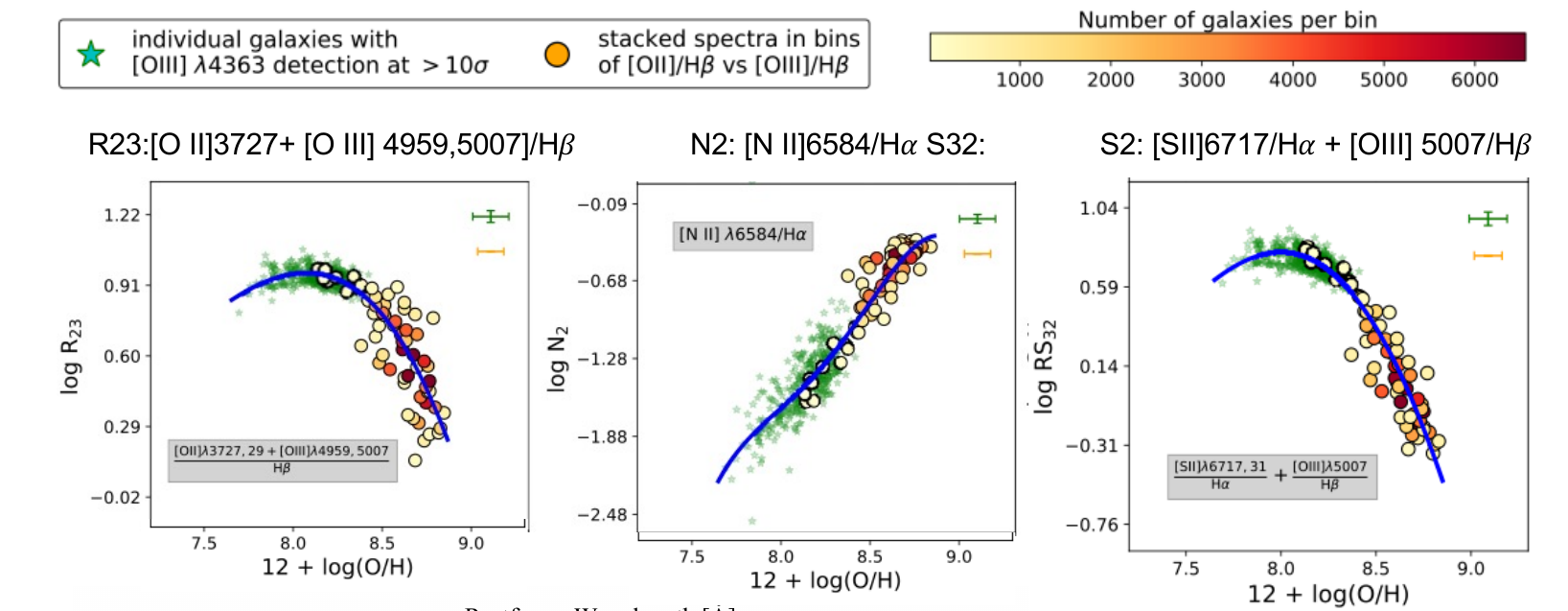}
\caption{\it Calibration of three popular strong line ratios using the collisionally-excited auroral line [O III] 4363 \AA\ for galaxy spectra in the Sloan Digital Sky Survey \citep{Curti2017}.}
\label{fig:sdss_cal}
\end{figure}

\begin{figure}
\center
\includegraphics[width=0.9\linewidth]{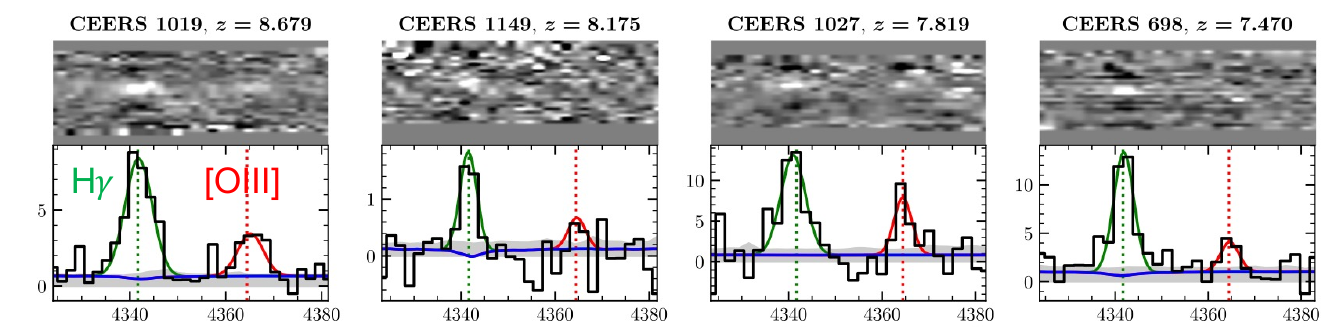} \\
\includegraphics[width=0.5\linewidth]{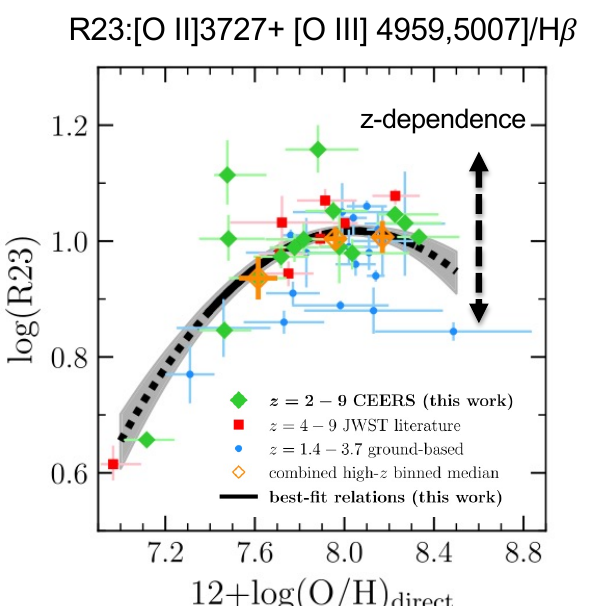}
\caption{\it Application of the collisional excited line method in the reionisation era. (Top) Detection of [O III] 4363 \AA\ at z$\simeq$7-9. (Bottom) The R23 line ratio calibration as seen at various redshifts. The arrow shows the possible redshift dependence which is strongest over 0 $ < z < 2$ \citep{Sanders2024}.}
\label{fig:z8_calib}
\end{figure}

We can estimate the abundances of various species in two ways, via collisionally-excited lines (CEL) or recombination lines (RL). As the collisionally-excited lines are usually much stronger, this is the common method using auroral lines of high principal quantum number $n$ e.g. [O III] 4363 \AA\ to determine $T_e$ and other doublet line ratios e.g. [O II] 3727 \AA\ to determine $n_e$. Although the RL method is less sensitive to $T_e$, it is currently only practical for certain species (C,N,O) in nearby galaxies. To illustrate the CEL method, Figure~\ref{fig:sdss_cal} shows the auroral calibration for three popular strong line ratios for SDSS survey galaxies. It can be seen that some of the line ratios are double-valued in terms of the derived abundance \citep{Curti2017}. The real breakthrough with JWST has been the ability to identify such auroral lines at high redshift which has enabled us to bypass assuming such local $T_e, n_e$ calibrations for the various strong emission line ratios. \citet{Sanders2024} first applied this to a sample of 16 galaxies at $z>6$ which, together with examples in the literature, provided a dataset of 46 galaxies over all redshifts. As expected, there is evidence of a redshift-dependence in the calibration implying $T_e$ changes, although most of the evolution occurs over 0 $<z<$2 (Figure~\ref{fig:z8_calib}, see also \citet{Chakraborty2025}).

\begin{figure}
\center
\includegraphics[width=0.32\linewidth]{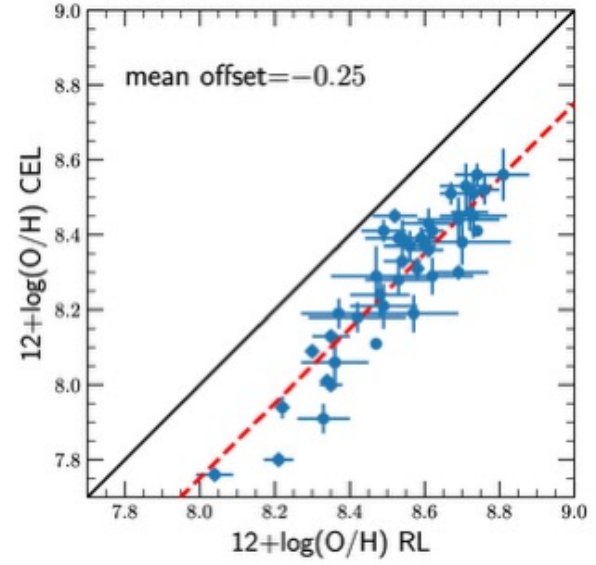}
\includegraphics[width=0.32\linewidth]{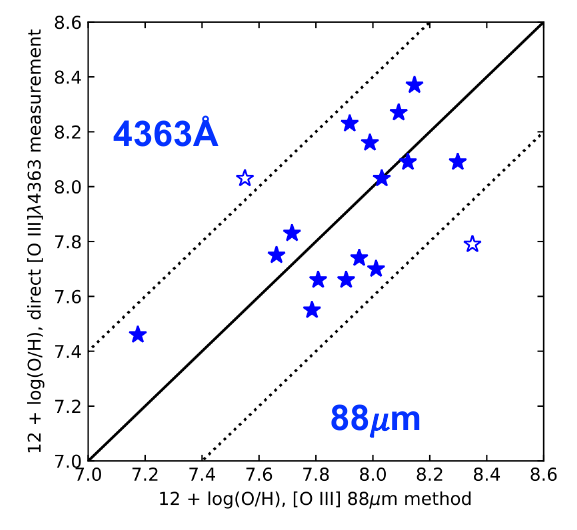}
\includegraphics[width=0.32\linewidth]{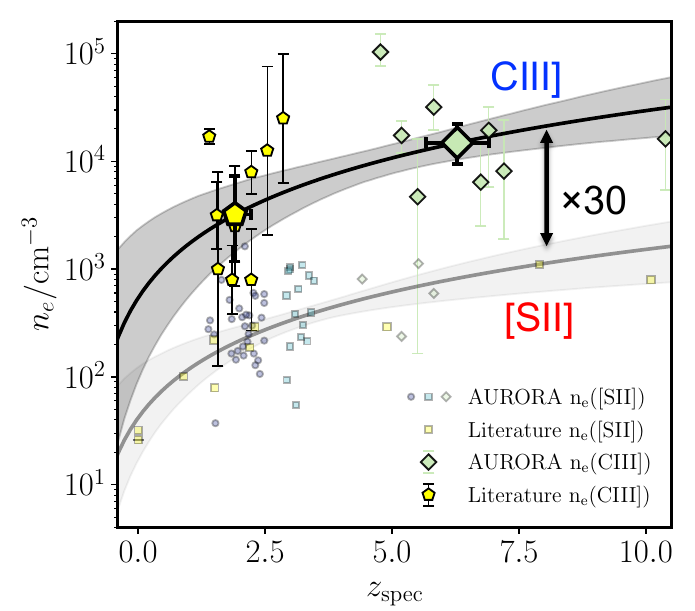}
\caption{\it Cautionary results when considering oxygen abundances derived from the CEL method: (Left) A puzzling discrepancy between the CEL and RL abundances for local galaxies may arise from fluctuations in $T_e$ in the ISM, in which case using weak auroral lines may bias selection to high $T_e$ regions \citep{Garcia-Rojas2007}. (Centre) The scatter seen between complementary calibrations based on far-IR [O III] 88$\mu$m and optical [O III 4363 \AA\ lines may support this hypothesis \citep{Jones2020}. (Right) Evidence for a multi-phase ISM is also found from inconsistent $n_e$ values using different doublet lines sensitive to the ionisation state of the gas \citep{Topping2025}.}
    \label{fig:multi_ism}
\end{figure}

Unfortunately, when we compare the CEL and RL abundances for nearby galaxies, they disagree by about 0.25 dex (left panel of Figure~\ref{fig:multi_ism}). Although the origin of this discrepancy is unclear, it may imply there are spatial fluctuations in $T_e$ within the ISM of galaxies and thus when we detect weak [O III] 4363 \AA\ we may be biased to regions of high $T_e$ \citep{Garcia-Rojas2007, Mendez-Delgado2023}. ALMA may be able to assist, since there are sub-mm emission lines e.g. [O III] 88$\mu$m and 52$\mu$m which are much brighter and accessible at high redshift that can be used in combination to control such biases \citep{Jones2020}. Comparing Herschel [O III] 88$\mu$m and SDSS [O III] 4363 \AA\ calibrations there is a large scatter which might support this explanation (centre panel of Figure~\ref{fig:multi_ism} ).

Furthermore, for the small number of z$\simeq$6-7 galaxies for which both JWST [O II] 3727 and 4363 \AA\  and ALMA 88 $\mu$m and 52$\mu$m provide independent measures, there are discrepancies. The far-infrared ALMA data indicates low electron gas measures ($n_e\sim$100 cm$^{-3}$, $T_e\sim$8000K) in contrast to the JWST rest-frame optical lines ($n_e\sim$1000 cm$^{-3}$, $T_e\sim$13000K) thereby suggesting a multi-phase ISM \citep{Harikane2025}. Even within the optical data, there are inconsistencies between electron densities derived from doublets of various ionisation levels. The most commonly-used doublets [O II] 3727/29 and [S II] 6717/31 trace low ionisation gas ($\simeq$10 eV) whereas CIII] 1907/09 and [Si III] 18883/92 trace higher ionisation gas ($\simeq$16-24 eV). For a uniphase ISM, the $n_e$ estimates should agree. However, over 1$<z<7$, the CIII]-derived $n_e$ values are consistently $\simeq$30 times larger also suggestive of a highly-structured multi-phase ISM, thereby complicating the interpretation of result based on integrated spectroscopic data (right panel of Figure~\ref{fig:multi_ism}, \citet{Topping2025}).

Perhaps understandably, much early work on chemical evolution has focused on estimating the metallicities of galaxies at the redshift frontier $z>$10.  The remarkable NIRSpec prism and grating spectra of GN-z11 at $z=10.6$ has yielded many surprises. Although the oxygen abundance is approximately 12\% solar, the N/O ratio is 4 times solar \citep{Bunker2023}. This is based on prominent lines of NIV and NIII] which are rarely seen in galaxy spectra (c.f. Figure~\ref{fig:gnz11} ). The origin of this N/O excess is currently unclear. Other N-excess sources located by JWST have abundance patterns resembling those seen in the stars in local globular clusters. A possible explanation is preferential N-enrichment following ejecta from super-massive ($>10^3 M_{\odot}$) stars that collide in dense hot star-forming regions \citep{Cameron2023}. At the time of writing, oxygen abundances for the first swath of $z>10$ spectra range from 3-12\% solar \citep{Castellano2024}.

\begin{figure}
\center
\includegraphics[width=0.9\linewidth]{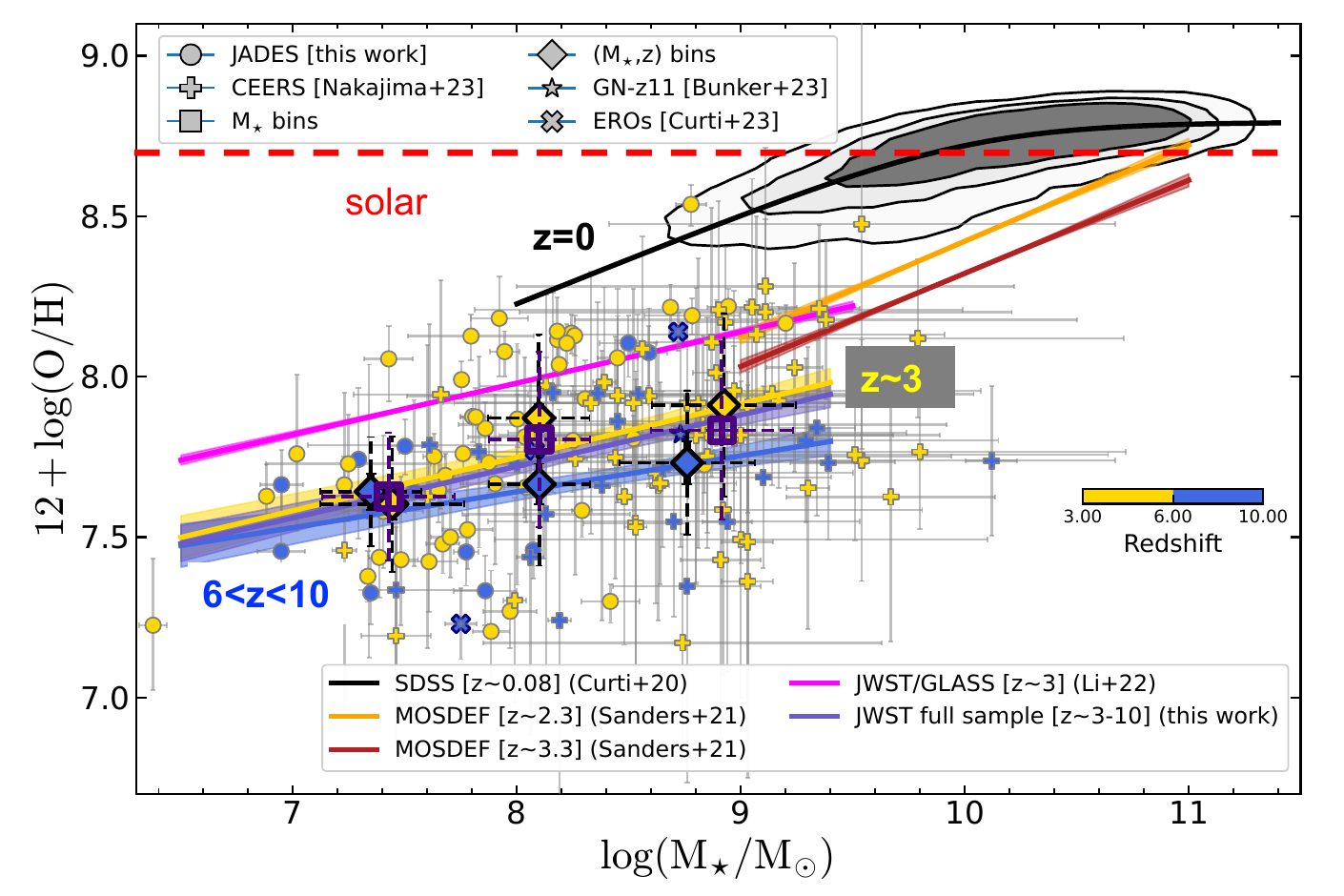}
\caption{\it A compendium of results tracing the evolving mass-metallicity relation in [O/H] for 146 galaxies in the JADES survey over 3$<z<$10 calibrated using local dwarfs \citep{Curti2024}. The figure includes earlier high redshift results by \citet{Nakajima2023} and \citet{Morishita2024} calibrated using appropriate [O III] 4363 \AA\ data. Various fits to the relation at z=0 (SDSS, \citet{Curti2020}, z=3 and 6$<z<$10 are indicated in yellow and blue respectively, and present a convincing evolutionary trend although there is much scatter in the data and differences in various fits to the data at z=3. The solar [O/H] abundance is shown by red dashed line.} 
\label{fig:mmr}
\end{figure}

In searching for global trends of chemical enrichment over the full redshift baseline, early work exploiting the growing JWST spectroscopic archive has confirmed a declining metallicity trend with redshift. Since luminous/massive galaxies are more enriched than their sub-luminous, less massive cohorts, the results are usually presented in the context of the gas-phase {\bf\it mass-metallicity relation} (MMR). This relation has been known at low redshift for decades and can be understood as arising from the increased gravitional potential of massive galaxies which can retain their enriched supernova products more readily than their less massive counterparts. Campaigns based on the growing amount of archival data show broadly consistent trends in the evolving mass-metallicity relation (MMR), with a mean decline in [O/H] across the relation of 0.8 dex from z=0 to z=3 and a less marked decline of perhaps 0.2 dex to z=10 (\citet{Curti2024, Chakraborty2025}, see Figure~\ref{fig:mmr}). The overlap in stellar mass across the full redshift range is not ideal since reionisation era galaxies have stellar masses in the range $10^7$ to $10^9 M_{\odot}$, a range only explored locally with metal-poor systems such as ``Green Pea" and ``Blueberry" galaxies whose context with respect to normal galaxies is not fully understood. In fact, there is evidence the slope of the MMR flattens at such low masses both at high redshift and locally, a trend that simulations fail to match and maybe suggesting inadequate feedback prescription in the models.

\begin{figure}
\center
\includegraphics[width=0.6\linewidth]{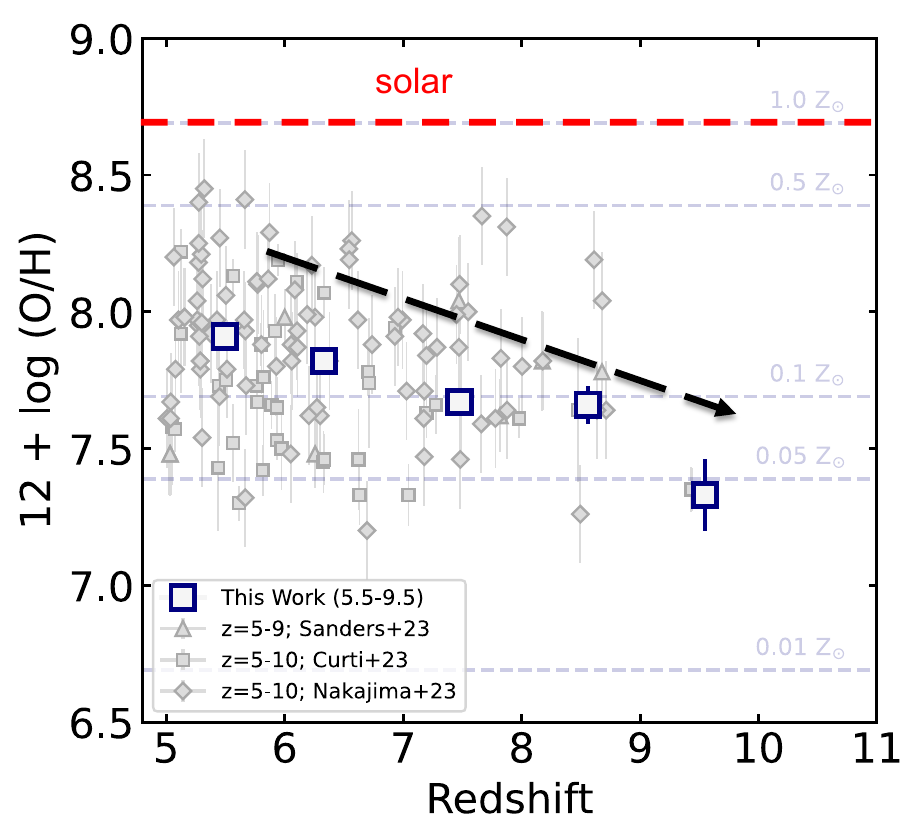}
\caption{\it Redshift-dependent metallicities derived from spectral stacks utilising the O32 metallicity diagnostic calibrated by intermediate redshift [O III] 4363 \AA\ measures of $T_e$ from \citet{Sanders2024} \citep{Roberts-Borsani2024}. Similar results have been presented by \citet{Heintz2025} and \citet{Hayes2025}.}
\label{fig:stack_z}
\end{figure}

Stacked spectra can also be useful in providing insight into the systematic redshift-dependent trends although compiling samples drawn from inhomogeneous surveys in the data archive risks introducing various biases. The most obvious arises from the need for an accurate spectroscopic redshift which may limit selection to more luminous or line-emitting sources which could, for example, be undergoing bursts. A striking glimpse of a systematic decrease in the oxygen abundance with increasing redshift has nonetheless been provided by \citet{Roberts-Borsani2024}, \citet{Heintz2025} \& \citet{Hayes2025} (Figure~\ref{fig:stack_z}). With care such stacking studies may ultimately be useful in tracing the early enrichment rate and may offer the best statistical estimate of when the first pristine sources emerged (see Lecture 3).

\section{Recognising AGN in the Reionisation Era}
\label{sec: 5}

Although other, more authoritative, lecturers will cover the topic of high redshift quasars and black hole growth, I will finish this second lecture with a purely observational discussion on how we might locate and conduct a census of AGN in the reionisation era. However, I will not address the thorny question of how to estimate their black hole masses and hence the various implications for early black hole growth.

AGN are important to locate at early times for two reasons. Firstly, we need to understand whether their non-thermal radiation provides an important component for cosmic reionisation (Lecture 1). Secondly we need to understand how quasars with black hole masses of $\simeq 10^9 M_{\odot}$ can assemble by redshifts z$\simeq$7-7.5. JWST is pioneering three ways to locate reionisation-era AGN.

\begin{enumerate}
\item{} Rest-frame optical spectra sampling permitted recombination lines whose broad line widths contrast with those of forbidden nebular lines; a key challenge is ruling out stellar-driven winds.

\item{} X-ray detection and follow-up spectroscopy: individual X-ray detections at high redshift are usually impractical but stacking may be useful and some claims are based on strongly-lensed sources.

\item{} Time variability: this method has the advantage of probing sub-luminous sources beyond reach of spectroscopy but, due to (1+z) time dilation, an
extended time baseline is required.    
\end{enumerate}

\subsection{Broad-line sources}

\begin{figure}
\center
\includegraphics[width=0.9\linewidth]{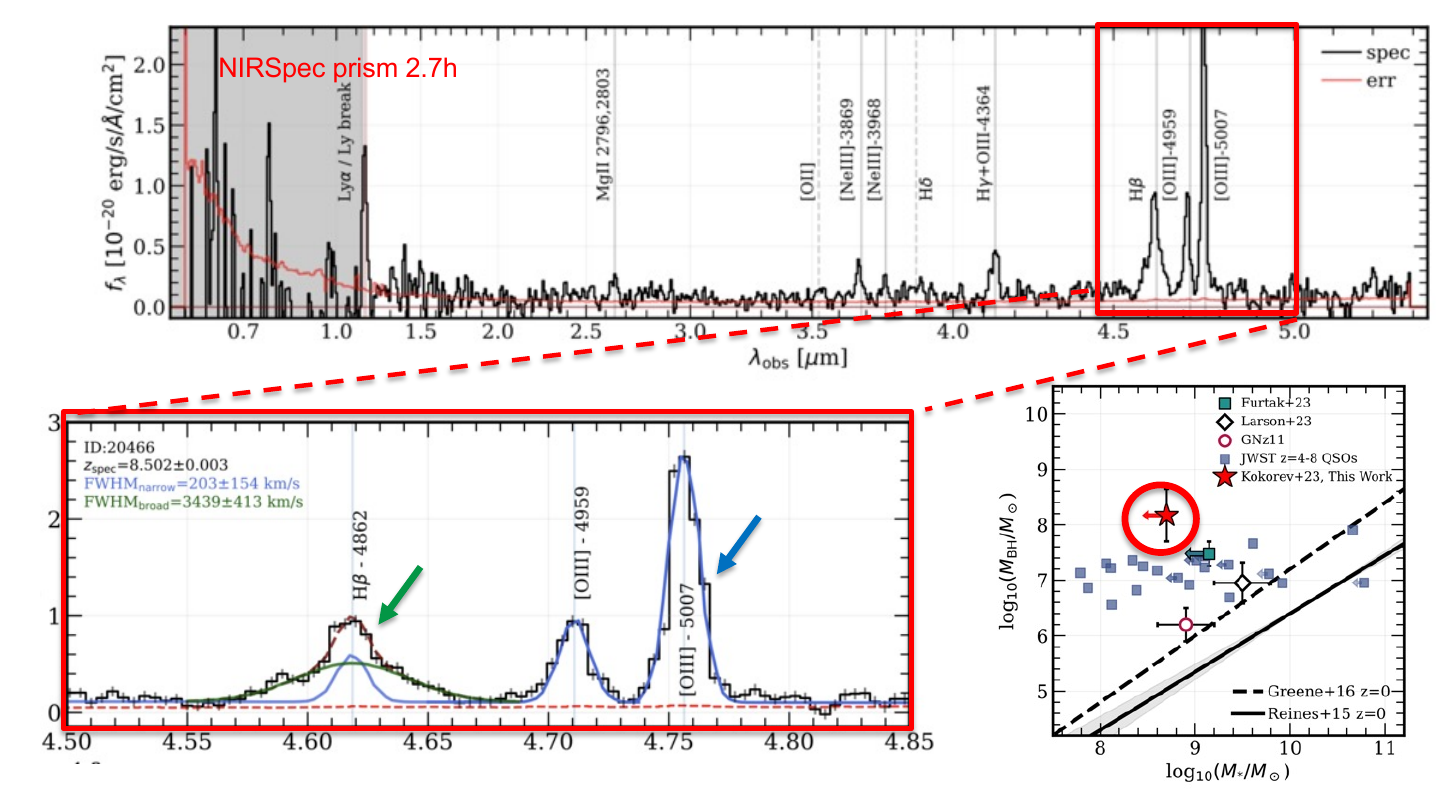}
\caption{\it A convincing case for a high mass black hole in a z=8.50 galaxy deep in the reionisation era \citep{Kokorev2023}. The spectrum fulfills the usual AGN criteria with exceptionally broad permitted lines (H$\beta$) and narrow forbidden lines ([O III]). A BH mass of $\simeq 10^8 M_{\odot}$ inferred from local reverberation mapping relations assuming virial equilibrium would, if correct, imply $\simeq$30\% of the stellar mass of the galaxy placing it well above the local $M_{BH} - M_{\ast}$ relation (inset panel). }
\label{fig:blagn}
\end{figure}

Figure~\ref{fig:blagn} shows a promising example of a broad-line AGN at a spectroscopic redshift of z=8.5 from the UNCOVER survey \citep{Kokorev2023}. Balmer H$\beta$ has a line width of $\sigma$=3400 km sec$^{-1}$ which contrasts with a much smaller $\sigma$=200 km sec$^{-1}$ for  [O III] emission. Such a broad H$\beta$ line would be difficult to explain solely in terms of gaseous ouflows. Using scaling relations adopted from lower redshift reverberation mapping measurements, a compact broad line region in virial equilibrium would imply a black hole mass of $M_{BH}$ = 1.5 10$^8 M_{\odot}$ which represents a staggering $\simeq$30\% of the stellar mass of the galaxy! To reconcile the observations with the local BH - stellar mass relation (see sub-panel in Figure~\ref{fig:blagn}) one would have to argue either that the BH mass is overestimated by a factor of 2 dex or the stellar mass similarly underestimated. For most reasonable BH seed masses, one would conclude the BH is assembling more rapidly than the galaxy itself. Early spectroscopic searches locating broad lines in the redshift range $4 < z < 7$ located a significantly higher fraction of low luminosity AGN than pre-JWST studies \citep{Harikane2023}. As many as 5-10\% of the targeted galaxy population showed some evidence of AGN activity with implied BH masses of $10^{6-7} M_{\odot}$ compared to 1\% locally. However, there is much uncertainty regarding the line width threshold that should be adopted as an AGN criterion given outflows may be more energetic in physically smaller galaxies with higher star formation rates at high redshift.

Soon after the high abundance of low luminosity AGN was claimed at z$\sim$5, a distinct category of broad-line sources termed "little red dots" (LRDs) was discovered from a blind NIRCam grism slitless JWST survey \citep{Matthee2024}. While these do not constitute the entire population of broad line cases (and not all LRDs show broad lines, \citet {Greene2024}), their puzzling nature has led to many studies. Unfortunately, there is no agreed definition of a LRD. Initially located as broad H$\alpha$ emitters in a slitless survey with compact red morphologies, they were considered to be dust-attenuated AGN. Indeed, in many cases the flux observed in the broad H$\alpha$ component is consistent with the fraction of the unresolved flux in the NIRCam images.

\begin{figure}
\center
\includegraphics[width=0.9\linewidth]{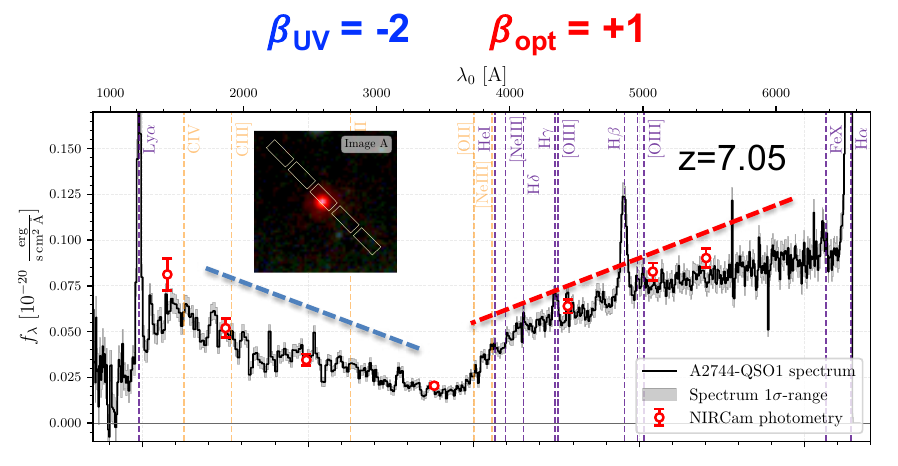}
\caption{\it The puzzling V-shaped spectral energy distribution of the population of compact "Little Red Dots" first identified by \citet{Matthee2024}. The steep rest-frame UV continuum slope is characteristic of star-forming galaxies while the red optical slope was originally interpreted either in terms of a dust-enshrouded AGN or an old stellar population. Broad emission lines suggest the AGN hypothesis \citep{Greene2024} while the occasional detection of a Balmer break suggest a stellar explanation consistent also with weak detections at mid-IR and sub-mm wavelengths \citep{Baggen2024}. This particular example is gravitationally lensed by a foreground cluster indicating that the compact core has a diameter of $<$ 60pc \citep{Furtak2023}. }
\label{fig:lrd}
\end{figure}

Only later was it discovered that many of these so-called LRDS have an unusual V-shaped SED (Figure~\ref{fig:lrd}) with a rest-frame UV slope $\beta_{UV}\simeq$-2 (characteristic of star-forming galaxies) and a red rest-frame optical slope $\beta_{opt}\simeq$+1 suggestive of a dust-enshrouded system. However, a dusty AGN hypothesis runs into difficulties in explaining the weak or absent MIRI/ALMA fluxes \citep{Williams2024, Casey2025} and no X-ray detection in stacked data \citep{Sacchi2025}. The presence of a Balmer break in the spectra of several examples led to claims that the optical component of the SED might be more easily explained via an old stellar population \citep{Baggen2024}. However, a stellar-only model would lead to very high stellar densities given the compact nature \citep{Leung2024}. Indeed, the gravitationally-lensed LRD at z=7.05 shown in Figure~\ref{fig:lrd} remains unresolved suggesting a physical diameter of less than 60 pc and a low mass host \citep{Furtak2023}. Its Balmer break is inconsistent with a stellar origin \citep{Ji2025} and variability has been recently detected \citep{D'Eugenio2025} indicative of an AGN surrounded by dense gas.

At the time of these lectures, the jury was still out on the nature of LRDs. Photometric searches based on the unique V-shaped SEDs has located several hundred examples, mostly beyond $z\simeq$4 \citep{Kocevski2025}. Attempts to explain the complicated SED via a AGN hypothesis require embedding the BH in an extended dust cloud with large grains, thereby enabling little extinction in the UV and a lower dust temperature to satisfy MIRI non-detections \citep{Li2025}. However, a pure stellar non-AGN model requires extraordinary densities within the compact red core.  Since the lectures, it has been proposed that the rest-frame optical continuum may arise from a dense gaseous photosphere that can produce a steep Balmer decrement and also scatter and broaden the recombination lines by various processes thus significantly reducing the implied BH mass \citep{Naidu2025, Brazzini2025}. Regardless of the correct interpretation (even if there is only one hypothesis as LRDs show a diversity of properties), they would appear to be a minor contributor to cosmic reionisation.

\subsection{X-ray AGN?}

\begin{figure}
\center
\includegraphics[width=0.9\linewidth]{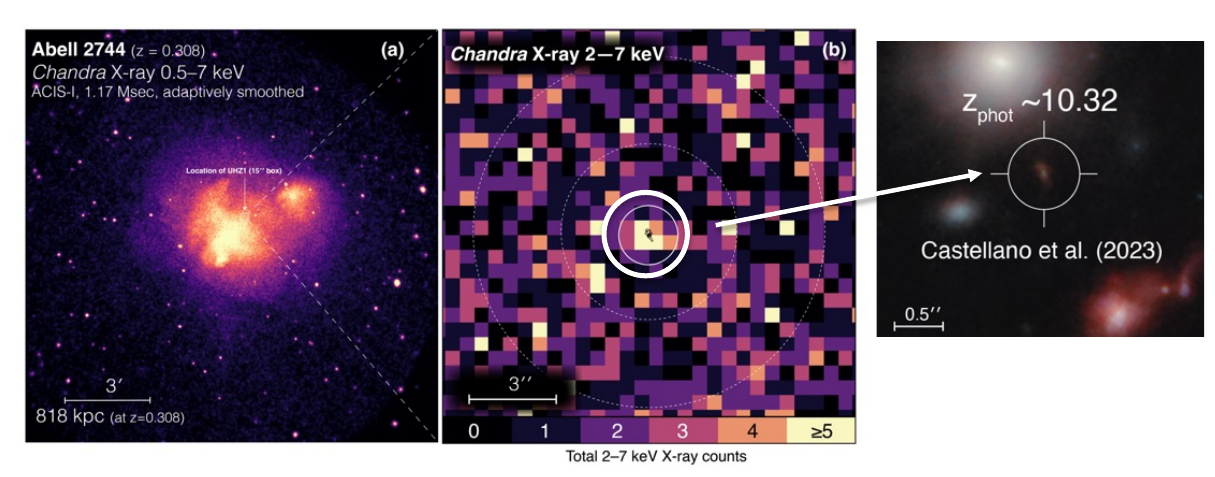}
\caption{\it Evidence for a gravitationally-lensed X-ray emitting AGN at $z > 10$ \citep{Bogdan2024, Kovacs2024}. The massive lensing cluster Abell 2744 has a hot X-ray emitting intracluster medium but a 4$\sigma$ X-ray feature coincides with the position of a spectroscopically-confirmed background z=10.07 galaxy dubbed UHZ-1 \citep{Goulding2023}, although its spectrum shows no evidence of AGN activity. The same authors have located a second lensed X-ray candidate GHz9 in the same cluster and estimate both black hole masses are in the range 4 $\times$ 10$^7$ to 2 $\times$ 10$^8 M_{\odot}$ supporting non-stellar BH seeds\citep{Natarajan2024}.}
\label{fig:xray_agn}
\end{figure}

Even the powerful Chandra X-ray satellite does not have the sensitivity to detect individual supermassive black holes (SMBH) in the reionisation era and, in the past, only stacked images of $z\simeq$6 Lyman break galaxies have been used to constrain the present of AGN. Nonetheless, two recent claims have been made for X-ray detected AGN at redshifts z$>10$ which, if correct, would imply the existence of SMBHs at a very early cosmic age ($\simeq$450 Myr) and hard to reconcile with a stellar BH seed (Figure~\ref{fig:xray_agn})

In both cases, the detections are point X-ray sources seen through a foreground lensing cluster Abell 2744 which overlap in angular position with a $z>$10 galaxy \citep{Bogdan2024,Kovacs2024}. Since Abell 2744 is a massive cluster with a hot X-ray emitting intracluster medium, the key question in both cases is whether the X-ray feature is merely a fluctuation in the cluster emission and thus represents a chance alignment with a background galaxy. UHZ-1 is seen in X-rays as a 4$\sigma$ blip in the cluster emission aligned with a galaxy with a photometric redshift of z=10.3. A subsequent JWST spectrum reveals a redshift z=10.07 \citep{Goulding2023} but no broad-line features nor any particularly convincing high ionisation lines. Remarkably, the same authors located a second 4-5$\sigma$ X-ray blip in the same cluster. GHz9 is a z=10.15 galaxy which shows some high ionisation lines (CIV, He II, CIII]) \citep{Napolitano2024} but its angular position is not exactly coincident with the X-ray blip and there is a low redshift galaxy close by. To derive the claimed BH masses of $M_{BH}\simeq4.10^7 M_{\odot}$(UHZ-1) and $2.10^8 M_{\odot}$ (GHz9) they adopt a bolometric luminosity which, whilst uncertain, is naturally very large when placed at $z\simeq$10. While the potential importance of finding such SMBHs at $z\simeq$10 would undoubtedly be very profound \citep{Natarajan2024}, the absence of convincing spectroscopic features to support the AGN claim is a concern.

Although unrelated to X-ray emission, evidence has also been presented for AGN activity in the spectra of other $z>10$ galaxies, mostly through high ionisation lines, outflows and high density gas \citep{Maiolino2024}. However, follow up campaigns with MIRI have not, thus far, provided confirmation via broad H$\alpha$ emission \citep{Alvarez-Marquez2025}.

\subsection{Time variable sources}

In principle, locating AGN by photometric variability has some distinct advantages over the previously-described methods. Foremost one can penetrate to much fainter luminosities than through methods requiring spectroscopic diagnostics so, whilst not all AGN necessarily display variability, a more complete census is possible even at the highest redshifts. Contaminants such a Type II supernovae can be readily identified either from multi-epoch observations or, exploiting JWST's superlative image quality, by location within the host galaxy (if visible). The challenge lies in establishing an appropriate baseline of multi-epoch visits over several years given time dilation scales as (1 + z) and typical rest-frame fluctuations of $\simeq$months. 

\begin{figure}
\center
\includegraphics[width=0.9\linewidth]{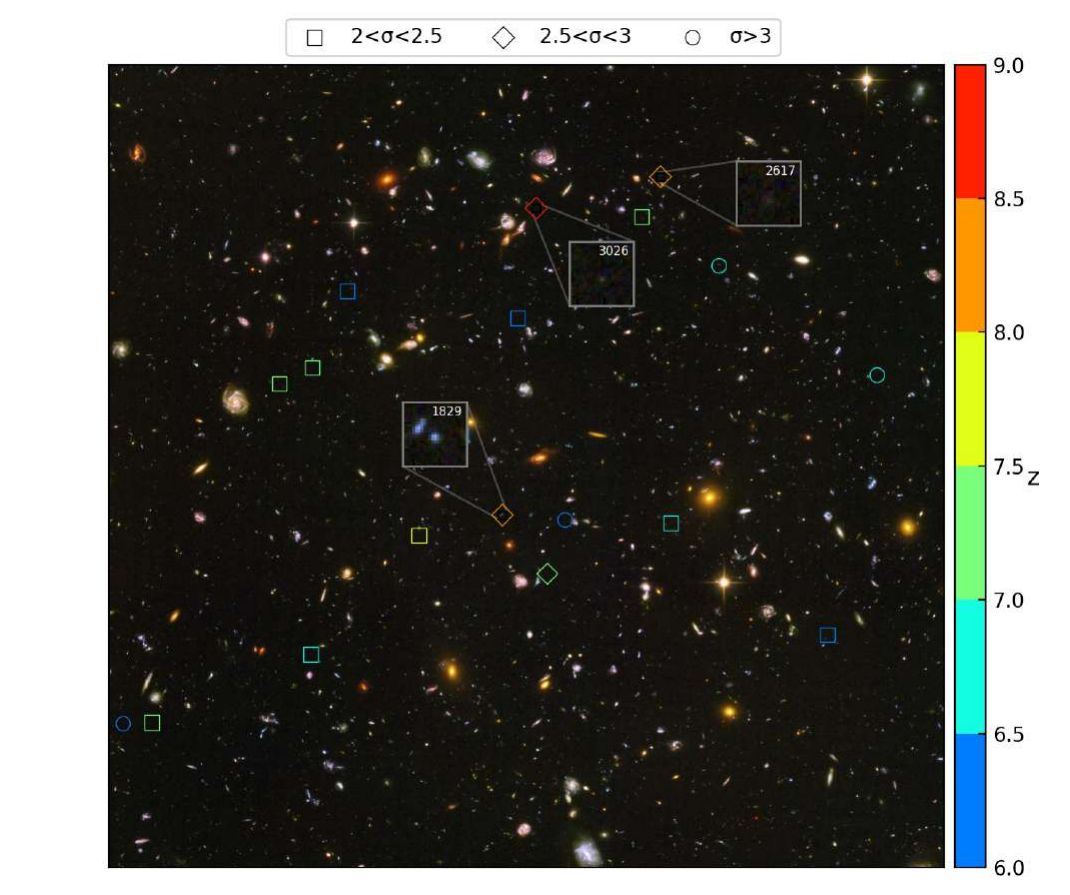}
\caption{\it Time variability as a method for searching for AGN in the reionisation era. The image shows variable sources detected to luminosities of $M_{UV}\simeq$-17.0 by comparing HST WFC3/IR images of the Hubble Ultra Deep Field taken with similar filters from 2009, 2012 and 2023 - a rest-frame baseline of $\simeq$ 1 year for z$\simeq$7 sources. Careful photometric comparisons and corrections for non-variable AGN (from intermediate redshift X-ray studies) and luminosity-dependent incompletess, already suggest a higher volume density of AGN at $z\simeq$6-7 than that adopted in contemporary numerical simulations \citep{Cammelli2025}. Given the extended lifetime of JWST, time variable studies are clearly an important and independent method to probe the presence of AGN at high redshift.}
\label{fig:udf_var}
\end{figure}

A first attempt at exploiting this method involved revisiting the Hubble Ultra Deep Field (UDF) with WFC3/IR a decade after the original campaigns in 2009 and 2012 \citep{Hayes2024}. 208 (116) sources showed variability at more than 2.5 (3.0) $\sigma$ down to a luminosity of $M_{UV}\simeq -17$, of which 8 (4) have photometric redshifts beyond z$\simeq$6 (Figure~\ref{fig:udf_var}). By comparing the fraction of $z < 5$ variables found in a UDF catalog of X-ray AGN, volume-based corrections can be made to include 
non-variable AGN as well as redshift-dependent incompleteness. The resulting integrated number density of BH at $z\simeq$6-7, whilst extremely uncertain as it is only based on a few sources, provides a surprisingly stringent constraint on some models of BH growth exceeding by a large factor the number adopted in current numerical simulations \citep{Cammelli2025}. Given JWST's extended operational lifetime, extending this method is clearly promising, particularly givenn its potential to locate sub-luminous AGN beyond reach of spectroscopic surveys.

\section{Summary}
\label{sec: 6}

In this second lecture we have reviewed our physical understanding of galaxies in the reionisation era. 

Firstly, we discussed the pros and cons of using spectroscopy and photometry to derive various diagnostics including (i) the UV continuum slope which is sensitive to the stellar age, dust content and nebular continuuum and (ii) line emission which gauges the star formation rate and the intrinsic ionisation production rate, and fitting spectral energy distributions for deriving all of the above. JWST's early photometric samples were larger and more complete whereas early spectroscopic samples have been drawn from different surveys and may be affected by redshift-selection biases. Nonetheless, it seems clear the precision of the spectroscopic data is much greater e.g. for UV continuum slopes and line flux measures, and thus will ultimately represent the way ahead.

Highlights of where JWST has provided new opportunities in the reionisation era include providing access to the Balmer H$\alpha$ and H$\beta$ line so measures of the star formation rate can be directly compared with UV and other estimates, as well providing measures of the intrinsic Lyman $\alpha$ flux as a guide to the state of the local IGM. Access to MIRI has been shown to be important in deriving robust stellar masses and ages free from the ``outshining" effect of a secondary burst of star formation. Spectroscopy has also revealed cases where nebular continuum from a population of supermassive stars may dominate the UV luminosity, although such cases are not common.

However, the most important advance lies in tracing the chemical evolution of the interstellar gas. The temperature-sensitive auroral line [O III]4363 \AA\ has been detected in $\simeq$ 100 $z>5$ galaxies yielding direct [O/H] abundances and a vital calibration for larger samples. By tracing the mass-metallicity relation and through spectral stacks in redshift bins, we see the first evidence for a declining [O/H] abundance with increasing redshift. Nonetheless, when it comes to precise measurements, we are seeing evidence for a multi-phase ISM in
comparing various electron density diagnostics as well as in comparisons of [O III] 4363 \AA\ temperatures with similarly sensitive ALMA lines.

Finally, I touched briefly on searches for AGN activity, focusing primarily on the abundance without commenting in detail on the BH masses and various seed/growth mechanisms. It seems clear that low luminosity AGN are more abundant than pre-JWST estimates and there are many convincing cases of broad-line AGN at $z\simeq$7-8. A new population of compact LRDs appear to be AGN surrounded by dense gaseous photospheres which become rare at $z<4.$ Although there are claims for high mass BHs at $z>$10, the evidence is not yet compelling in my opinion.

\section{Recommended Reading on Galaxies in Reionisation Era}
\label{sec: 7}

\noindent{\bf Major Review Articles}

\begin{enumerate}
\item{\cite{Conroy2013} - stellar population synthesis models}
\item{\cite{Madau2014} - classic paper on cosmic star formation history}
\item{\cite{Stark2016} - pre-JWST review of galaxies in the first Gigayear}
\item{\cite{Maiolino2019} - chemical evolution and abundance determination methods}
\item{\cite{Ouchi2020} - analyses of Lyman alpha emitting galaxies}
\item{\cite{Robertson2022} - review of early JWST projects and science goals}
\item{\cite{Stark2025}} - a valuable early review of JWST observations of galaxies
\item{\cite{Curti2025}} - a valuable early review of JWST spectroscopic studies
\end{enumerate}

\noindent{\bf Pre-JWST articles}

\begin{enumerate}
\item{\cite{Ellis2013} - Hubble Ultra Deep Field galaxies to z$\simeq$11}
\item{\cite{Oesch2018} - pre-JWST census of star-forming galaxies to z$\simeq$11}
\item{\cite{Stark2017} - ground-based spectroscopy beyond z$\simeq$7}
\item{\cite{Roberts-Borsani2016} - Spitzer-excess galaxies with Lyman alpha emission}
\end{enumerate}

\noindent{\bf JWST articles (changing monthly!)}

\begin{enumerate}
\item{\cite{TWang2024} - MIRI-based stellar masses}
\item{\cite{Topping2024} - UV continuum slopes at high z}
\item{\cite{Katz2024} - role of nebular continuum}
\item{\cite{Sanders2024} - high redshift electron temperature calibrations}
\item{\citet{Cameron2023} - nitrogen excesses in $z>10$ galaxies}
\item{\cite{Curti2024} - evolving mass-metallicity relation}
\item{\citet{Leung2024} - the puzzling SED of Little Red Dots}
\item{\citet{Naidu2025} - the physical nature of Little Red Dots}

\end{enumerate}
\newpage
\begin{center}
{\bf \Large  Lecture 3: The Redshift Frontier and Cosmic Dawn}
\end{center}

\medskip
\setcounter{section}{0}

\section{Introduction}
\label{sec:1}

\begin{figure}[hbt!]
\center
\includegraphics[width=\textwidth]{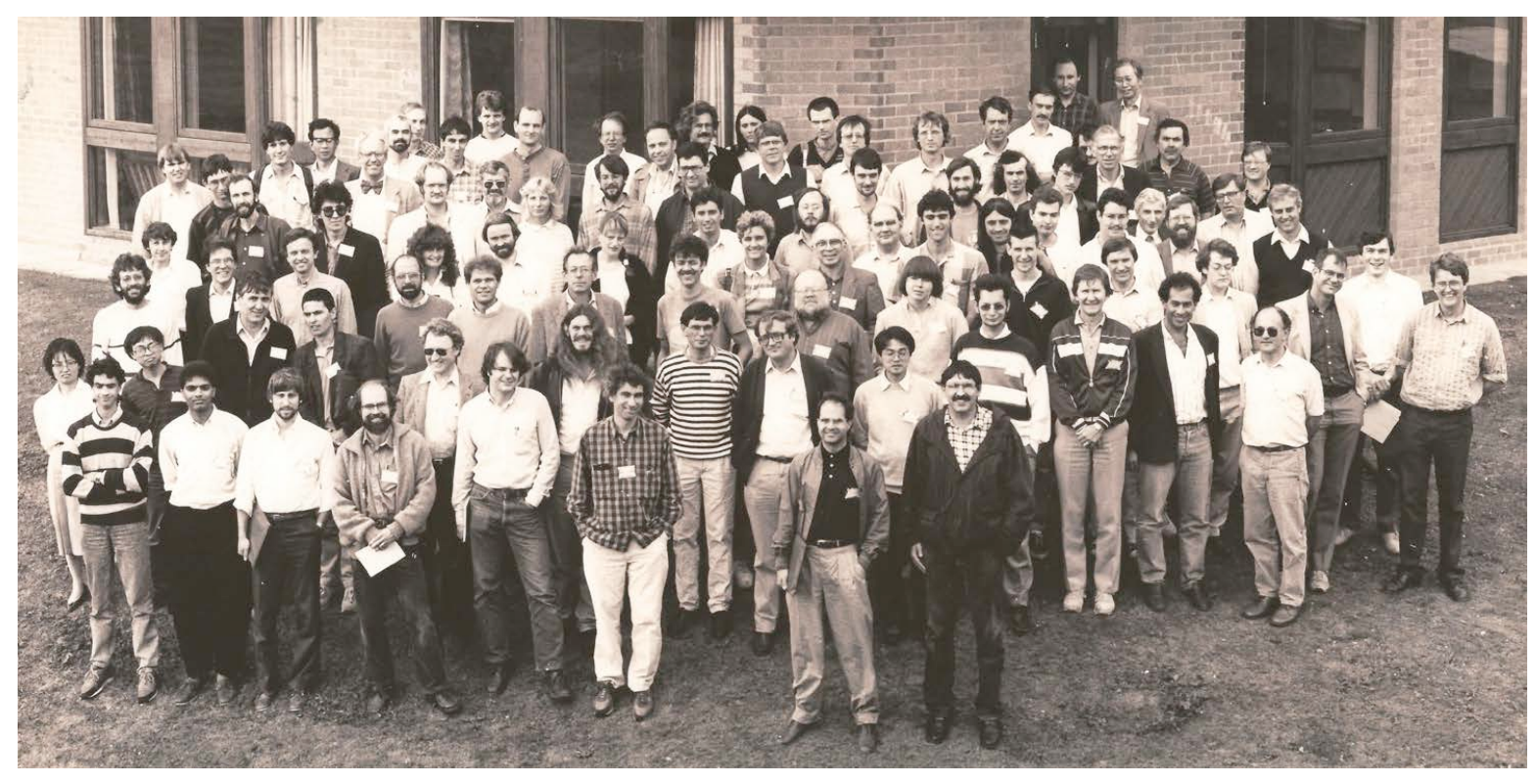}
\caption{\it Conference photograph for the {\it Epoch of Galaxy Formation} meeting at Durham in 1988. (Colour photography hadn't been invented at that time!) Many luminaries, some now sadly departed, were present. See text for the results of a vote as to the redshift of when galaxies formed.}
\label{fig:durham88}
\end{figure}

In this final lecture we'll explore what has been learned so far from JWST in pushing the redshift frontiers towards `cosmic dawn' when the first chemically pristine stellar systems emerged from darkness. Something of a `Holy Grail' in discussing the longer term prospects of JWST observations, we are currently guided largely by numerical simulations in attempting to recognise such first generation galaxies. As in earlier lectures, I wish to emphasise that the topic of searching for primaeval galaxies dates back over 50 years. With colleagues I organised a conference at Durham in 1988 entitled {\it The Epoch of Galaxy Formation}, (\citet{Frenk1989}, Figure~\ref{fig:durham88}) at the conclusion of which a vote was held on when galaxies formed. The precise definition was the redshift at which a typical ($L^{\ast}$) galaxy had assembled 50\% of its stars. Most voted for a redshift within 1$ < z < $5 and only a handful conceived of earlier galaxies. At the time of the 1988 conference, the most distant spectroscopically-confirmed galaxy was 3C326.1 at z=1.82 \citep{McCarthy1987}.

\begin{figure}
\center
\includegraphics[width=0.55\textwidth]{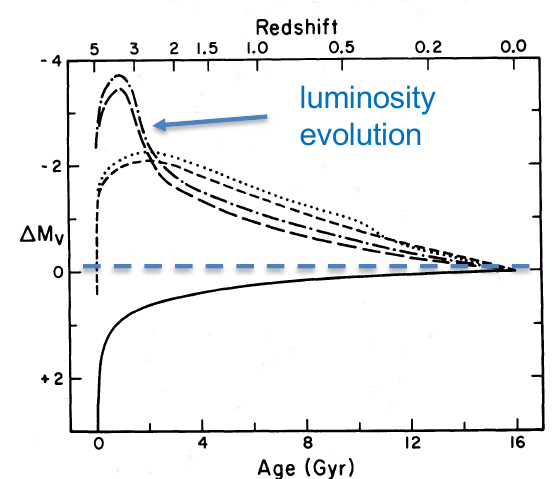}
\includegraphics[width=0.4\textwidth]{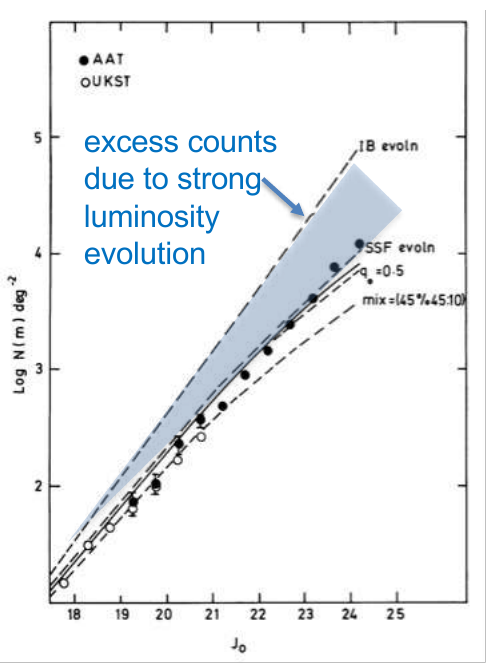}
\caption{\it Evolutionary predictions for primaeval galaxies from Beatrice Tinsley \citep{Tinsley1980a}. (Left) Since early type galaxies presently contain a uniform population of old stars, she conjectured they formed in an "initial burst" at redshift z=3 and thus would be spectacularly luminous and detectable as an excess in deep galaxy counts. (Right) Early galaxy counts \citep{Peterson1979} to 25th magnitude (black dots) found only a modest excess compared to various non-evolving models in a flat ($q_0$=0.5) universe indicating a slow star formation history (SSF) rathe than one with an intense initial burst (IB).}
\label{fig:tinsley_evol}
\end{figure}

Motivated by observations that supported the view that the oldest stars in the Milky Way had formed during a single monolithic collapse \citep{Eggen1962}, theorists predicted that primordial galaxies would be undergoing energetic initial bursts of star formation and therefore be spectacularly luminous. \citet{Partridge1967} estimated that during a brief period of $\simeq10^8$ years, they would attain luminosities $\simeq$700 times those of present-day galaxies with gas clouds glowing in Lyman $\alpha$ emission at redshifts z$\simeq$10-20. As such they should be visible at near-infrared wavelengths. Since the concept of hierarchical assembly was not appreciated at the time, a flaw in the calculation was the assumption that galaxies evolved in isolation with constant mass. During 1977 - 1980 Beatrice Tinsley did more than anyone to place the subject on a quantitative basis. A pioneer of stellar population modelling (see Lecture 2), she made detailed galaxy count predictions that motivated observers. As the light of elliptical galaxies is dominated today by old red giants, she predicted their precursors would be blue, luminous and compact, and detectable as an excess signal in number - magnitude counts \citep{Tinsley1980a}. Faint galaxy counts derived from deep photographic plates taken with the newly-available 4-metre class telescopes failed to see such a strong excess down to blue magnitudes of B$_J \simeq $25 (\citet{Peterson1979}, Figure~\ref{fig:tinsley_evol}). 

Nonetheless observers continued to search for luminous primaeval galaxies in the redshift range $z\simeq$2-5 and candidates were proposed whose star formation rates deduced from strong Lyman $\alpha$ emission would imply the formation of a massive galaxy on a timescale of less than a Gyr \citep{Djorgovski1985,McCarthy1987}. Meanwhile deeper galaxy counts and the first faint galaxy redshift surveys enabled by multi-object spectrographs (Lecture 2) suggested only modest evolution in galaxy luminosities with the bulk of galaxies to B$_J$=24 lying below a redshift z$\simeq$1 with no high redshift tail ({\citet{Broadhurst1988, Colless1993, Glazebrook1995, Lilly1995}, Figure~\ref{fig:redshift_surveys}). The resolution of this disagreement was explained by the Cold Dark Matter (CDM) model of structure formation which postulated a continued hierarchical assembly of galactic mass rather than an intense burst of initial star formation following monolithic collapse. Detailed calculations by \citet{Baron1987} were able to reconcile the modest evolution seen in the faintest galaxy counts as well as the absence of a high redshift tail in the early redshift surveys.

\begin{figure}
\center
\includegraphics[width=\textwidth]{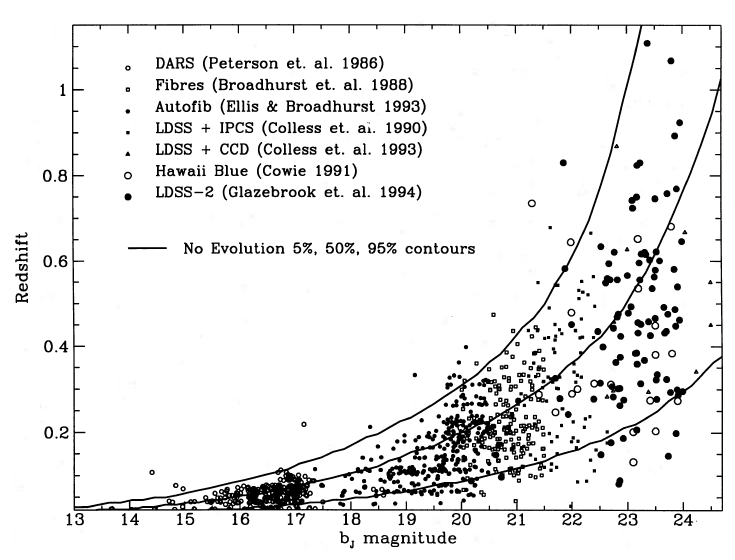}
\caption{\it The status of faint galaxy redshift surveys in the early 1990s, prior to the onset of 8-10 metre class telescopes (\citet{Glazebrook1995} with other campaigns labeled). Despite a significant excess of galaxies to $B_J$=24 over no evolution models, few $z>$1 galaxies associated with luminous early galaxies were found. The paradox of an excess number but no increased redshift range was explained via a decline in the star formation rate for sub-luminous galaxies (also referred to as `downsizing'.) Luminous primaeval galaxies were also shown to be inconsistent with the hierarchical nature of galaxy assembly predicted in cold dark matter models.}
\label{fig:redshift_surveys}
\end{figure}

\section{The JWST Census to $z\simeq14$}
\label{sec: 2}

As discussed in Lecture 2 (see Figure 22), prior to the launch of JWST, there was a fierce debate in the community as to the rate of decline in the comoving volume density of star forming galaxies beyond a redshift z$\simeq$8. \citet{Oesch2018} argued for a rapid decline consistent with theoretical models based on a constant efficiency of star formation from the baryonic mass associated with growing dark matter halos. \citet{McLeod2016, McLeod2021} argued for a more gradual decline in number density. Part of the discrepancy was simply the paucity of reliable HST candidates beyond z=8, of which only 3 had spectroscopic redshifts. Nonetheless, the two decline rates would lead to an order of magnitude difference in the number of candidates beyond z=11 expected with JWST. Within only a few weeks following the commencement of JWST science operations, the discrepancy was resolved with numerous groups offering candidates in early release images to redshifts z$\simeq$13 and beyond (\citet{Naidu2022, Castellano2023, Finkelstein2023, Harikane2023a, Bradley2023, Donnan2023, Bouwens2023, McLeod2024} Figure~\ref{fig:jwst_census}).

\begin{figure}
\center
\includegraphics[width=\textwidth]{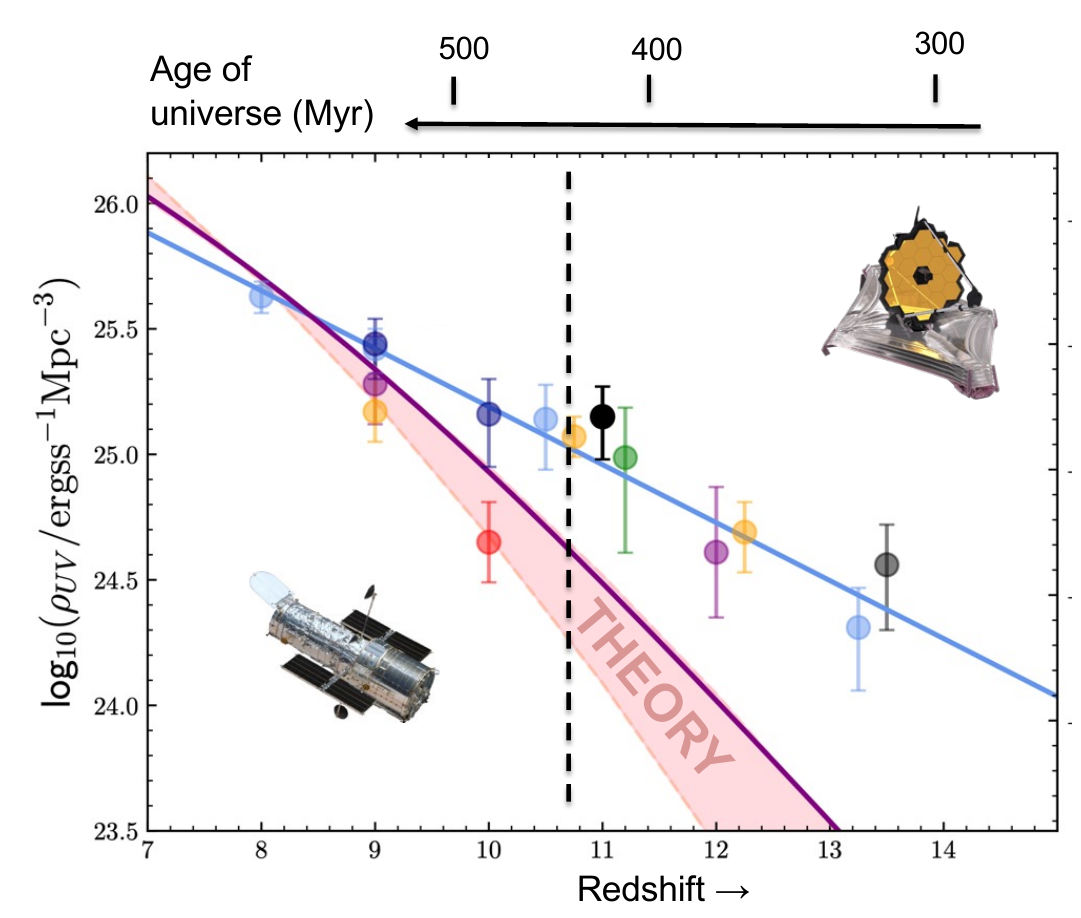}
\caption{\it Early compilation of the rate of decline of the comoving UV luminosity density from star-forming galaxies located in the first public JWST datasets (\citet{Donnan2023}, see text for similar articles). The black dashed line indicates the frontier achieved with HST and the band marked "Theory" broadly indicates the prediction for various models assuming a constant star formation efficiency given the cosmological baryon fraction in dark matter halos assembling in standard $\Lambda$CDM.}
\label{fig:jwst_census}
\end{figure}

Subsequent to this early glimpse of galaxies seen when the universe was only 300 Myr old, much wider field imaging was undertaken by the PRIMER project sampling 400 arcmin$^2$, an area 10 times that of the early release images discussed above. This yielded over 2500 candidates beyond a photometric redshift z=8.5 charting a smooth decline to redshift z=14 \citep{Donnan2024}. This was complemented by the JADES Origins Field, an ultra-deep exposure in two small fields purposely designed to explore the redshift range 12$<z<$20 with 14 JWST filters \citep{Robertson2024}. It located 8 new galaxies in the redshift range 11.5$ < z < $14.4. Although no galaxies were detected beyond z=15, the absence in the small survey area of 9.05 arcmin$^2$ is consistent with the rate of decline seen at lower redshifts (see Figure~\ref{fig:primer_jof}). 

\begin{figure}
\center
\includegraphics[width=0.55\textwidth]{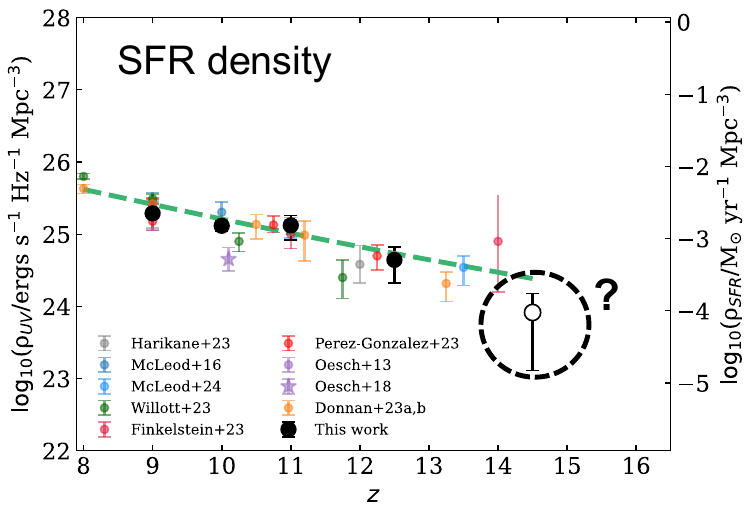}
\includegraphics[width=0.40\textwidth]{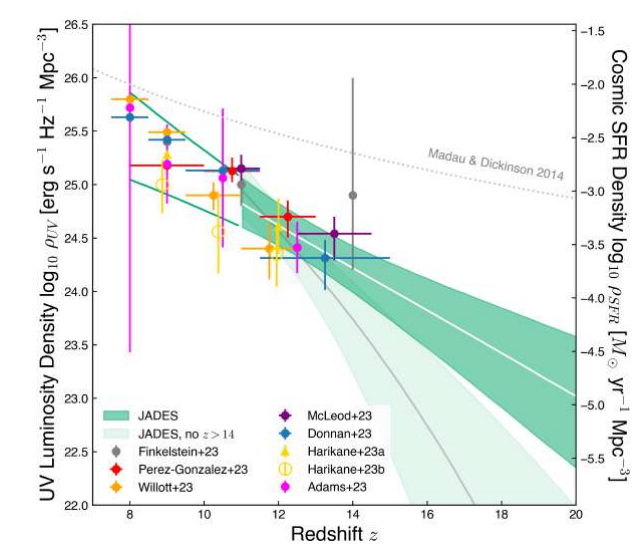}
\caption{\it Wide and deep: (Left) UV luminosity density versus redshift for the panoramic PRIMER survey (incorporating other publicly available datasets) totalling 400 arcmin$^2$ with over 2500 galazxies beyond z=8.5 \citep{Donnan2024}. The survey demonstrates clear continuity in star-forming galaxies out to z=14 with a tantalising 1$\sigma$ decline thereafter. (Right) Equivalent figure for the ultradeep JADES Origins Field sampling 9.05 arcmin$^2$ with the goal of probing the redshift range 12$ < z < $20. Although 8 galaxies were located beyond z=11.5, no convincing z $>$ 15 galaxies were found \citep{Robertson2024}.}
\label{fig:primer_jof}
\end{figure}

Since these early imaging campaigns relied on photometric redshifts, it is interesting to ask how reliable are those redshift estimates now we have extensive spectroscopic data. Figure~\ref{fig:photo_specz} shows a recent comparison where it can be seen there are very few outliers from a one-to-one relation. Looking more closely, however, it is seen that photometric redshifts are systematically overestimates by $\Delta z$ = 0.10-0.15. This can be well-understood by {\it in situ} hydrogen absorption which damps the blueward edge of the continuum near Ly$\alpha$ (see inset, \citet{Heintz2024}). Indeed, this effect has been used to estimate the redshift-dependent IGM neutral fraction $x_{HI}$ \citep{Mason2025}, although others have expressed concern on the quantitative accuracy of such estimates with low resolution prism spectra \citep{Huberty2025}.

One prominent high redshift candidate that did turn out to be a low redshift interloper is instructive to discuss. \citet{Donnan2023} published a fairly convincing z=16.6 candidated based on a Lyman break between prominent signals in filters longward of F200W and absent shortward of F150W. However, a spectrum secured by \citet{Arrabal-Haro2023} revealed the candidate to be at z=4.912 with intense emission lines of [O III] and H$\alpha$ contributing significantly to fluxes at F200W and F277W (see also \citet{Naidu2022a}).  This represents an important cautionary tale indicating that the redshift range 16 $<z<$ 18 is particularly tricky with broad-band photometric selection given the possibility of strong optical emission lines arising from z$\simeq$5 galaxies at 2-4 $\mu$m. Certainly, medium-band photometry will be helpful.

\begin{figure}
\center
\includegraphics[width=\textwidth]{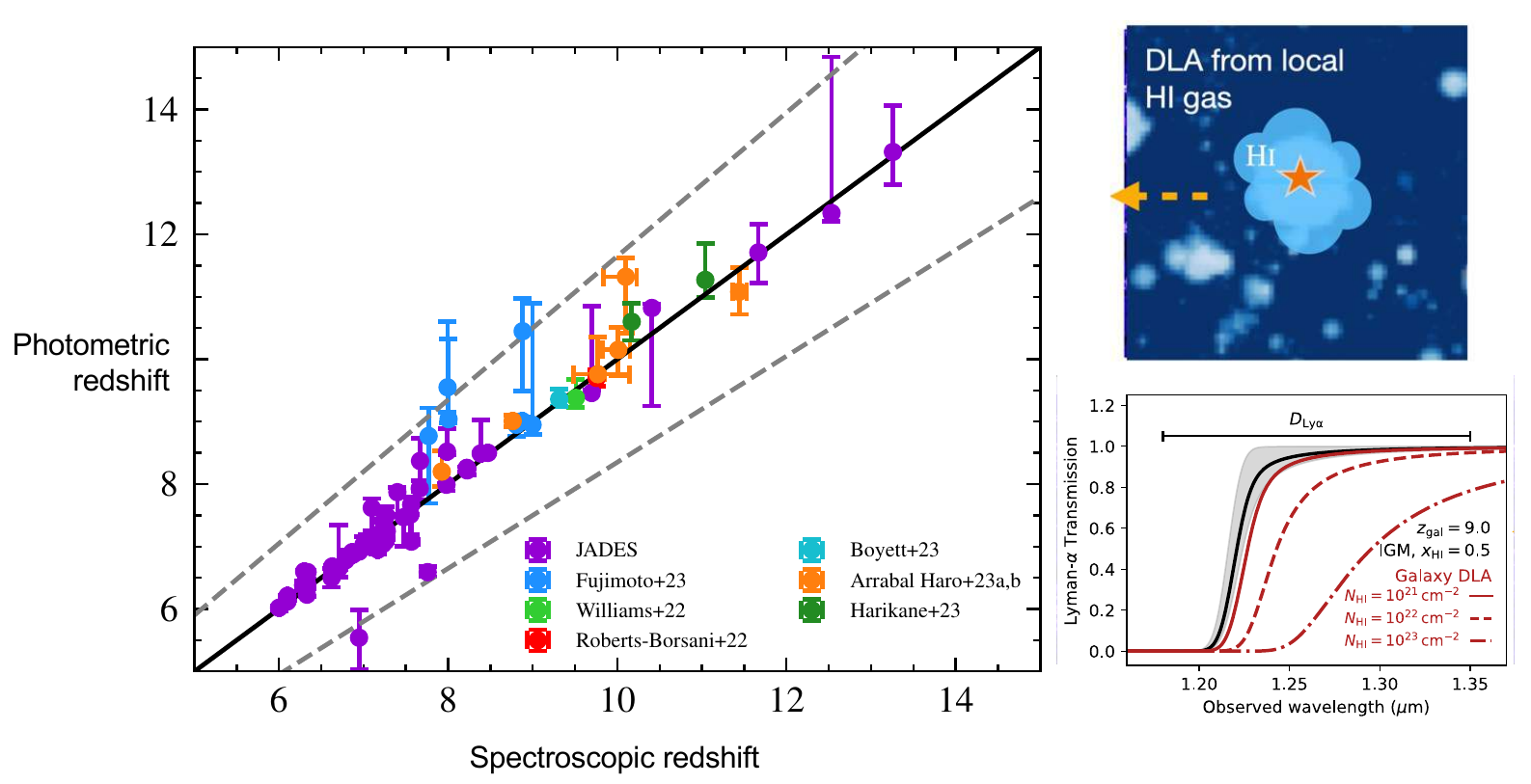}
\caption{\it A recent comparison between JWST photometric and spectroscopic redshifts (Courtesy C. Donnan) . In general the agreement is excellent. The small bias of $\Delta\,z\simeq$ 0.10-0.15 arises due to {\it in situ} hydrogen absorption (right panels) which damps the Lyman limit, the primarily photometric redshift indicator \citep{Heintz2024}. }
\label{fig:photo_specz}
\end{figure}

The PRIMER data charted a smooth decline to $z\simeq$14 and hinted a marginal drop in number density beyond \citep{Donnan2024}, so it is natural to ask if any promising candidates have been found beyond. The most recent attempt at the time of writing is that from the GLIMPSE survey \citep{Kokorev2025} which proposed five $z>15.9$ candidates based on deep NIRCam photometry. Several of these candidates lie at the periphery of the imaged areas which may indicate less reliable photometry. The implied comoving number density would also be in significant tension with the upper limit provided by the JADES Origin Field \citep{Robertson2024}. Although the JOF is a small area, this discrepancy is nonetheless a concern. Other $z>$15 searches since the Lectures include those by \citet{Asada2025} and \citet{Weibel2025}.

\section{The Mystery of Superluminous z$>$10 galaxies}
\label{sec: 3}

Clearly the JWST census beyond $z\simeq$8 is in disagrement with models of constant star formation efficiency based on the baryonic mass associated with growing dark matter halos in $\Lambda$CDM. Before attempting to understand this, we will first consider what we can learn from spectroscopy of these luminous sources. Within a few weeks of the start of science operations, the awesome power of JWST's spectrographs became apparent. Figure~\ref{fig:hiz_spectra} shows the "poster child" from late 2022, a spectrum of GN-z11 \citep{Bunker2023} showing an abundance of emission features for the most distant source confirmed with HST by \citet{Oesch2016}. The figure also shows the JWST spectrum of JADES-GS-z14.0, the highest redshift object at the time of the Saas-Fee lectures \citep{Carniani2024}, whose precise redshift of z=14.1 was later determined using ALMA \citep{Heintz2025a} \footnote{Unsurprisingly, in a rapidly developing field, since the Lectures the spectroscopic redshift record has been broken and currently stands at z=14.4 \citep{Naidu2025a}}. Both sources are spectacularly luminous ($M_{UV}\simeq$-21) with star formation rates of $\simeq$20 $M_{\odot}$ yr$^{-1}$ yet are physically compact with radii of only 100-200 pc. Similar quality spectroscopic data has been presented by \citet{Curtis-Lake2023, Harikane2023a, Hainline2024}. 

\begin{figure}
\center
\includegraphics[width=\textwidth]{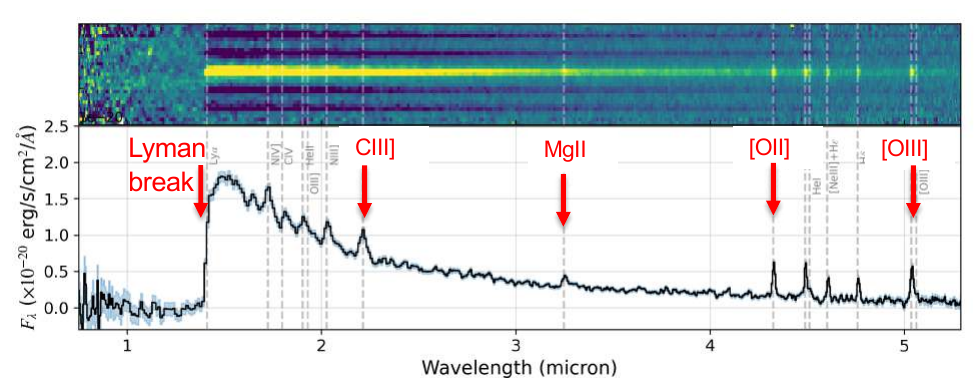}
\smallskip
\includegraphics[width=\textwidth]{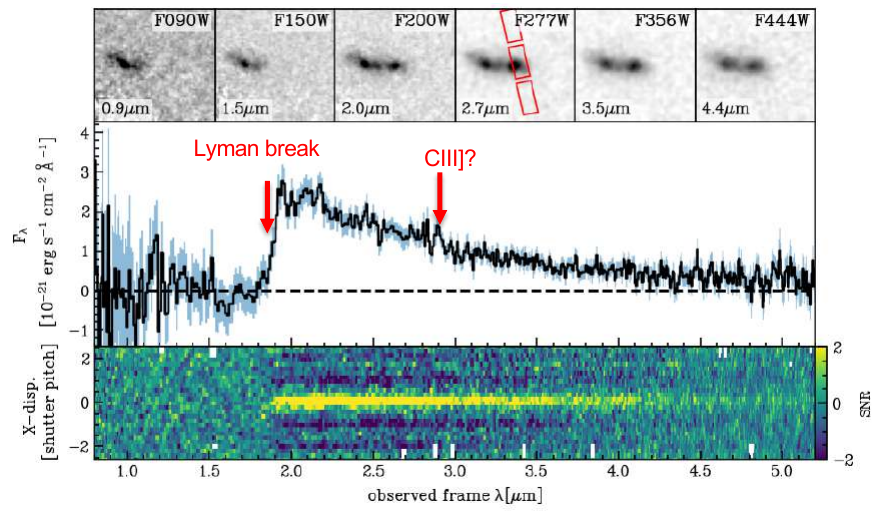}
\caption{\it The remarkable promise of JWST through detailed spectra of two z$>$10 galaxies.
(Top) The "poster child" JWST spectrum of GN-z11 at z=10.6 \citep{Bunker2023}, a galaxy whose high redshift was originally claimed via a Lyman break with HST \citep{Oesch2016}. Numerous emission lines are present with only the most prominent highlighted. (Bottom) JWST spectrum of GS-z14 \citep{Carniani2024}. Originally disfavoured as a high redshift candidate by the JADES team as it is 0.4 arcsec from a foreground z$\simeq$3.5 galaxy, perseverance paid off in securing the redshift record holder at the time of the Saas-Fee meeting. Both sources are remarkably luminous and compact, yet forming stars at over 20 M$_{\odot}$ yr$^{-1}$.}
\label{fig:hiz_spectra}
\end{figure}

\citet{Carniani2024} summarised the surprising luminosities of the spectroscopically confirmed $z>$10 sources in Figure~\ref{fig:carniani_lz}. The two sources shown in Figure~\ref{fig:hiz_spectra} are more luminous than most systems down to redshifts z$\simeq$8. Given the small cosmic volumes explored to locate such sources, it is tempting to contemplate we are witnessing some new, time-specific, physical processes responsible for these luminous sources, perhaps even related to the approach to `cosmic dawn'. Unsurprisingly, many hypotheses have been put forward to explain these extreme luminosities. Broadly speaking they fall into three categories:

\begin{figure}
\center
\includegraphics[width=\textwidth]{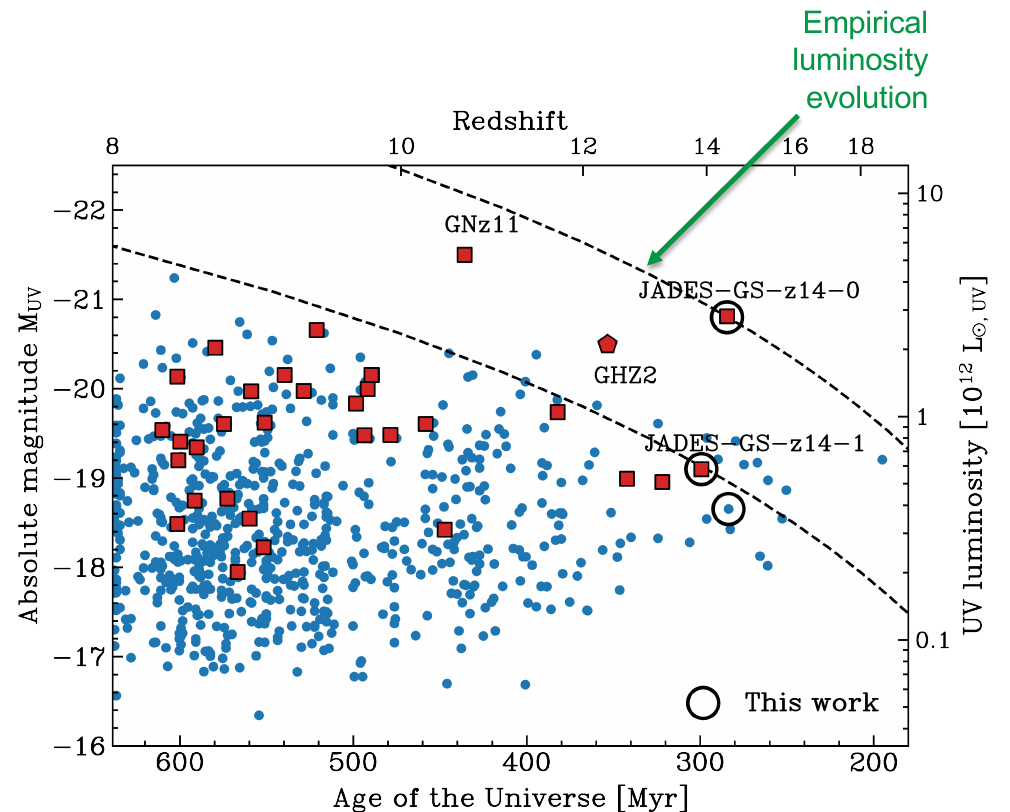}
\caption{\it UV absolute magnitude versus redshift for $z>8$ galaxies from \citet{Carniani2024}. Blue symbols represents sources with photometric redshifts, red squares have spectroscopic confirmation. Selected spectroscopic sources above z=10 are highlighted. The dashed curves indicate semi-empirical evolutionary trajectories for halos of a given cosmic abundance. Such analyses have led to numerous explanations for these extraordinarily luminous early galaxies.}
\label{fig:carniani_lz}
\end{figure}

\begin{enumerate}
    \item {The absence of dust or processes which regulate star formation in very early sources so that there has been insufficient time for galaxies to assemble a critical mass \citep{Ferrara2023, Ferrara2025, Dekel2023}}
    \item {Star formation is burst-like in early systems and thus there is a bias towards finding galaxies at the peak of a duty cycle of activity. The sources are then unlikely to sustain such high star formation rates for very long \citep{Mason2023,Shen2023, Gelli2024, Kravtsov2024}}
    \item {A top-heavy initial stellar mass function which might arise naturally in low metallicity systems where metal cooling processes whereby gas clouds fragment into smaller masses are less prevalent \citep{Harikane2023,Menon2024,Schaerer2025}}.
\end{enumerate}

The no-dust hypothesis proposed by \citet{Ferrara2023} posits that the increasing specific star formation rate (sSFR) at $z>$10 leads to radiation pressure-driven outflows that expel any dust. By considering the fraction of such critical outflows derived from the sSFRS distribution, it is possible to match the redshift-dependent cosmic star formation rate density in the context of limited data on outflow velocities. However, as we saw in Lecture 2, there is no significant evidence of a steepening in the UV continuum slope $\beta$ at $z>$10 \citep{Heintz2025, Saxena2024}. Indeed, if anything, there is evidence for {\it redder} slopes which may arise from an increased contribution from nebular continuum emission.

Many have considered that the young ages inferred for many high redshift galaxies are suggestive of burst-like behaviour that ``outshines" older stellar populations (Lecture 2) . \citet{Endsley2024} presented a detailed analysis of line emission (albeit inferred from photometric data) that supported the view that low luminosity galaxies show weak spectral features consistent with a downturn in the star formation rate whereas their luminous, stronger line counterparts are ``up-scattered" by bursts. \citet{Gelli2024} has developed a predictive bursting model by introducing a scatter $\sigma_{UV}$ in the UV luminosity - halo mass relation (Figure~\ref{fig:bursts}). Incorporating such a scatter increases the visibility of sub-luminous systems and can go someway to reproducing the $z>10$ observations (see also \citet{Mason2023,Shen2023, Kravtsov2024}).

\begin{figure}
\center
\includegraphics[width=\textwidth]{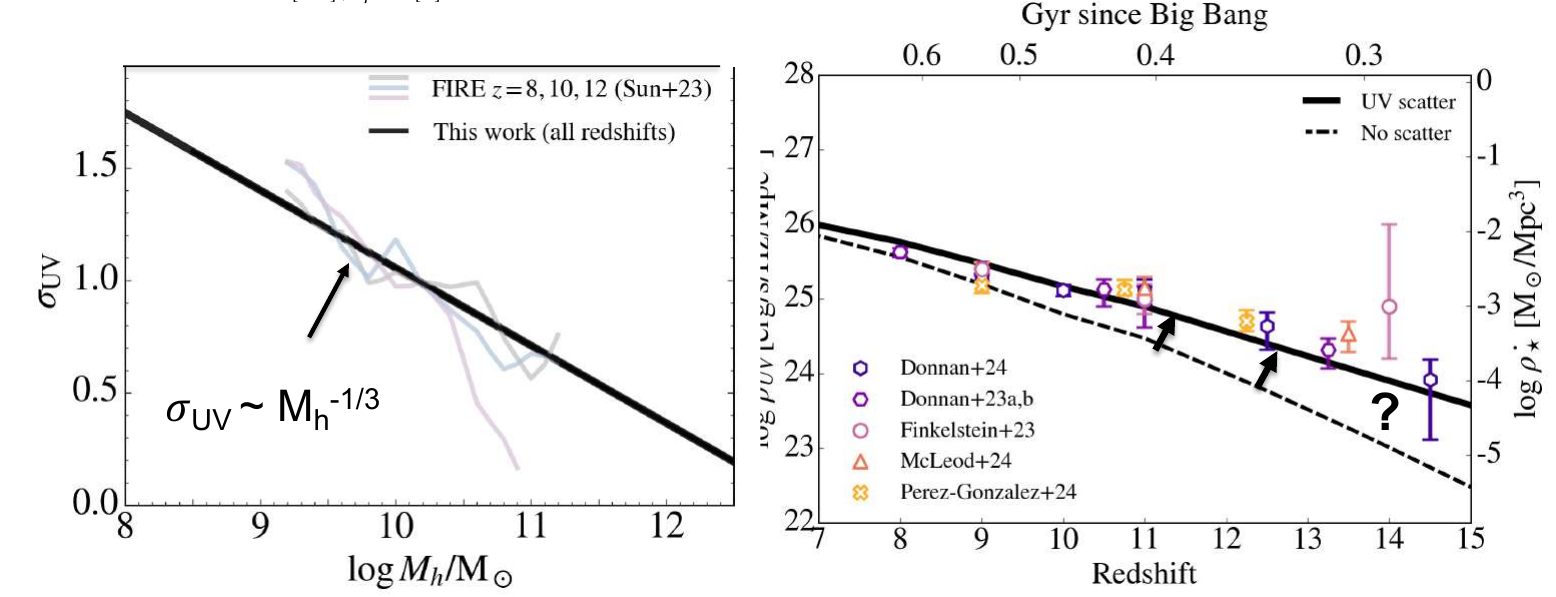}
\caption{\it Modelling how bursty star formation could explain the high luminosities of z$ > $10 galaxies as temporarily up-scattered lower luminosity systems \citep{Gelli2024}. (Left) FIRE simulations based on bursty star formation indicate the scatter $\sigma_{UV}$ in the $M_{UV}$ - $M_{halo}$ relation is larger for lower mass galaxies. (Right) Incorporating this relation goes some way to explaining the excesss luminosity but may be insufficient, on its own, to match the observations at the highest redshifts.}
\label{fig:bursts}
\end{figure}

Finally let us consider the impact of a top-heavy stellar initial mass function (IMF). There is some evidence of super-massive (100-400 $M_{\odot}$) stars right on our doorstep in the Large Magellanic cloud. \citet{Schaerer2025} have shown how their presence can provide a huge boost to both the stellar and nebular UV emission for short periods ($<$10 Myr. Their models predict UV luminosity boosts of $\times$6 and harder radiation fields with $\xi_{ion}$ increased by $\times$1.5 (Figure~\ref{fig:imf}). A by-product of super-massive stars would be the presence of high ionisation lines such as He II. While there are some examples of He II line emission, its presence is certainly not ubiquitous in z$>$10 galaxies.

\begin{figure}
\center
\includegraphics[width=0.8\textwidth]{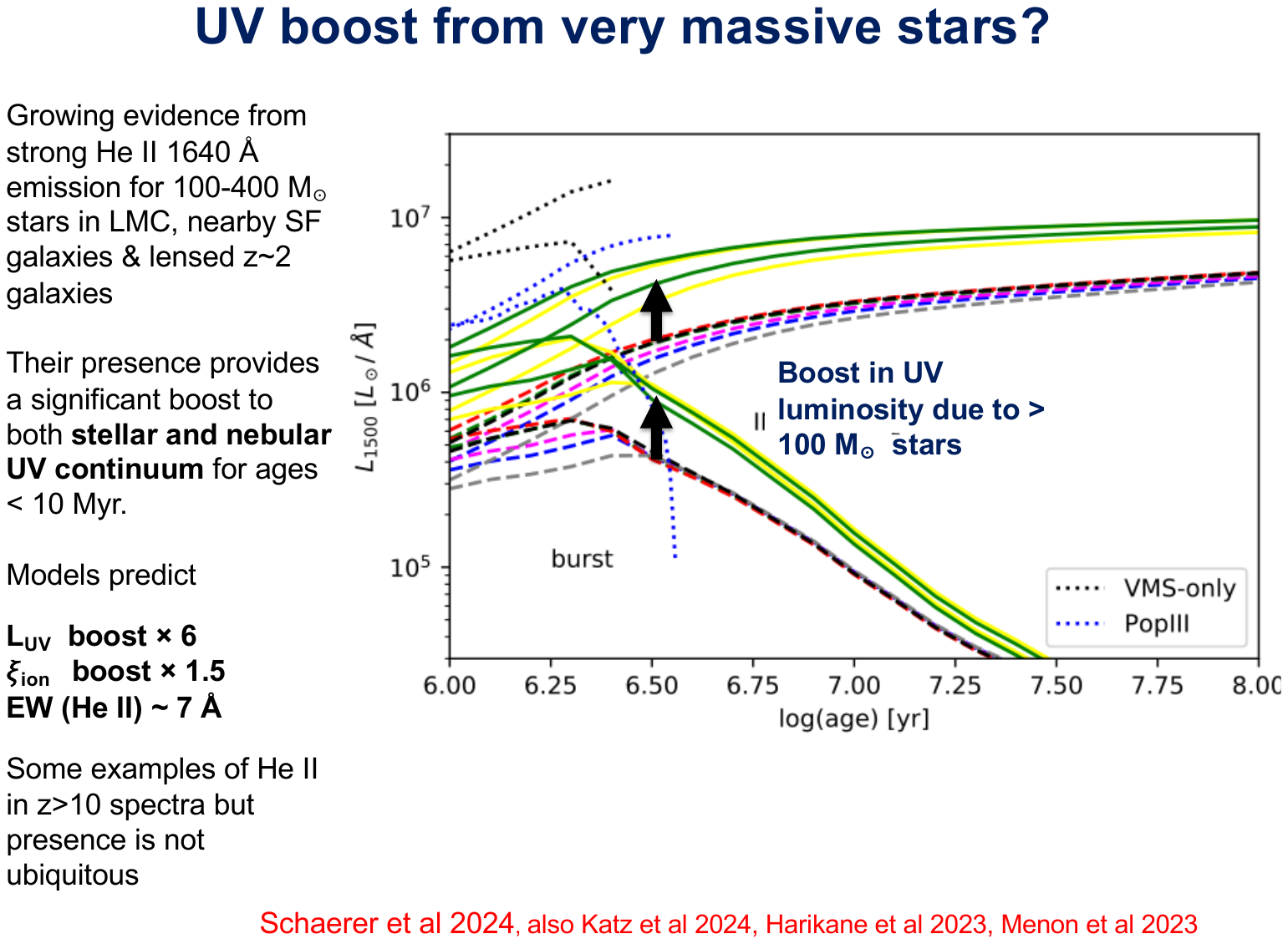}
\caption{\it The dependence of the time-dependent UV 1500 \AA\ monochromatic luminosity on various assumed stellar initial mass functions \citep{Schaerer2025}. Upper curves represent models with constant star formation and the lower ones assume a single burst. The significant boost in luminosity arising from changing the IMF from a classic Salpeter slope of $\alpha_2$=-2.35 above 0.5 $M_{\odot}$ to one with a flatter $\alpha$=2.0 slope that maximises the contribution of massive stars is shown for both star formation histories by vertical arrows for mass functions extending to 150 $M_{\odot}$ (yellow curves) and 300 $M_{\odot}$ (green ones). All models include nebular emission.}
\label{fig:imf}
\end{figure}

Some more creative ways to explore the puzzle have also been proposed. \citet{Shuntov2025} have examined the {\it clustering} of galaxies in the reionisation era since this is linked to their halo masses. Connecting the results with their UV luminosities provides a measure of the star formation efficiency for a given halo mass. They claim no significant evolution in this efficiency during the reionisation era. An independent measure of burstyness can also be gleaned by comparing estimates of star formation rates derived independently from H$\alpha$ and the UV continuum given these diagnostic probe different timescales as discussed in Lecture 2 (Perry et al, in prepn.).

So what are we to make of all these explanations? Whilst theorists often enthusiastically propose a single explanation for a puzzling observational result, experience suggests it's often a combination of physical effects. The dust-free solution is the only one that's readily testable and there seems little correlation between extreme luminosities and steep UV continuum slopes. As an example, the most luminous galaxy of all (GNz-11) has $\beta$=-2.36. Up-scattering from bursty star formation can certainly contribute to an excess star formation rate density at $z>$10 but models suggest this becomes harder at $z>12$ (Figure~\ref{fig:bursts}). Finally, top heavy IMFs are a good bet given the spectroscopic evidence for very massive stars inferred to explain super-solar N/O abundances (\citet{Cameron2023}, Lecture 2). 

One final new development might further support the bursting hypothesis. Spectroscopic follow up of candidates located in the HST pure parallel BoRG survey is also finding super-luminous examples at 7.5$< z < $8.5 (Figure~\ref{fig:borg}, \citet{Roberts-Borsani2022, Roberts-Borsani2025} suggesting the z$>$10 era is not so ``special". Of course at this later time, finding more massive and energetic sources might not be so surprising. Nonetheless, the wide area covered in the HST survey provides a more representative estimate of the bright end of the luminosity function in the reionisation era \citep{Rojas-Ruiz2025}. Mounting evidence from rest-frame optical spectra in these later examples suggests similar physical conditions to those seen at $z>10$, including periods of stochastic star-formation. If confirmed with deeper and higher resolution data, this may lead to re-examination of those explanations for the $z>$10 "blue monsters" that appeal to a brief, very early phase of galaxy assembly.

\begin{figure}
\center
\includegraphics[width=0.5\textwidth]{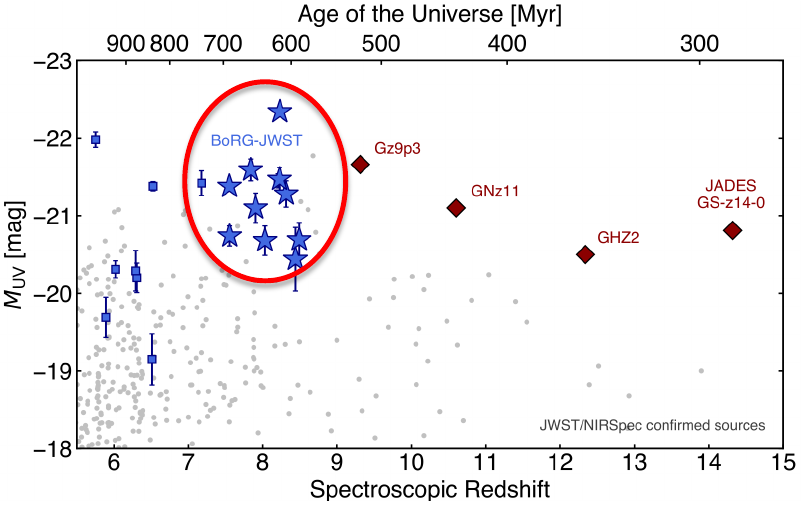}
\includegraphics[width=0.45\textwidth]{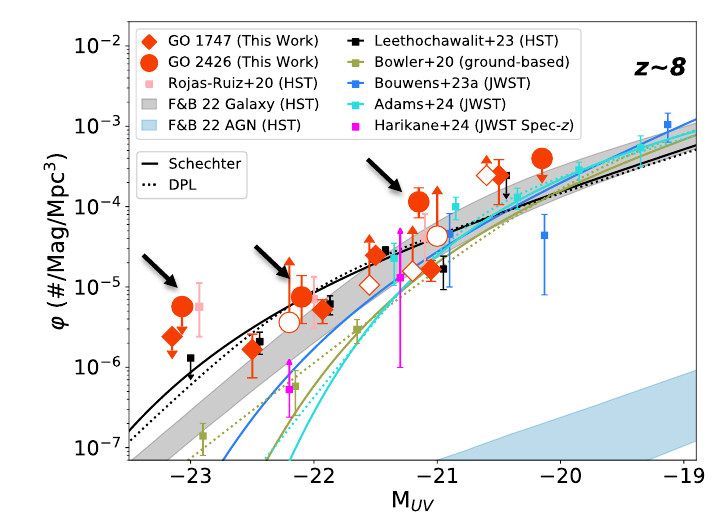}
\caption{\it Is $z>10$ the only special era? (Left) Spectroscopic follow-up of the HST pure parallel BoRG survey is finding similarly super-luminous galaxies at 7.5$ < z < $8.5 \citep{Roberts-Borsani2025}. (Right) The large original survey area ($\simeq10^3$ arcmin$^2$) and numerous sightlines which minimise cosmic variance enables a robust measure of the bright end of the luminosity function at $z\simeq$8 \citep{Rojas-Ruiz2025}.
}
\label{fig:borg}
\end{figure}

\section{Cosmic Dawn and Population III Stars}
\label{sec: 4}

How would we recognise a primordial galaxy? Valuable insight has followed decades of theoretic work predicting spectroscopic signatures for metal-free stellar populations (referred to as Population III stars, \citet{SChaerer2002,Schaerer2003}) as well as the nuclear products of pair instability supernovae arising from super-massive stars \citep{Heger2002}.

The formation of a star-forming galaxy requires that the gas cooling timescale is less than the dynamical timescale. Pop III stars would likely form from gas clouds which cooled via molecular hydrogen in halos with virial temperatures of $T_{vir}\sim$10$^3$ K. As atomic cooling is ineffective in such situations, the simulations suggest clouds will fragment with scales larger than those seen in present-day galaxies, thereby producing stars with masses $>$ 100 $M_{\odot}$. Such hot stars evolve rapidly in their gaseous environments, producing a dominant nebular continuum which weakens line emission. For these metal-free systems, the most prominent signature expected is an intense He II recombination line at 1640 \AA\ . Furthermore, stars in the mass range 140 $ <M/M_{\odot} < $ 250 are capable of exploding via a distinct pair instability supernova (PISN) mechanism. For these super-massive stars the core temperature is sufficient to enable pair exchanges between photons and electrons viz: 

\begin{equation}
    \gamma \rightleftharpoons e^+ + e^-
\end{equation}

This exchange reduces the core pressure enabling rapid collapse. As the lifetime of such massive stars is very short ($\simeq$2-5 Myr), the rapid decline in He II visibility represents a major challenge in recognising such primordial systems (Figure~\ref{fig:pop3_schaerer}). However, such PISN may leave characteristic metal signatures such as Fe excess abundance seen in z$>$10 galaxies \citep{Nakane2025}.

\begin{figure}
\center
\includegraphics[width=\textwidth]{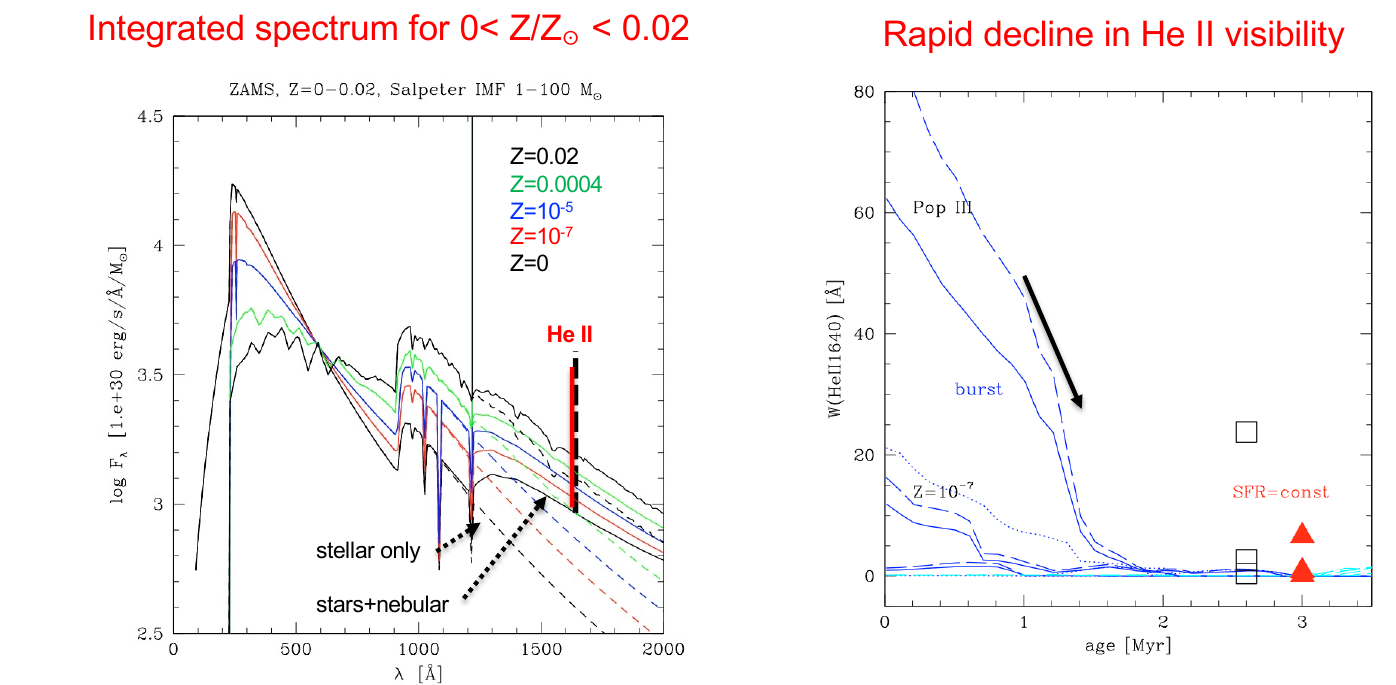}
\caption{\it Pioneering metal-free stellar population synthesis calculations from \citet{Schaerer2003} which have stood the test of time remarkably well. (Left) Integrated spectra for a range of metallicities from zero to 0.02 solar, demonstrating for the metal-free Z=0 case the relative contributions of stars and nebular continuum. The prominent emission line of He II 1640 \AA\ emerges as the most likely detectable feature for a Population III system.  (Right) The rapid decline in He II visibility over 2 Myr will make it challenging to identify such primordial systems.
}
\label{fig:pop3_schaerer}
\end{figure}

The major factor that governs the transition to enriched Pop II stars is a lower characteristic stellar mass $M_c$ associated with gas cooling with metals and dust as well as the redshift-dependent temperature of the cosmic microwave background $T_{CMB} \simeq 2.7 (1 + z)$ deg K which acts as a temperature floor, e.g. T=60 K at z$\simeq$25. Once $M_c$ is reached, cooling cannot continue and Jeans' instability leads to a single star or cluster of that mass. The timescale of this transition depends on the supernova enrichment history and complexities due to blast wave physics, cosmological accretion, halo mergers and radiative feedback which can suppress H$_2$ cooling. Clearly hydrodynamical simulations are required to address these many issues. As an early example, \citet{Wise2014} studied the metallicity evolution for 10$^6 M_{\odot}$ halos over 15 $< z <$7 and claimed a critical metallicity of $Z_{crit}\sim 10^{-3}$ for enabling lower mass star formation. The work demonstrated a halo might reach this threshold simply after a single pair instability supernova (PISN). The complete transition from Pop III to Pop II then follows over several hundred Myr. Pockets of Pop III stars can co-exist with Pop II populations throughout the reionisation era and possibly even to lower redshifts. The topic is well reviewed by \citet{Bromm2004, Bromm2011} and most recently by \citet{Klessen2023}.

Following the launch of JWST, a number of articles have focused specifically on how to recognise Pop III systems and here I focus on the zoom simulations by \citet{Katz2023} for a 3. 10$^8 M_{\odot}$ halo with sub-parsec resolution using on-the-fly radiative transfer, non-equilibrium chemistry and cooling formalisms (Figure~\ref{fig:pop3_katz}, see also \citet{Nakajima2022, Zier2025, Lecroq2025, Rusta2025}). As can be seen, metal enrichment in C, N, O and Fe proceeds almost instantaneously after the first PISN.  Pop III star formation can also co-exist with Pop II over several hundred Myr (down to $z\simeq$10 in this particular halo). We can use these simulations to check the credibility of several JWST claims for extremely metal-poor systems.

\begin{figure}
\center
\includegraphics[width=\textwidth]{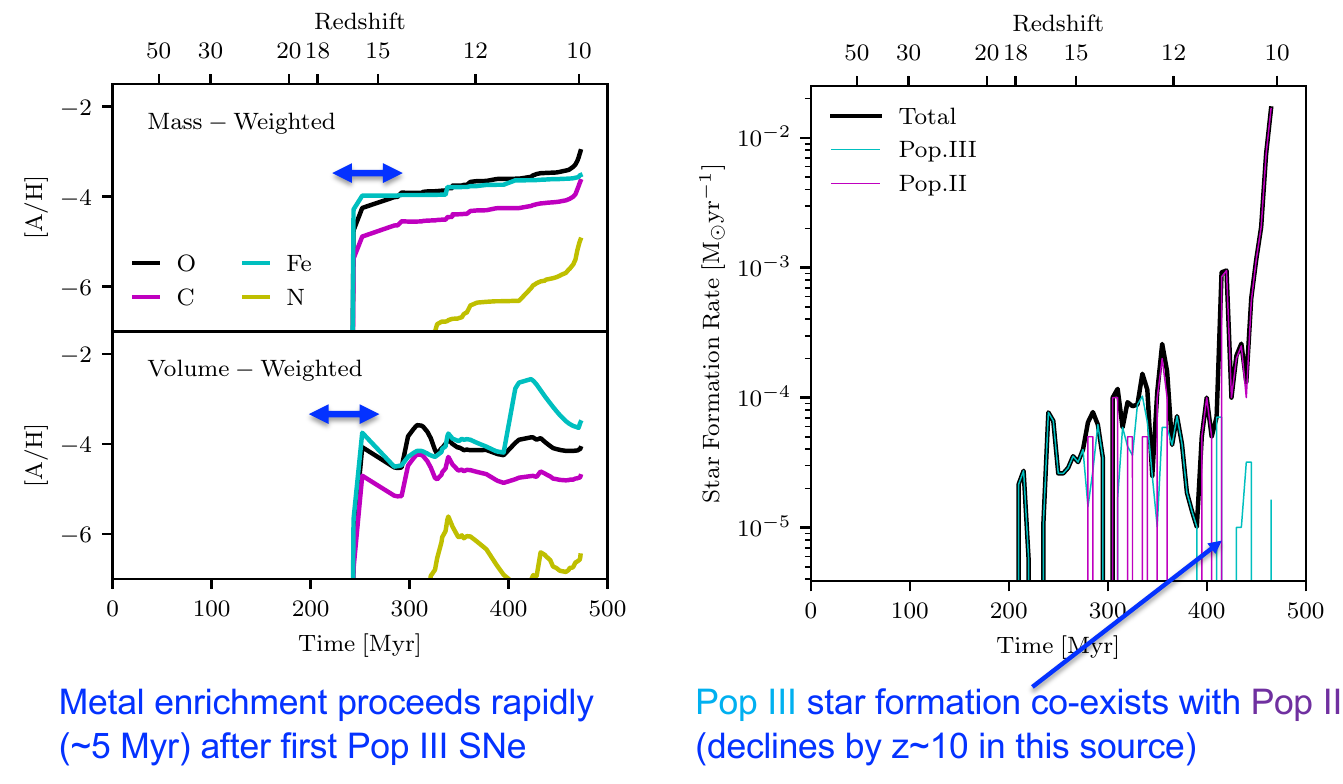}
\caption{\it Hydrodynamical simulation of the Pop III to Pop II transition incorporating radiative transfer, non-equilibrium chemistry and cooling for a 3. 10$^8 M_{\odot}$ halo \citep{Katz2023}. (Left) Metal-enrichment proceeds extremely rapidly ($<$5 Myr) following the first supernovae making any discovery of genuine metal-free populations very challenging observationally. (Right) However Pop III regions can co-exist with enriched stellar populations over many 100 Myr.}
\label{fig:pop3_katz}
\end{figure}

Figure~\ref{fig:pop3_candidates} shows two claimed Pop III candidates that have received widespread discussion at the time of the Lectures. \citet{Maiolino2024a} has located an isolated He II 1640 \AA\ emitter in the vicinity of GNz=11 at z=10.6. No other lines are observed and the stellar mass of the associated population is estimated to be $\simeq 6 \times 10^5 M_{\odot}$. \citet{Vanzella2023} have used the NIRSpec integral field mode to study a highly-magnified compact source, LAP1-B at z=6.6 straddling the critical line of the Frontier Field cluster MACS0416. At this lower redshift, Balmer lines and a tentative He II 1640 \AA\ are detected and the absence of [O III] emission places an impressive upper metallicity limit of $<$ 0.3\% solar. More recent observations of this source are discussed by \citet{Nakajima2025,Visbal2025}. There are several other Pop III claims in the current literature worthy of investigation, including \citet{XWang2024, Maiolino2025,Morishita2025}. However, here I focus on the first two claims.

\begin{figure}
\center
\includegraphics[width=0.8\textwidth]{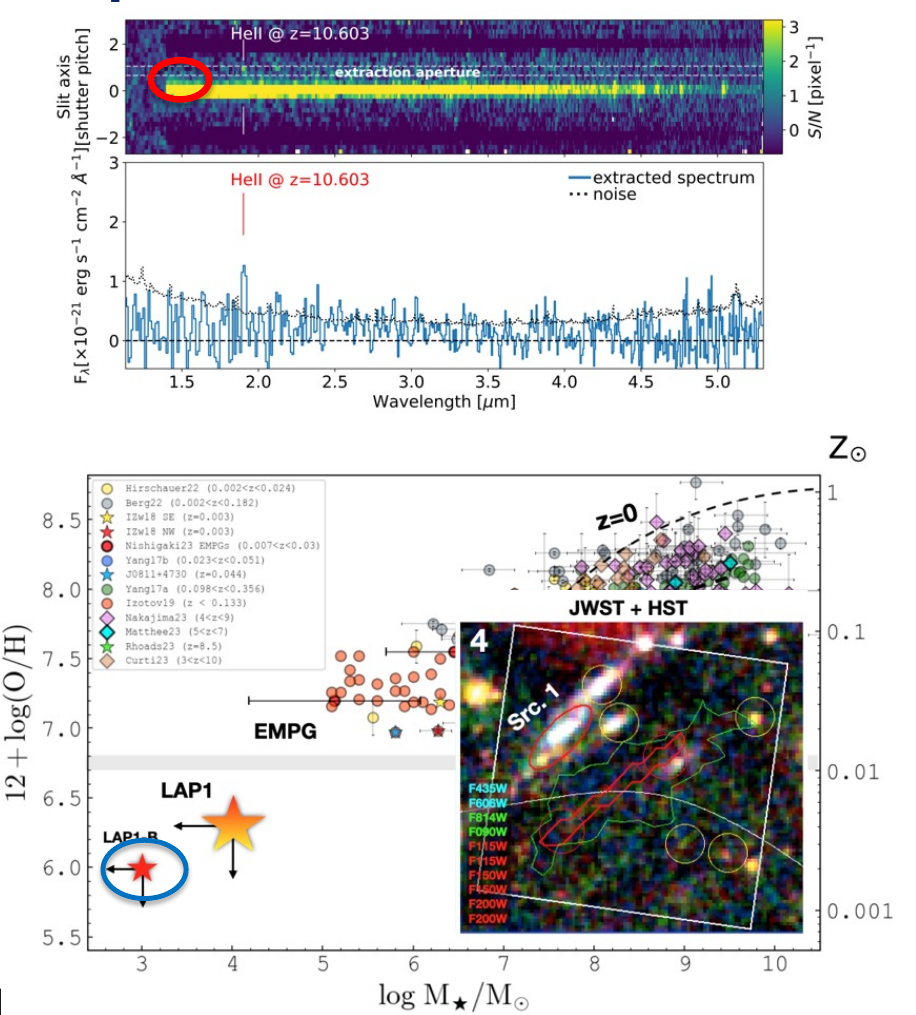}
\caption{\it Two well-discussed recent claims of promising Pop III candidates from JWST observations. (Top) An isolated emission line, ostensibly He II 1640 \AA\ alongside the spectrum of GN-z11 at a redshift z=10.6 \citep{Maiolino2024a}. The absence of any continuum suggests a rest-frame equivalent width of 170 \AA\ and no other emission lines are visible. (Bottom) NIRCam IFS data for a highly-magnified ($\mu>$100) galaxy (red elongated contour) straddling the critical line of the lensing cluster MACS0416. Lyman and Balmer emission lines are detected and a possible He II but no [O III] leading to an upper limit on the gas-phase metallicity of 0.3\% solar \citep{Vanzella2023, Nakajima2025}. }
\label{fig:pop3_candidates}
\end{figure}

The burden of proof for a Pop III claim is very high! According to the \citet{Katz2023} models, we require He II to be particularly intense in comparison to H$\alpha$ in order to capture the brief lifetime associated with such metal-free massive stars (specifically log He II/H$\alpha >$ -0.5).  Likewise, [O III] 5007 \AA , the most prominent metal line accessible at high redshift remains weak initially so any claim for metal-free halos will necessitate exquisitely deep JWST exposures. Katz et al recommend log [O III]/H$\beta <$ -1.5. Figure ~\ref{fig:katz_candidates} shows these constraints as shaded areas in the time evolution of He II/H$\alpha$ and the [O III]/H$\beta$ vs metallicity plot. Vanzella's lensed source LAP1-B certainly fulfills the He II criterion but deeper data would be needed to match the [O III]/H$\beta$ constraint. At the time of writing the sole detection of He II in the object associated with GNz-11 could be promising in terms of a high He II/Balmer line ratio, but deeper data is required to place a meaningful upper limit on O/H. The candidate proposed by \citet{Morishita2025} is also very promising but at lower redshift.

\begin{figure}
\center
\includegraphics[width=\textwidth]{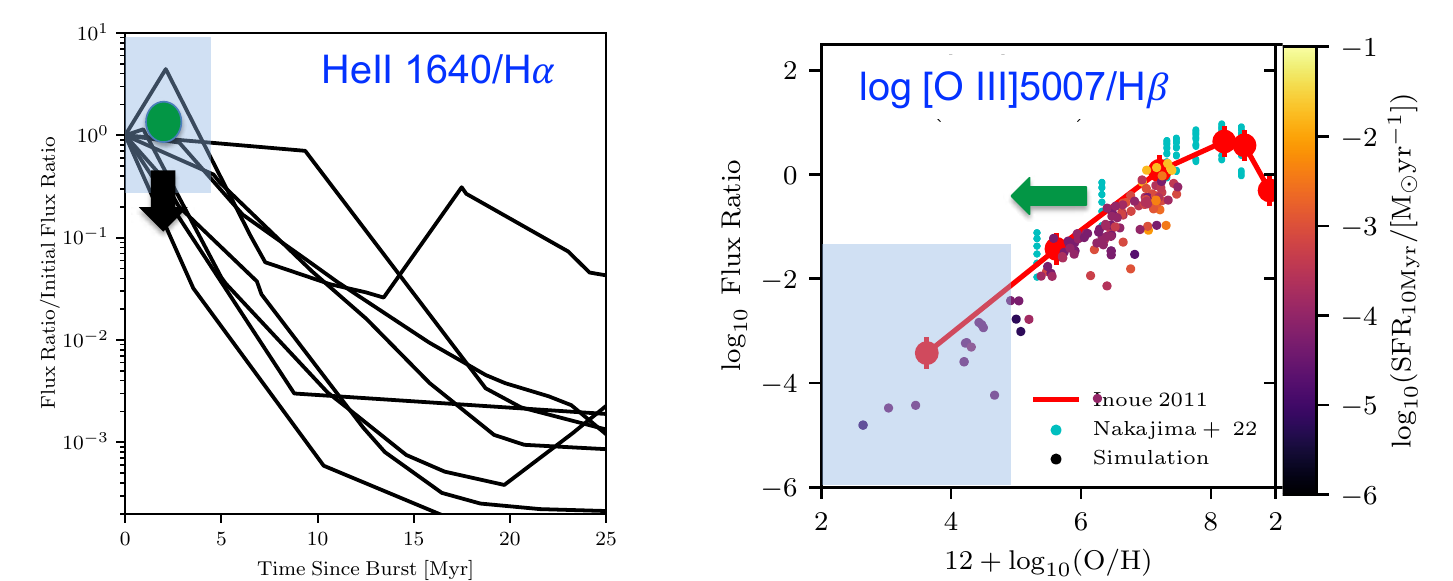}
\caption{\it Comparison of the properties of the two claimed Pop III candidates discussed in Figure~\ref{fig:pop3_candidates} in the context of two Pop III criteria proposed by \citet{Katz2023}.
In both panels these criteria are indicated by the shaded blue region where convincing Pop III candidates should lie. Green symbols refer to the observational data for the lensed source proposed by \citet{Vanzella2023}, black symbols to the He II detection of \citet{Maiolino2024a}. (Left) The prominence of He II should approximate that of H$\alpha$ as the ratio decays rapidly after an initial metal-free burst. Vanzella et al's He II identification is tentative but would be within range; Maiolino et al see no hydrogen emission so in principle their source is also in range. (Right) [O III] / H$\beta$ should be lower than 1\% for a promising case. Again Vanzella et al are close with such a weak [OIII] but there is no meaningful constraint from Maiolino et al. }
\label{fig:katz_candidates}
\end{figure}

So what are the prospects for locating `cosmic dawn' with JWST? I'm of the opinion that there may never be an unambiguous discovery of a chemically-pristine galaxy. The time window for metal-free stellar population is $<$5 Myr so catching an example would be extremely fortuitous. We may find examples embedded in chemically-enriched regions at lower redshift but I suspect this is not the "Holy Grail" that JWST was hoped to deliver! More likely we may deduce the birth of starlight indirectly from tracing the continued decline of the cosmic star formation rate density and gas phase metallicity from more extensive surveys. Right now the absence of convincing $z>15$ candidates might optimistically suggest we can anticipate achieving such a statistical result during the lifetime of JWST.

\section{The Promise of 21cm Surveys}
\label{sec: 5}

Despite considerable technical challenges, charting the redshift-dependent 21 cm signal of neutral hydrogen with radio interferometers offers the exciting prospect of complementing the rapid progress being made in studies of the early universe with JWST. \cite{Liu2020} document an exciting array of ``pathfinder" facilities prior to the deployment of the Square Kilometre Array (SKA) of which the low frequency component under construction in Western Australia is the most relevant for high redshift surveys. The current facilities of interest include the Giant Metrewave Radio Telescope (GMRT), a 45 metre dish in Pune, India surveying at frequencies above 50 MhZ, the Murchison Widefield Array (MWA), a 3km interferometer in Western Australia ($>$70 MHz), the Low Frequency Array (LOFAR), a widely distributed European interferometer centered on the Netherlands (1-=90 MHz) and the Hydrogen Epoch of Reionisation Array (HERA), a 3 km interferometer in South Africa ($>$50 MHz). At the redshifts of interest for cosmic dawn (15$ < z < $25 corresponding to 21cm transitions at 100 to 60 MHZ), each of these pathfinders is attempting to overcome foreground signals and thereby demonstrate the potential of SKA due to begin science operations later this decade (2028 at the time of writing).

\begin{figure}
\center
\includegraphics[width=\textwidth]{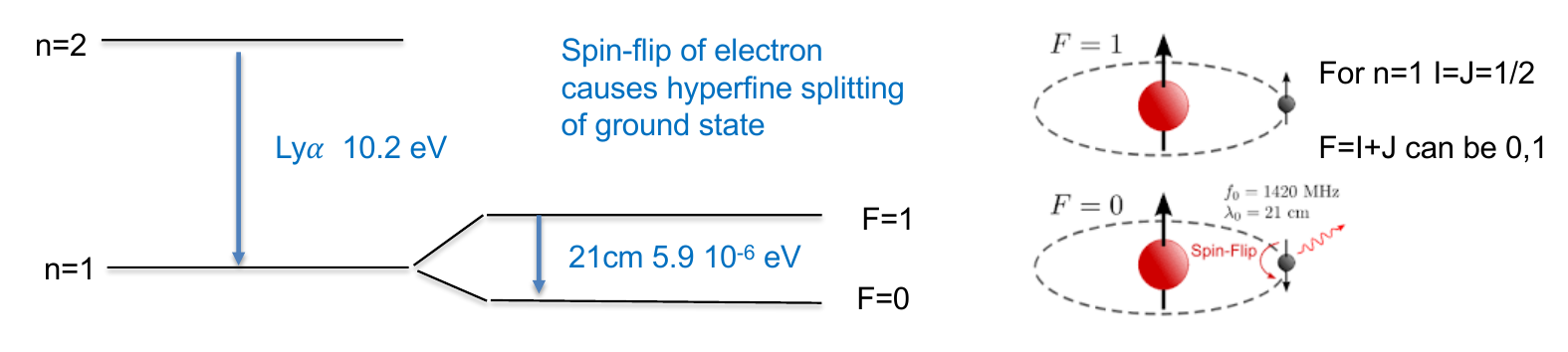}
\caption{\it Physical origin of the 21cm of neutral hydrogen via the hyperfine splitting of the n=1 ground state via the two oppositing spin states of the electron, corresponding to a total angular momentum quantum number F=I+J=0 or 1. Compared to Ly$\alpha$, the 21cm line is a sensitive measure of the neutral hydrogen throughout the reionisation era and beyond.}
\label{fig:21cm}
\end{figure}

Radio wavelength studies of the 21cm line of neutral hydrogen have a number of advantages over those of Lyman $\alpha$. The latter is a resonant line and even a small neutral fraction by volume ($x_{HI}\simeq 10^{-3}$) renders the IGM opaque to detection. Prior to the onset of reionisation, the temperature of the IGM $T_{IGM}$ is too low to collisionally excite the n=2 level of hydrogen so we must use a weaker line.  Predicted by \citet{vandeHulst1945} following a suggestion by Jan Oort, the 21cm line was first detected (from the window of a Harvard office!) in 1951 \citep{Ewen1951}. The 21cm transition arises from the spin flip of the electron in the hydrogen atom causing a hyperfine splitting of the ground state with an energy gap of 5.9 10$^{-6}$ eV. Figure~\ref{fig:21cm} illustrates some of the key characteristics of this remarkable atomic transition.

Radio astronomers have their own nomenclature for studies of this line. The {\it spin temperature} $T_s$ is defined via

\begin{equation}
    n_{F=1}/n_{F=0} \propto exp (- T_{\ast}/T_{S})  
\end{equation}

where $T_S = (h \nu_{21} / k) = 68 mK$

The associated optical depth $\tau_{21}$ is only $\simeq$1\% so the line can readily be traced even in a fully neutral IGM, a major advantage over Ly$\alpha$.

Radio astronomers also define the {\it brightness temperature} $T_B$ according to the line intensity $I_{\nu}$ in the Rayleigh-Jeans limit (where the 21cm frequency is less than the peak of the CMB black body emission at that redshift), whence:

\begin{equation}
    I_{\nu} = 2 k T_{B} \nu^3 c^{-2}
\end{equation}

Note that $T_S$ can deviate from the kinetic temperature of the gas due to the presence of radiation.

In the short term, the main utility of SKA 21cm studies will be in tomographic studies of the neutral and ionised regions during the later stages of reionisation. This relates primarily to the topics of Lecture 1 and will not be expanded further here, except to say a particularly promising technique will be the cross-correlation between fluctuations in the 21cm emission and the spatial distribution of various sources (e.g. galaxies and AGN) \citep{Gagnon-Hartman2025}. This will offer a promising causal connection between sources of ionising photons and the nature of the IGM. 

\begin{figure}
\center
\includegraphics[width=\textwidth]{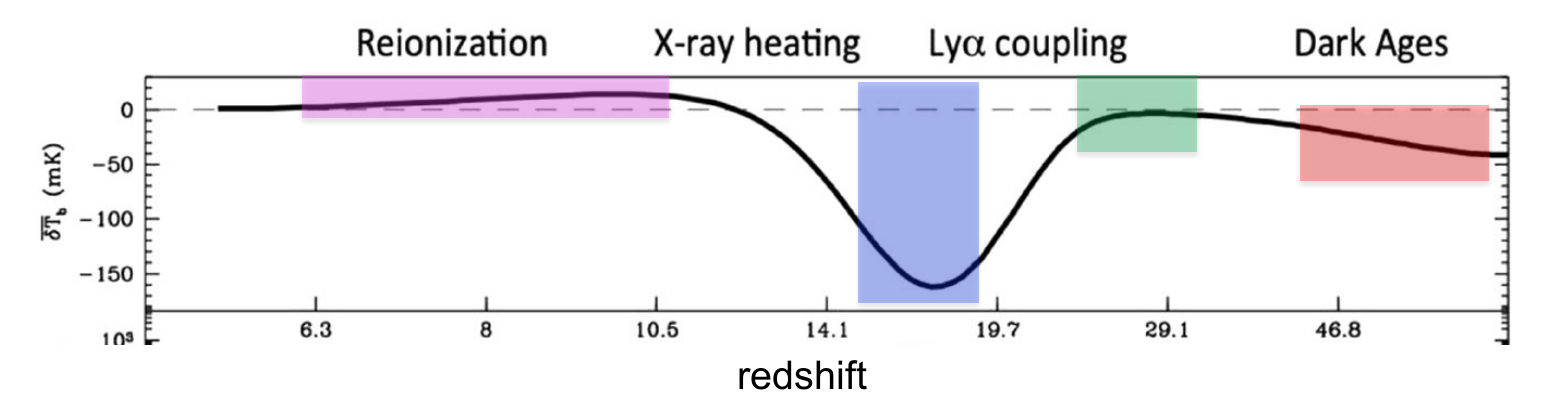}
\caption{\it Evolution of the differential brightness temperature $\delta\,T_B$ of the 21cm line of neutral hydrogen with respect to the background microwave background during the transition from the dark ages through cosmic dawn to the reionisation era. A negative differential indicates the line is seen in absorption and a positive one indicates an emission line. Of particular significance is the strong absorption which occurs when Ly$\alpha$ photons propagate from gas clouds heated by young stars (see Figure~\ref{fig:wf_effect}). This offers a promising marker of `cosmic dawn' \citep{Koopmans2021}.}
\label{fig:21cm_evol}
\end{figure}

However, for the present topic of `cosmic dawn', Figure~\ref{fig:21cm_evol} illustrates how the brightness temperature $T_B$ is expected to evolve during this key era (see \citet{Pritchard2012, Koopmans2021} for details). Starting from the highest redshifts in the dark ages, the cosmic expansion cools the gas at a rate ($T_{gas} \propto a(t)^{-3}$) faster than the background radiation ($T_{rad} \propto a(t)^{-1}$), while atomic collisions maintain a spin temperate $T_S$ equal to the gas temperature. In this case the 21cm line is seen in absorption against the CMB. As the universe expands, the decreasing gas density renders atomic collisions ineffective so the spin temperature equals the radiation temperature and there is no 21cm signal. When starlight emerges, the IGM is heated, so the spin temperature matches the gas temperature and the 21cm signal is seen once again in absorption. Eventually $T_S$ exceeds $T_{rad}$ as the radiation continues to cool. This leads to 21cm emission throughout the reionisation era.

\begin{figure}
\center
\includegraphics[width=0.3\textwidth]{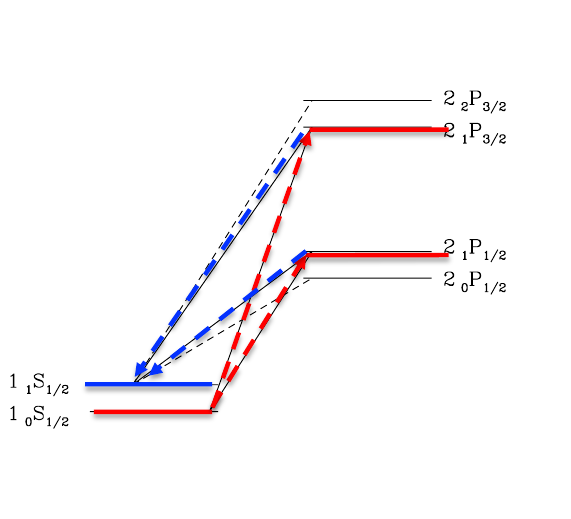}
\includegraphics[width=0.65\textwidth]{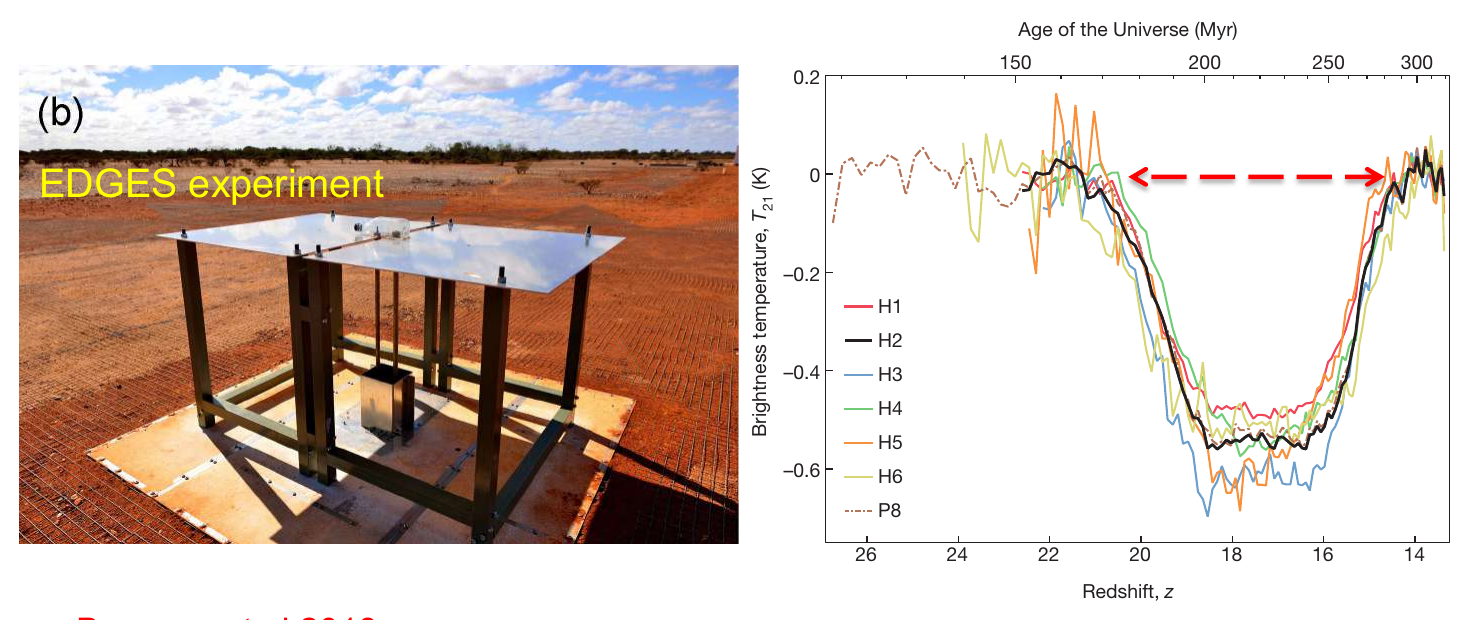}
\caption{\it Importance of the Wouthuysen-Field effect as a potential 21cm absorption signature of cosmic dawn. (Left) Resonant scattering of Ly$\alpha$ photons from the first stars can excite hydrogen from a hyperfine n=1 state to an equivalent n=2 state. The selection rule does not permit a transition from F=0 to 0, but from either n=2 level a Ly$\alpha$ photon can be emitted and return the atom into the other hyperfine state and thus induce the spin-flip necessary for a 21cm absorption. (Centre and Right) Claimed detection of all-sky 21cm absorption at z$\simeq$18 from the EDGES experiment \citep{Bowman2018}. This potentially exciting result may be consistent with early JWST galaxy data but has not been reproduced by other 21cm experiments.}
\label{fig:wf_effect}
\end{figure}

Although the spin flip that produces either 21cm emission or absorption is initially produced by collisional processes (e.g. H+H, H+p, H+e), at `cosmic dawn' Lyman $\alpha$ photons can induce an additional process first evaluated independently by Siegfried Wouthuysen \citep{Wouthuysen1952} and George Field \citep{Field1958}. This remarkable effect illustrated in Figure~\ref{fig:wf_effect} offers a distinct indicator of `cosmic dawn'. Resonant scattering of Ly$\alpha$ can excite hydrogen from a hyperfine n=1 ground state to a hyperfine n=2 state. However, while the dipole selection rule permits quantum changes $\Delta F$= 0,1 the specific transition F= 0 $\rightarrow$ 0 is not permitted. However, from either n=2 level a Lyman $\alpha$ photon can be emitted, and by the same selection rule, return the atom into the other hyperfine state. In this way the arrival of Lyman $\alpha$ photons from hydrogen heated by the first stars can induce the spin flip necessary for a 21 cm absorption signal against the CMB.

Since this would be a generic signal at a given redshift across the sky, there have been various attempts to detect it prior to SKA. The most widely-discussed result is a claimed detected at 78 MHz by the EDGES all-sky receiver in Western Australia (\citet{Bowman2018}, Figure~\ref{fig:wf_effect}). Skeptics have argued this absorption signal over 15$ < z < $20 is surprisingly deep and may be an artefact of incorrect subtraction of foreground synchrotron and free-free emission (c.f. \citet{Hills2018}). An independent radio telescope, SARAS3 in India, has failed to reproduce the EDGES signal \citep{Singh2022}. Despite these controversies and challenges in removing foregrounds, we can be optimistic that this is a second, independent, probe of the natural of the IGM at early cosmic times which is sensitive to the birth of starlight through a remarkable coupling of Lyman $\alpha$ and the 21 cm transition. It is encouraging that the lifetime of JWST will overlap with almost a decade of SKA observations so synergistic work will hopefully pinpoint cosmic dawn.

\section{Summary}
\label{sec: 6}

In this lecture we discussed the emerging census of galaxies at redshifts z$>$10. The frontier has been extended from z=11.1 with HST (415 Myr after the Big Bang) to z=14.4 (290 Myr). Although a mere 125 Myr earlier in cosmic time, both the increased data content and results have been impressive. Specifically

\begin{enumerate}
    \item {Prior to JWST, we only had one spectroscopically confirmed z$>$10 galaxy (GN-z11} whereas at the time of writing we have over 40. Early exploration of the the 15$< z <$ 20 redshift range has failed to find credible candidates which may indicate a tantalising downturn in the population towards `cosmic dawn'.
    \item {The abundance and particularly the luminosity of these sources exceeed predictions from models based on the assumption of constant star-formation efficiency in the standard cosmological model. Although the origin of this excess is not yet agreed, some combination of burst-like activity and the presence of super-massive stars in metal-poor halos are promising explanations.}
\end{enumerate}

\medskip

We also discussed the practicality of recognising a primordial galaxy containing purely metal-free Population III stars. 

\medskip

\begin{enumerate}
    \item {Simulations suggest a genuine Pop III stellar system occupies a very brief phase, lasting $<$3-5 Myr, prior to enrichment from pair instability and Type II supernovae. According to contemporary models, the burden of proof in locating one with the expected ratios of He II, Balmer and oxygen lines will be observationally very challenging. Pop III systems can also be embedded in otherwise enriched galaxies through the reionisation era.}
    \item {The epoch of `cosmic dawn' may more easily be identified statistically via comprehensive measures of the declining gas-phase metallicities or careful ``age-dating" of lower redshift systems. We also discussed the Wouthuysen-Field effect whereby Lyman $\alpha$ photons from gas clouds heated by the first generation of stars can induce an all-sky 21cm absorption signature against the microwave background. SKA-Low and related pathfinders may soon be able to confirm the redshift of this feature.}
\end{enumerate}

\medskip

Finally, as a parting thought, will our work on galaxy formation and reionisation be complete with an estimate of when cosmic dawn occurred? Possibly not if individual super-massive stars could exist at yet higher redshifts leading to an earlier ionised era. Conceivably the universe may then briefly recombine and return to darkness. Such ``two-stage" reionisation histories were popular in the early 2000s and may not be ruled out by JWST during its lifetime. Perhaps we can then continue a search for such a brief period of ionised gas with SKA-Low in the redshift range 20$ < z < $30?

\section{Recommended Reading on the Redshift Frontier}
\label{sec: 7}

\noindent{\bf Classic Articles}

\begin{enumerate}
\item{\cite{Partridge1967} - pioneering paper discussing visibility of Ly$\alpha$ in primaeval galaxies (PGs)}
\item{\cite{Baron1987} - primaeval galaxies in CDM cosmology}
\item{\cite{Pritchet1994} - historic review of attempts to find PGs}
\item{\cite{Bromm2011} - theory of PGs}
\item{\cite{Klessen2023} - theory of Pop III stars}
\item{\cite{Pritchard2012} - prospect of 21cm surveys}
\end{enumerate}

\noindent{\bf JWST-relevant Predictions}

\begin{enumerate}
\item{\cite{Schaerer2003} - visionary article on recognising Pop III stars}
\item{\cite{Inoue2011} - visibility of metal-free galaxies}
\item{\cite{Wise2014} - early numerical simulation of PGs}
\item{\cite{Katz2023} - recent simulations in context of JWST data}
\item{\cite{Koopmans2021} - 21cm observations at high redshift}
\end{enumerate}

\noindent{\bf Observations}

\begin{enumerate}
\item{\cite{Oesch2016} - pioneering HST redshift measurement for GN-z11}
\item{\cite{Bunker2023} - JWST spectrum of GN-z11}
\item{\cite{Donnan2024} - wide-field census of star-forming galaxies beyond z=10}
\item{\cite{Robertson2024} - deep field census of star-forming galaxies beyond z=11.5}
\item{\cite{Carniani2024} - discussion of super-luminous galaxies beyond z=10}
\item{\cite{Vanzella2023} - extremely low metallicity lensing system}
\item{\cite{Maiolino2024a} - claimed Pop III candidate associated with GN-z11}

\end{enumerate}

\section{Epilogue}
\label{sec: 8}

It probably escaped the attention of most Saas-Fee participants that this is the second time I have given a course of lectures at Saas-Fee. I gave a series of nine(!) lectures entitled {\it Observations of the High Redshift Universe} in 2006. The text of those lectures can be found on the arXiv \citep{Ellis2008} and are, of course, published in the relevant Saas-Fee volume 36 entitled {\it First Light in the Universe}. I thought it would be amusing (and humiliating for me...) to compare what was written about future prospects in 2006 with today's reality!

\medskip

Here are two examples:

\medskip

\noindent{\bf On cosmic reionisation:} At the time the most distant galaxy was IOK-1 at z=6.96 \citep{Iye2006} and so observationally we hadn't truly entered the reionisation era. Accordingly I wrote: 

\medskip

{\it It seems an impossible task to give an authoritative observational account of how to probe this era. So many issues are complete imponderables! When did reionisation occur? Was it a gradual event made possible by a complex time sequence of sources, or was there a spectacular synchronised moment? Can we conceive of an initial event, followed by recombination and a second phase? What were the sources responsible? And what is the precise process by which photons escape the sources.}  

Reading this 19 years later, I think we can all agree we have made remarkable progress with several hundred spectroscopically-studied galaxies in the reionisation era from 7 $< z <$14 (an era completely uncharted in 2006) which has given us a fairly well-defined history of reionisation although the escape fraction challenge remains. We've also encountered new puzzles, including the presence of what appear to be supermassive black holes at early epochs and luminous galaxies at $z>10$. The difference is the exquisite observational and computational tools we have now to make further progress.

\medskip

\noindent{\bf On future prospects:} I wrote the following embarrassing prophecy. 

\medskip

{\it I take out my crystal ball and consider the likely progress we can expect with current and near-term facilities including JWST (due to be launched in 2013) and a new generation of extremely large telescopes..due to be operational from 2016. The exciting redshift range 7 $ < z < $12 will be the province of improved drop-out searches using the (near-infrared) instrument WFC3 to be installed on HST in 2008-9. A major stumbling block, even at z=5-6, is efficient spectroscopic follow-up. Candidates may be found in abundance but how will they be confirmed? It is quite likely that, by 2013, the redshift range containing the earliest galactic sources, estimated at present to be 10 $ < z < $ 20, will have been refined sufficiently by instruments on our existing 8-10 metre class telescopes. JWST and future ELTs will focus on..the chemical maturity of the most luminous sources found at hight redshift.}  

Although I was pretty good on predicting we'd reach z=11 with HST's WFC3, clearly I was over-optimistic by at least a decade with JWST and the US-based ELTs. It may seem depressing to some that what's written about the future in 2006 is almost as apt in some respects today, but I think we have renewed optimism now having seen the spectacular quality of the JWST data and the astonishing progress made in only 2.5 years.

\bibliographystyle{Harvard}
\bibliography{reference}

\end{document}